\documentclass[aps,pra,twocolumn,superscriptaddress,10pt,article,nofootinbib,showpacs]{revtex4-1}

\usepackage{amsfonts}
\usepackage{amsmath}
\usepackage{amsthm}
\usepackage{braket}
\usepackage{graphicx}
\usepackage{amssymb}
\usepackage{dsfont}
\usepackage{mathrsfs}
\usepackage{sidecap}
\usepackage[T1]{fontenc}
\usepackage[utf8]{inputenc}
\usepackage{float}
\usepackage{eufrak}
\usepackage{leftidx}
\usepackage{mathtools}
\usepackage[usenames,dvipsnames]{color}
\usepackage[normalem]{ulem}
\usepackage[hidelinks]{hyperref}

\definecolor{blue(munsell)}{rgb}{0.0, 0.5, 0.69}

\begin{document}

\title{Three-dimensional superconductors with hybrid higher order topology}

\author{Nick Bultinck}
\affiliation{Department of Physics, Princeton University, Princeton, NJ 08544, USA}
\author{B. Andrei Bernevig}
\affiliation{Department of Physics, Princeton University, Princeton, NJ 08544, USA}
\affiliation{Physics Department, Freie Universitat Berlin, Arnimallee 14, 14195 Berlin, Germany}
\affiliation{Max Planck Institute of Microstructure Physics, 06120 Halle, Germany}
\author{Michael P. Zaletel}
\affiliation{Department of Physics, University of California, Berkeley, California 94720, USA}
\affiliation{Department of Physics, Princeton University, Princeton, NJ 08544, USA}

\begin{abstract}
We consider three dimensional superconductors in class DIII with a four-fold rotation axis and inversion symmetry. It is shown that such systems can exhibit higher order topology with helical Majorana hinge modes. In the case of even-parity superconductors we show that higher order topological superconductors can be obtained by adding a small pairing with the appropriate $C_4$ symmetry implementation to a topological insulator. We also show that a hybrid case is possible, where Majorana surface cones resulting from non-trivial strong topology coexist with helical hinge modes. We propose a bulk invariant detecting this hybrid scenario, and numerically analyse a tight binding model exhibiting both Majorana cones and hinge modes.
\end{abstract}

\maketitle

Since the discovery of the quantum spin Hall effect \cite{qsh,HgTe} tremendous progress has been made in our understanding of topological quantum phases of matter. A systematic theoretical understanding of topological band structures for both insulators and BdG superconductors with time reversal and particle-hole symmetry has been obtained in every dimension \cite{kitaev2,ryu,ryu2}. The common physical feature of these topological band structures is that they have gapless boundary states, which cannot be realized  as independent local lattice systems, i.e. without the presence of the higher dimensional bulk. A fruitful interplay with experiment has resulted in the prediction and discovery of many materials realizing these topologial phases \cite{HgTe,Molenkamp,Hsieh,Xia,chu,hor,park,Roth,LiuHughes,Knez,Knez2,QianXiaofeng,CavaHasan,HorCava,Roushan,STM,Sasagawa,Seo,Ong,Okada,QiKun}. 

The original periodic table of topological phases places the band insulators and BdG superconductors in ten Altland-Zirnbauer symmetry classes \cite{az}, based on their properties related to time reversal, particle-hole and chiral symmetry. However, it has been shown that also lattice symmetries can play a decisive role in the formation of topological band structures, leading to so-called topological crystalline insulators and superconductors \cite{fucrystalline,fureview,hsin,ProdanBernevig,VishwanathYi,TurnerMong,mong,FangGilbertBernevig,Duan,FangChen,Matthew,VanLeeuwen,FangFu,yingchun,slager,Hourglass}. By now, many crystalline topological insulators have also been observed in experiment \cite{Dziawa,Tanaka,Xu,Wei,sato,MaYi}. The crystalline topological phases exhibit gapless boundary states, provided that the boundary surface respects the spatial symmetry protecting the phase.

Recently, it was realized that crystalline topological phases can have gapless modes not only the boundary of a sample, but also on the corners or the hinges. For example, in Refs. \cite{quadrupole} the concept of a quadrupole model was introduced, where mirror symmetries protect fractional charges on the corners of the sample and a boundary polarization. In Ref. \cite{hughes}, it was shown how a 3D crystalline phase can exhibit chiral modes on the hinges of the sample. Topological phases that exhibit such fractionally charged or gapless modes on corners or hinges were dubbed `higher order topological phases' \cite{hoti}. Higher order topological phases have also been discussed in superconducting systems, where they for example give rise to Majorana states bound to the corners \cite{teo,bth,zhu}. Recently, higher order topological insulators were proposed experimentally in Refs. \cite{bismuth,SerraGarcia,Peterson,huber,thomale,wieder}.

Developing a complete theoretical understanding of crystalline and higher order topological band structures is currently a very active line of research \cite{TQC,cano,bradlyn,po2,watanabe,Tiantian}. Especially in the case of insulating systems \cite{Alexandradinata,Gilbert,Po,Fang2,song,xue,benalcazarli} and two-fold spatial symmetries  \cite{Shiozaki,langbehn,Shapourian,Geier,Khalaf} substantial progress has been made. In Ref. \cite{fang}, a general subclass of rotation symmetric crystalline topological superconductors exhibiting edge states in two and three dimensions was discussed. 

The study of crystalline phases has also been extended to interacting systems. An intuitive picture of crystalline phases as the stacking of lower-dimensional strong topological phases was put forward in Refs. \cite{hermele1,hermele2,Cheng}. The stacking picture provides a physical interpretation for the proposed classification of interacting crystalline phases in bosonic systems of Ref. \cite{else}, where the gapless surface, hinge or corner modes correspond to the boundary modes of the lower dimensional stacked systems. Recently, explicit spin models exhibiting corner modes were constructed \cite{devakul,dubinkin}. See also the recent paper Ref. \cite{Litinski}, where the effect of interactions on superconducting higher order topological phases is discussed.

In this work we focus on time reversal symmetric three-dimensional superconducting systems with $C_4$ symmetry. The non-trivial higher order phases we aim to study are physically distinguished from trivial phases by the presence of helical Majorana hinge modes, where the left and right moving Majorana modes $\gamma_L$ and $\gamma_R$ transform under time reversal as $\gamma_R\rightarrow \gamma_L\, ,\; \gamma_L\rightarrow -\gamma_R$. In a recent work \cite{hughes2}, higher order topological superconductors that break both $C_4$ and time reversal $\mathcal{T}$ but preserve $C_4\mathcal{T}$ were studied. It was found that such higher order topological superconductors also exhibit hinge modes. However, the hinge modes in Ref. \cite{hughes2} are chiral and are therefore different from the helical hinge modes we find in this work. 

We first discuss how recent findings on higher order topology in 3D insulators generalize to superconductors with time-reversal, $C_4$ and inversion symmetry. We define band invariants that detect the presence of helical Majorana hinge modes on samples with open boundaries. For even-parity superconductors we show how a non-trivial higher order topological superconductor can be obtained by combining a three dimensional topological insulator with a small pairing that has the appropriate symmetry properties. Because we start with a bulk insulating material the pairing only affects the low-energy modes on the boundary. It is subsequently shown that helical hinge modes can also robustly coexist with boundary Majorana cones due to a large difference in their crystal momentum.
This coexistence, which we call hybrid higher order topology, has not been discussed before in the literature and does not yet exist in higher order topological insulators, though the possibility is a clear consequence of the stacking and packing picture \cite{hermele1,hermele2}.

 A bulk criterion for hybrid boundary modes is presented for weak-pairing odd-parity superconductors. For such odd-parity superconductors, a small pairing is added to a gapless particle-number conserving Hamiltonian, and the resulting higher order topology depends on the Fermi surface properties. We  construct a tight binding model realizing hybrid higher order topology and numerically obtain both the Majorana cones and the helical hinge modes.

We end with a discussion of our results and possible future directions. The supplementary material contains calculations for a continuum model realizing helical hinge modes, more detailed arguments about the higher order bulk invariant, and the real space version of the tight binding model exhibiting hybrid higher order topology. We start by reviewing some general aspects of BdG superconductors with time reversal and $C_4$ symmetries.

\begin{center}
\textbf{Superconductors with time reversal and $C_4$ symmetry}
\end{center}

In this section we discuss the symmetries relevant for this work, and at the same time set the notation. Consider a general translationally invariant BdG superconductor in momentum space $\hat{H} = \sum_{\textbf{k}} \Psi^\dagger_{\textbf{k}}H(\textbf{k})\Psi_{\textbf{k}}$, with 
\begin{eqnarray}
H(\textbf{k}) &=&\left(\begin{matrix}h(\textbf{k}) & \Delta(\textbf{k}) \\ \Delta^\dagger(\textbf{k}) & -h^T(-\textbf{k}) \end{matrix}\right)\, ,\label{bdg} \\ 
\Psi_\textbf{k} &=& \left(c_{\textbf{k},1},\dots, c_{\textbf{k},n},c_{-\textbf{k},1}^\dagger ,\dots ,c_{-\textbf{k},n}^\dagger\right)^T\, .
\end{eqnarray}
Such a Hamiltonian always has particle-hole symmetry of the form $\mathcal{C}^\dagger H(-\textbf{k})\mathcal{C} = -H(\textbf{k})$, where $\mathcal{C} = \tau^xK$. $\tau^x$ is the Pauli $x$-matrix in Nambu space, and $K$ denotes the complex conjugation operator. Particle hole `symmetry' follows from the fact that any nonvanishing pairing satisfies $\Delta^T(-\textbf{k}) = -\Delta(\textbf{k})$. The particle-hole symmetry satisfies $\mathcal{C}^2=\mathds{1}$. 

We require the BdG Hamiltonian to be invariant under time reversal symmetry, which is an anti-unitary operator acting on the annihilation operators as $c_{\textbf{r},\alpha} \rightarrow \sum_{\beta=1}^n \left(U_T\right)_{\alpha\beta} c_{\textbf{r},\beta}$, where $U_T$ is a unitary matrix. In this work, we are interested in the case where $U_TU_T^*=-\mathds{1}$. Because we want to study higher order phases with helical hinge modes, the BdG Hamiltonian cannot have spin SU$(2)$ symmetry. We will also assume that there is no spin U$(1)$ symmetry, which means that we are considering class DIII of the Altland-Zirnbauer classification \cite{az,ryu}. In momentum space, the time reversal invariance implies $\mathcal{T}^\dagger H(\textbf{k})\mathcal{T} = H(-\textbf{k})$, with $\mathcal{T}\equiv (U_T\oplus U_T^*)K$.

Next, we also require the superconductors to be invariant under a $C_4$ rotation symmetry, where without loss of generality we take the rotation axis along the $z$-direction. The $C_4$ rotation is defined to act on the annihilation operators as 
\begin{equation}\label{eq:rotC4}
c_{(x,y,z),\alpha}\rightarrow (-1)^{r(x+y)}\sum_{\beta=1}^n \left(U_R\right)_{\alpha,\beta}c_{(y,-x,z),\beta}\, ,
\end{equation}
where $x,y,z,r\in\mathbb{Z}$ and $U_R$ is a unitary matrix. This action of $C_4$ implies that the rotation axis goes through the sites with coordinates $(0,0,z)$, and that all orbitals lie on the vertices of the cubic lattice (Wyckoff position $1a$ \cite{bilbao}). Note that in principle one could also consider the rotation axis going through the points with coordinates $(1/2,1/2,z)$ (Wyckoff position $1b$ \cite{bilbao}). A rotation around $1b$ is equivalent to a rotation around $1a$ followed by a translation over one lattice vector. This is consistent with the fact that an AB sublattice symmetry breaking term (such as for example a staggered chemical potential $\sum_{x,y,\alpha}(-1)^{x+y}\mu\, c^\dagger_{(x,y),\alpha}c_{(x,y),\alpha}$) breaks the $1b$ rotation symmetry, but not the $1a$ rotation symmetry. A topological insulator or superconductor protected by a $1b$ rotation symmetry is therefore not expected to exhibit hinge modes, because by breaking translation symmetry one removes the protecting symmetry. For this reason, we consider a rotation axis centered at $1a$. 

Note that $r=0$ and $r=1$ in Eq. \eqref{eq:rotC4} represent two very different $C_4$ symmetry actions. In particular, with $R$ the generator of $C_4$ and $T^x$ ($T^y$) the translation operator in the $x$ ($y$) direction, it holds that $RT^x = (-1)^rT^yR$. By redefining, say, $T^y$ to be a translation in the $y$ direction followed by a gauge transformation $c_{(x,y),\alpha}\rightarrow (-)^r c_{(x,y),\alpha}$, the space group commutation relations become the same for $r=0$ and $r=1$. However, in band theory there is a prefered translation operator $T^y$ to define crystal momentum in the $y$-direction, so we will distinguish between the cases $r=0$ and $r=1$. Going to momentum space, the rotation symmetry \eqref{eq:rotC4} implies
$\mathcal{R}^\dagger H(\textbf{k})\mathcal{R} = H(\textbf{Rk}+\textbf{b})$, with $\mathcal{R}=U_R\oplus U_R^*$. Here we introduced the notation $\textbf{Rk}=(k_y,-k_x,k_z)$ and $\textbf{b}=r(\pi,\pi,0)$. Below we will only explicitly consider the case with $r=0$, but our results can be generalized to the case $r=1$.

Now that we have discussed all the main symmetries separately, let's consider the interplay between them. As a first step, we forget about particle-hole symmetry and just consider the normal part of the BdG Hamiltonian. For a $C_4$ symmetry it holds that $U_R^4=\alpha\mathds{1}$, with $\alpha\in U(1)$. We now also make the assumption that $U_TU_R = \beta U_R^*U_T$, with $\beta\in U(1)$. One can always redefine $U_R$ as $\beta^{-1/2}U_R$ such that $U_TU_R=U_R^*U_T$ and $U_R^4=\alpha'\mathds{1}$, where $\alpha'=\alpha/\beta^2$. In redefining $U_R$ with a phase, the symmetry properties of the normal part, i.e. $U_T^\dagger h(\textbf{k})^*U_T = h(-\textbf{k})$ and $U_R^\dagger h(\textbf{k})U_R = h(\textbf{Rk}+\textbf{b})$, remain unaltered. Because $U_TU_R^{4}=U_R^{4*}U_T$, it follows that $\alpha'=\pm1$. In this paper we are considering spinful fermions, so we are going to focus on the case $U_R^4=-\mathds{1}$.

Now we reintroduce the pairing, and therefore the particle-hole symmetry. In general, to preserve $C_4$ symmetry the pairing can transform as $U_R^\dagger\Delta(\textbf{k})U_R^* = e^{im\pi/2}\Delta(\textbf{Rk})$, where $m\in\{0,1,2,3\}$ \cite{fang}. Now consider the equality $U_T^\dagger U_R^T\Delta(\textbf{k})^*U_RU_T^* = U_R^\dagger U_T^\dagger \Delta(\textbf{k})^*U_T^*U_R^*$, which follows from $U_TU_R=U_R^*U_T$. Evaluating both sides of the equality shows that $e^{im\pi/2}=e^{-im\pi/2}$, from which we conclude that $m=0$ or $m=2$, which was also proven in Ref. \cite{fang}. For the case $U_R^\dagger\Delta(\textbf{k})U_R^*=\Delta(\textbf{Rk})$, the $C_4$ rotation matrix of the BdG Hamiltonian is given by $\mathcal{R}=U_R\oplus U_R^*$. However, when $U_R^\dagger\Delta(\textbf{k})U_R^*=-\Delta(\textbf{Rk})$, the BdG $C_4$ rotation matrix is $\mathcal{R}=iU_R\oplus\left(iU_R\right)^*$. In the first case ($n=0$), we have that $[\mathcal{T},\mathcal{R}]=0$, while in the second case ($n=2$), we have $\{\mathcal{T},\mathcal{R}\}=0$, where $\{\cdot\}$ denotes the anti-commutator. We will consider both cases in the discussion of higher order topology in the following sections.

\begin{center}
\textbf{Higher order topology with helical hinge modes}
\end{center}

We consider 3D superconductors in class DIII. The corresponding band structures are characterized by a strong index $N$, which is integer valued \cite{ryu,kitaev2,ryu2}. In this section we start with the case $N=0$. The $C_4$ symmetry satisfies $\mathcal{R}^4=-\mathds{1}$ and $[\mathcal{T},\mathcal{R}]=0$ or $\{\mathcal{T},\mathcal{R}\}=0$. To simplify the discussion of higher order topology associated with the $C_4$ rotation symmetry, we assume that the BdG Hamiltonian also has an inversion symmetry given by $\mathcal{I}^\dagger H(\textbf{k})\mathcal{I}=H(-\textbf{k})$, where $\mathcal{I}\equiv U_I\oplus U_I^*$ satisfies $\mathcal{I}^2=\mathds{1}$, $[\mathcal{T},\mathcal{I}]=0$ and $[\mathcal{R},\mathcal{I}]=0$. The inversion symmetry allows us to define the Fu-Kane invariant $\nu$ of time reversal invariant band structures with $\mathcal{T}^2=-\mathds{1}$, which is given by the number of occupied Kramers pairs with negative inversion eigenvalues at the time-reversal invariant momenta (TRIM) modulo $2$ \cite{fukane}. In three dimensions, the Fu-Kane invariant for superconductors satisfies $\nu = N$ mod $2$ \cite{fuberg}.

The intuitive idea underlying the $C_4$ higher order band invariants is to calculate the strong invariant in different $C_4$ subspaces. We do this as follows. At the four TRIM fixed by $C_4$ in 3D we can label the Kramers pairs by their $C_4$ eigenvalues. In the case $[\mathcal{T},\mathcal{R}]=0$, a Kramers pair carries $C_4$ eigenvalues $(e^{i\pi/4},e^{-i\pi/4})$ or $(e^{i3\pi/4},e^{-i3\pi/4})$. When $\{\mathcal{T},\mathcal{R}\}=0$, the Kramers pairs at TRIM fixed by $C_4$ can be labeled by the eigenvalues $(e^{i\pi/4},e^{i3\pi/4})$ or $(e^{-i\pi/4},e^{-i3\pi/4})$. We first discuss the case with $[\mathcal{T},\mathcal{R}]=0$. In analogy to the analysis of higher order topology in Bismuth \cite{bismuth}, we now define $\nu^{(1)}\in\{0,1\}$, which counts the number of occupied Kramers pairs with negative inversion eigenvalues and $C_4$ eigenvalues $(e^{i\pi/4},e^{-i\pi/4})$ at the four TRIM fixed under $C_4$, modulo 2. Similarly, we also define $\nu^{(3)}$ using the Kramers pairs at the four TRIM fixed under $C_4$ with $C_4$ eigenvalues $(e^{i3\pi/4},e^{-i3\pi/4})$. A non-trivial higher order superconductor with $[\mathcal{T},\mathcal{R}]=0$ is then characterized by
\begin{equation}\label{invariant1}
\nu^{(1)}=\nu^{(3)}=1\,.
\end{equation}
For $\{\mathcal{T},\mathcal{R}\}=0$, the non-trivial indices are similarly given by
\begin{equation}\label{TIcase}
\nu^{(+)}=\nu^{(-)}=1\,,
\end{equation}
where now the index $\nu^{+}$ ($\nu^{-}$) counts the parity of the number of occupied Kramers pairs at TRIM fixed under $C_4$ with negative inversion eigenvalues and $C_4$ eigenvalues $(e^{i\pi/4},e^{i3\pi/4})$ $\left(  (e^{-i\pi/4},e^{-i3\pi/4}) \right)$. In the supplementary material we show that, by analogy to the insulating case \cite{bismuth}, these indices indeed detect the presence of helical hinge modes. We do this by constructing a continuum model that has non-trivial higher order invariants and explicitly solve for the zero-energy states associated with the hinge modes. The definition of the band invariant implies a $\mathbb{Z}_2$ classification for higher order superconductors with $C_4$ and time reversal symmetry, which is consistent with the fact that a pair of helical Majorana hinge modes can be gapped without breaking the symmetry. 

Although the higher order band invariants discussed here are closely related to those defined in the context of higher topology in Bismuth \cite{bismuth}, there is an important physical distinction for boundary surfaces orthogonal to the $C_4$ rotation axis. This is because with open boundary conditions in all three directions there is no $C_4$ symmetric way to connect the gapless modes on the hinges parallel to the rotation axis along the hinges orthogonal to the rotation axis. This simple geometrical argument shows that the boundary surfaces orthogonal to the rotation axis have to be gapless, which is not the case for Bismuth.

We now present an explicit method to obtain a higher order topological superconductor in the symmetry class with $\{\mathcal{T},\mathcal{R}\}=0$ exhibiting non-trivial band indices as in Eq. \eqref{TIcase}. Our starting point is a $C_4$ and time reversal symmetric Hamiltonian $h(\textbf{k})$, which conserves particle number. Concretely, we assume that there exist matrices $U_R$ and $U_T$ such that $U_R^\dagger h(\textbf{k})U_R= h(\textbf{Rk})$ and $U_T^\dagger h(-\textbf{k})^*U_T = h(\textbf{k})$. For particle number conserving systems we can without loss of generality fix the phase of the $C_4$ rotation matrix $U_R$ such that $U_TU_R=U_TU_R^*$. As before, we also assume inversion symmetry such that $U_I^\dagger h(-\textbf{k})U_I = h(\textbf{k})$, with $U_I^2=\mathds{1}$, $U_TU_I=U_I^*U_T$ and $[U_R,U_I]=0$. Now take $h(\textbf{k})$ to be a 3D strong topological insulator. We claim that after adding an arbitrarily small pairing satisfying 
\begin{eqnarray}
U_I^\dagger \Delta(\textbf{k}) U_I^* &=& \Delta(-\textbf{k}) \label{invpairing}\\
U_R^\dagger\Delta(\textbf{k}) U_R^*&=&-\Delta(\textbf{Rk}) \label{signpairing}\, ,
\end{eqnarray}
the higher order band invariants for the resulting superconductor will be non-trivial, such that the system will have helical hinge modes on samples with open boundaries. Because of transformation property \eqref{signpairing}, the BdG rotation matrix is $\mathcal{R}=iU_R\oplus \left(iU_R\right)^*$. Combining this with the commutation relation $U_TU_R=U_R^*U_T$, we see that $\mathcal{R}$ and $\mathcal{T}$ anti-commute. In the supplementary material we show in detail that a 3D topological insulator combined with a small pairing transforming as in Eqs. \eqref{invpairing} and \eqref{signpairing} has non-trivial higher order invariants $\nu^{(+)}=\nu^{(-)}=1$, as defined in Eq. \eqref{TIcase}. The associated physical picture is that by breaking particle number conservation, the pairing will gap out the surface Dirac cones of the topological insulator, but because of the minus sign in Eq.\eqref{signpairing} the pairing-induced gap is forced to vanish along the hinges. Because the Fermi level is such that there is an energy gap in the bulk, the pairing does not significantly change the bulk modes. We want to point out that in recent work it was shown how a similar mechanism of combining a 2D topological insulator with the appropriate pairing results in a 2D higher order topological superconductor with Majorana-Kramers corner modes \cite{lu,yan,liu}. As we explain in more detail below, the momentum space Hamiltonians of the 2D systems studied in Refs. \cite{lu,yan,liu} are equivalent to the fixed $k_z=0$ or $k_z=\pi$ slices of our 3D Hamiltonians with helical hinge modes.

\begin{figure*}
a)
\includegraphics[scale=0.3]{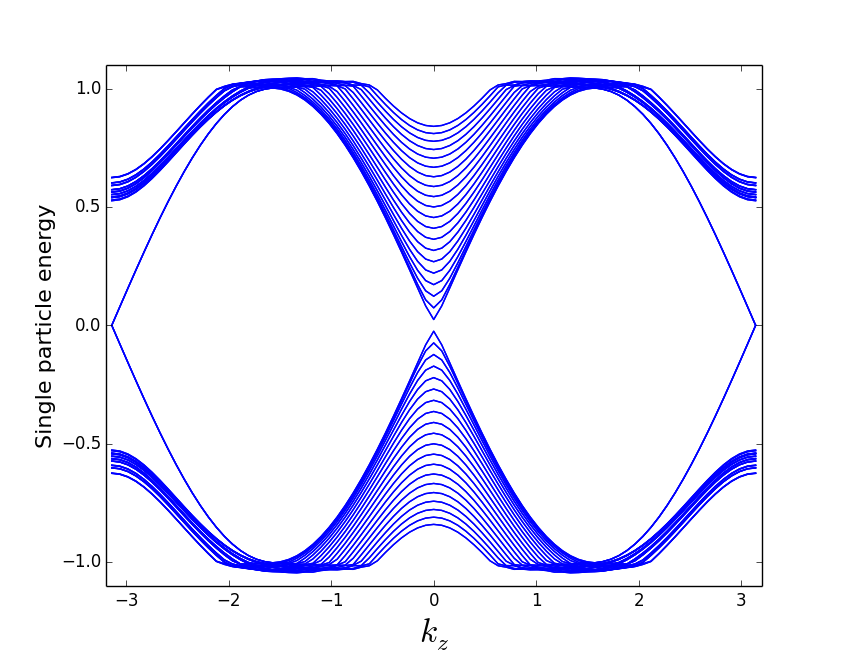}
b)
\includegraphics[scale=0.3]{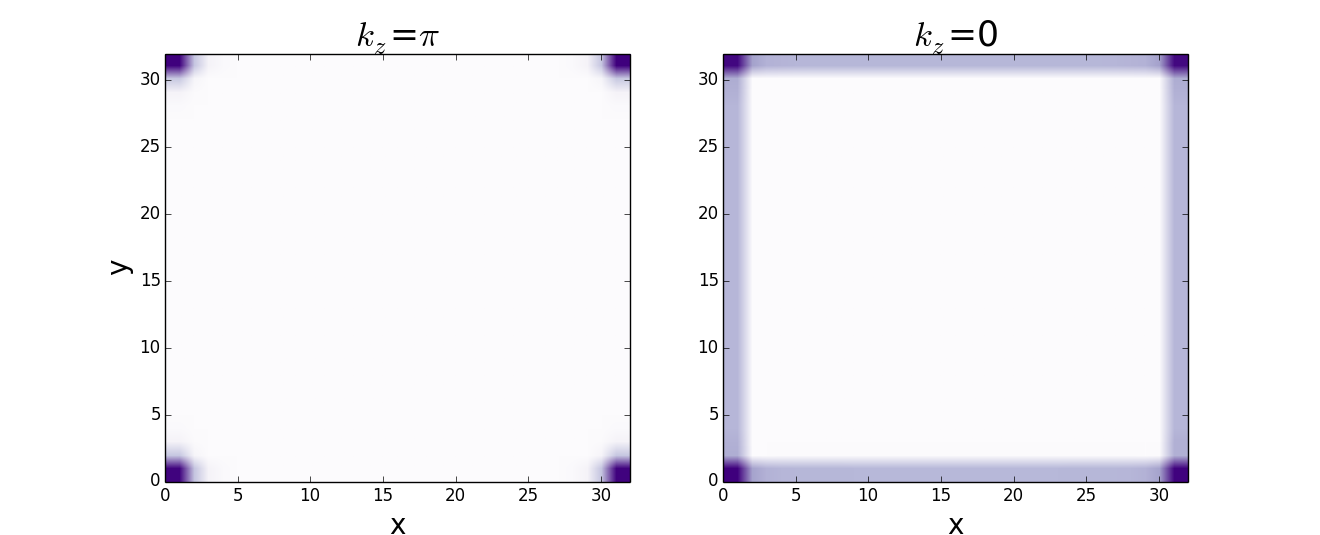}
\caption{Numerical results for the tight binding model with normal part given in Eq. \eqref{eq:normal} and pairing in Eq. \eqref{eq:pairing2}. The parameters were taken to be $t=\Delta=1$, $\mu_1=-2$, $\mu_2=\mu_3=2$, $a=0.6, b=0.3$ and $c=0.25$. All simulations were performed on a lattice of size $31\times 31$ in the $x$ and $y$-directions. (a) Dispersion with $k_z$ of the $80$ lowest energy bands. (b) The local zero-energy density of states as a function of the real space coordinates $x$ and $y$ both at momentum $k_z=\pi$ and $k_z=0$. At $k_z=0$ the density of states is observed along the entire surface, due to the Majorana surface cone, while at $k_z = \pi$, it is restricted to the hinge states. This momentum mismatch ensures the robustness of the hybrid state. \label{fig:tightb}}
\end{figure*}

\begin{center}
\textbf{Hybrid higher order topology}
\end{center}

We now discuss the occurrence of higher order topology in three-dimensional superconductors with a non-zero strong invariant $N$, and present band indices that detect this scenario. For this we will adopt a weak pairing picture, so our starting point is a particle number conserving Hamiltonian $h(\textbf{k})$ with time reversal $T$, inversion $I$ and four-fold rotation $R$ symmetries. In contrast to the gapped particle number conserving Hamiltonians considered in the previous section, here we will be working with Hamiltonians where the Fermi level is such that there are bulk Fermi surfaces. We denote the periodic part of the Bloch bands of $h(\textbf{k})$ as $|u_n(\textbf{k})\rangle$. We then add a small pairing and study the resulting 3D topological superconductor. We will use the weak pairing expression for the strong invariant $N$ \cite{qi}:
\begin{equation}\label{weakpairing}
N=\frac{1}{2}\sum_s \text{sgn}(\delta_s)C_s\, ,
\end{equation}
where the sum is over all Fermi surfaces of the normal part of the BdG Hamiltonian and $C_s$ is the Chern number of the corresponding Fermi surface. The orientation of the Fermi surface, determining the sign of the Chern number, is to be taken such that the vector normal to the Fermi surface $s$ is parallel to Fermi velocity vector $\textbf{v}_s=\nabla_{\textbf{k}}\epsilon_s(\textbf{k})$.  To explain the definition of $\delta_s$, we first define
\begin{equation}
\delta_{n,\textbf{k}}=\langle u_n(\textbf{k})|U_T^\dagger\Delta(\textbf{k})^\dagger|u_n(\textbf{k})\rangle\, .
\end{equation}
As reviewed in the supplementary material, the matrix $U_T^\dagger\Delta(\textbf{k})^\dagger$ is Hermitian, such that $\delta_{n,\textbf{k}}$ is real. Sgn($\delta_s$) is then simply sgn($\delta_{n,\textbf{k}}$) on the Fermi surface denoted by $s$, where $n$ is the band crossing the Fermi energy and $\textbf{k}$ is an arbitrary Fermi momentum on the surface (note $\text{sgn}(\delta_{s, \textbf{k}})$ is independent of $\textbf{k}$ if the pairing does not vanish on the Fermi surface). 

If a Fermi surface is degenerate, one has to add a small perturbation lifting this degeneracy in order to calculate the Fermi surface Chern numbers. Once the Chern numbers are obtained, the small perturbation is taken to zero and the result is independent of the choice of perturbation \cite{qi}. In the present context, the Fermi surfaces are always degenerate because of the symmetry $U_TU_I$.  Let us therefore consider the scenario where there is a two-fold degenerate Fermi surface, centered around the TRIM $\textbf{k}_c$ (the analysis of the situation with two degenerate Fermi surfaces centered around a pair of momenta $(\textbf{k}_c,-\textbf{k}_c)$ related by time reversal is analogous). We consider adding an inversion symmetry breaking term with infinitesimal strength $\epsilon$ to the Hamiltonian, which lifts the Fermi surface degeneracy, resulting in two separate and concentric `inner' and `outer' Fermi surfaces around $\textbf{k}_c$. We know that only by taking $\epsilon\rightarrow 0$, we recover the inversion symmetry. This implies that in the $\epsilon\rightarrow 0$ limit, inversion maps the `outer' Fermi surface around $\textbf{k}_c$ to the `inner' Fermi surface. Because the Chern number is odd under inversion, we conclude that the degenerate Fermi surface around $\textbf{k}_c$ will contribute two opposite Chern numbers $\pm C$ to the weak pairing invariant. This is consistent with the fact that with both inversion and time reversal symmetry, the trace of the Berry curvature matrix on the degenerate Fermi surface is zero. From this analysis we also see that in order for the sum in Eq. \eqref{weakpairing} to be non-zero, the pairing should be parity-odd, i.e. satisfy $U_I^\dagger\Delta(\textbf{k}) U_I^*=-\Delta(-\textbf{k})$. This parity requirement for the pairing has been pointed out before \cite{fuberg}.

A similar analysis as for inversion symmetry shows that only a pairing transforming under $C_4$ as $U_R^\dagger \Delta(\textbf{k})U_R^*=\Delta(\textbf{Rk})$ can give rise to a non-trivial strong invariant. To see this, let us assume that $U_R^\dagger \Delta(\textbf{k})U_R^*=-\Delta(\textbf{Rk})$ and again add the inversion breaking, but $C_4$ respecting, perturbation with strength $\epsilon$ to obtain non-degenerate Fermi surfaces. First, it immediately follows that there can be no Fermi surfaces centered around momenta invariant under $C_4$, because otherwise the pairing would not completely gap out this Fermi surface as a result of its sign-changing nature. Fermi surfaces centered around momenta related to each other by $C_4$ will all have the same Chern number because the Chern number is invariant under the orientation-preserving $C_4$ symmetry. Because the pairing on these Fermi surfaces has opposite sign, it is clear that the sum in Eq. \eqref{weakpairing} evaluates to zero when $U_R^\dagger \Delta(\textbf{k})U_R^*=-\Delta(\textbf{Rk})$. The $C_4$ transformation property of the pairing $U_R^\dagger \Delta(\textbf{k})U_R^*=\Delta(\textbf{Rk})$ compatible with a non-zero strong invariant puts us automatically in the symmetry class with $[\mathcal{T},\mathcal{R}]=0$. This is to be contrasted with Eq. \eqref{signpairing} of the previous section, where we considered superconductors with trivial strong invariant. 

We now formulate a bulk criterion for the coexistence of helical hinge modes with Majorana cones. First, we split up the higher order invariants defined in Eq.\eqref{invariant1} for the case $N=0$:
\begin{eqnarray}
\nu^{(1)}& = &\left(\nu^{(1)}_0+\nu^{(1)}_\pi\right) \text{ mod }2\, , \label{split1}\\
\nu^{(3)}&=&\left(\nu^{(3)}_0+\nu^{(3)}_\pi\right) \text{ mod }2\,, \label{split2}
\end{eqnarray}
where $\nu^{(i)}_{0}$ ($\nu^{(i)}_{\pi}$) contains the contributions from the TRIM fixed under $C_4$ in the $k_z=0$ ($k_z=\pi$) torus.

We now similarly isolate the different contributions to the strong invariant $N$. First we consider the situation where each Fermi surface encloses exactly one TRIM. In this case we define the invariants
\begin{eqnarray}
N_0 &=& \frac{1}{2}\sum_{s\in S_0}\text{sgn}(\delta_s)C_s\, ,\nonumber\\
N_\pi &=& \frac{1}{2}\sum_{s\in S_{\pi}}\text{sgn}(\delta_s)C_s\, ,
\end{eqnarray}
where $S_0$ ($S_\pi$) is the collection of Fermi surfaces surrounding any TRIM in the $k_z=0$ ($k_z=\pi$) torus. To generalize these invariants to arbitrary Fermi surface configurations, we recall that the authors of Ref. \cite{qi} showed that every non-degenerate Fermi surface enclosing a TRIM necessarily has odd Chern number. This implies that the Kramers degeneracies act as a source for the Berry flux. When a Fermi surface encloses multiple TRIM, one can always isolate the contributions from the different Kramers degeneracies to the total Chern number by considering smaller surfaces, each of which encloses only one TRIM. In the absence of inversion symmetry, there could also be Weyl nodes below or above the Fermi level \cite{weyl1,weyl2}, giving rise to Fermi surfaces with non-zero Chern number that do not enclose a TRIM. However, in this work our main focus is on inversion symmetric systems so we will not consider this possibility any further.

Having split up the invariants in Eqs. \eqref{split1} and \eqref{split2}, we now claim that with open boundaries in the $x$ and $y$-directions there will be a hinge mode at $k_z=n\in\{0,\pi \}$ when
\begin{equation}
N_n = 0\, ,\hspace{0.3 cm} \nu^{(1)}_n= \nu^{(3)}_n=1\, .
\end{equation}
To see this, consider taking the 2D $C_4$ symmetric higher order topological superconductors with Majorana-Kramers corner zero modes as studied in Refs. \cite{lu,yan,liu}, and stack them along the $z$-direction. After weakly coupling the layers in a way that respects translation in the $z$-direction, $C_4$ and time reversal, the resulting 3D momentum space BdG Hamiltonian will have band indices $ \nu^{(1)}_n= \nu^{(3)}_n=1$ at both $k_z=0$ and $k_z=\pi$ \cite{lu}, and also $N_n=0$ since the stacked layers are trivial 2D strong TSCs. As a result, there are Majorana-Kramers zero modes at $k_z=0$ and $k_z=\pi$, which will disperse when moving away from $k_z=0$ and $k_z=\pi$. The simulations further on in this section confirm that this dispersion with $k_z$ is indeed linear, as required for helical Majorana hinge modes.

The case of hybrid higher order topology will now occur when one $k_z$ torus leads to a helical hinge mode, while the other $k_z$ torus has a non-zero contribution $N_n$ to the strong invariant. If $N_n=1$, there will be one Majorana surface cone at $k_z=n$. This is because the BdG momentum space Hamiltonian at $k_z=n$ will correspond to a 2D strong TSC (which have a $\mathbb{Z}_2$ classification in 2D \cite{kitaev2,ryu2}), with helical Majorana boundary modes. 

The tight binding model is defined using the orbitals $\{|p_+\,\downarrow\rangle\,,|p_-\,\uparrow\rangle\,,\,|p_-\,\downarrow\rangle\,,\,|p_+\,\uparrow\rangle\,,\,|s \,\downarrow\rangle\,,\,|s\uparrow\rangle \}$ on the sites of a cubic lattice. The $C_4$ symmetry acts on these orbitals as $U_R=e^{i\pi\sigma^z/4}\oplus e^{-i3\pi\sigma^z/4}\oplus e^{i\pi\sigma^z/4}$, and the $C_4$ rotation matrix of the BdG Hamiltonian is $\mathcal{R}=U_R\oplus U_R^*$. The normal part of the BdG Hamiltonian is
\begin{equation}\label{eq:normal}
h(\textbf{k}) = \left(\begin{matrix}h_1(\textbf{k}) & A(\textbf{k}) & C(\textbf{k}) \\ A^\dagger(\textbf{k}) & h_2(\textbf{k}) & B(\textbf{k}) \\ C^\dagger(\textbf{k}) & B^\dagger(\textbf{k}) & h_3(\textbf{k})\end{matrix} \right)\, ,
\end{equation}
where each entry is a $2\times2$ matrix:
\begin{eqnarray}
h_1(\textbf{k}) &= &(t\cos k_x +t\cos k_y + t\cos k_z -\mu_1)\mathds{1}\nonumber\\
h_2(\textbf{k}) &= &(-t\cos k_x -t\cos k_y - t\cos k_z -\mu_2)\mathds{1}\nonumber\\
h_3(\textbf{k}) &= &(-t\cos k_x -t\cos k_y - t\cos k_z -\mu_3)\mathds{1}\nonumber\\
A(\textbf{k}) &  = & a(\cos k_x - \cos k_y)\mathds{1} \nonumber \\
B(\textbf{k}) &  = & b(\cos k_x - \cos k_y)\mathds{1} \nonumber\\
C(\textbf{k}) & = & c\left(\cos k_x + \cos k_y + \cos k_z\right)\mathds{1}\, .
\end{eqnarray}
The pairing is given by 
\begin{equation}\label{eq:pairing2}
\Delta(\textbf{k})=\Delta\mathds{1}_{3\times 3}\otimes (\sin k_x \mathds{1}-i\sin k_y\sigma^z+i\sin k_z\sigma^x)\, ,
\end{equation}
where $\mathds{1}_{3\times 3}$ is the three by three identity matrix (if no dimensions are specified, identity matrices are two-dimensional). All parameters in the BdG Hamiltonian are real numbers. The Hamiltonian has time reversal symmetry with $\mathcal{T}=(\tau^z\otimes\mathds{1}_{3\times 3}\otimes\sigma^y)K$, where $\tau^i$ are the Pauli matrices acting in Nambu space (the standard expression for time reversal $\mathcal{T}=i\sigma^yK$ can be recovered by a gauge transformation where the orbitals are multiplied by $e^{i\pi/4}$). The symmetry operators $\mathcal{R}=U_R\oplus U^*_R$ and $\mathcal{T}=(\tau^z\otimes\mathds{1}_{3\times 3}\otimes\sigma^y)K$ commute, which as we mentioned above is a necessary condition for non-trivial hybrid higher order topology. Because $h(\textbf{k})$ is even and the pairing $\Delta(\textbf{k})$ is odd in $\textbf{k}$, the inversion matrix is given by $\mathcal{I}=\tau^z$. Note that our BdG Hamiltonian actually breaks \emph{physical} inversion symmetry, under which the $s$ and $p$ orbitals have opposite parity. This breaking of physical inversion symmetry could occur spontaneously as the BdG Hamiltonian is a mean-field approximation to an interacting model.  Although physical inversion symmetry is broken, there is enough symmetry left in the BdG Hamiltonian to define the alternative inversion symmetry matrix $\mathcal{I}=\tau^z$, which is sufficient to define the higher order band invariants. In Appendix \ref{realspace} we lay out the details of the real space version of the tight binding model.

To understand the physics of the BdG tight binding model, we first consider the case where $a=b=c=0$. Then the Hamiltonian is a direct sum of three independent BdG Hamiltonians. Each of the three decoupled BdG Hamiltonians is closely related to the $B$-phase of $^3$He, which realizes a strong TSC. The decoupled BdG Hamiltonians have an energy gap as long as $\Delta\neq0$, $|\mu_i|\neq 3t$ and $|\mu_i|\neq t$, and a non-zero strong invariant when $-3t<\mu_i<-t$ and $t<\mu_i<3t$ for $i=1,2,3$. We now fix the chemical potentials of the decoupled Hamiltonians to be $\mu_1=\mu_3=-2t$ and $\mu_2=2t$. As a result, $h_1(\textbf{k})$ has a two-fold degenerate hole-like Fermi surface around $R=(\pi,\pi,\pi)$, $h_2(\textbf{k})=-h_1(\textbf{k})$ has a two-fold degenerate electron-like Fermi surface around $R=(\pi,\pi,\pi)$, and $h_3(\textbf{k})$ has a two-fold degenerate electron-like Fermi surface around $\Gamma=(0,0,0)$. Because $h_1(\textbf{k})=-h_2(\textbf{k})$ and the pairing is the same for these two decoupled BdG Hamiltonians, the resulting TSCs have opposite strong invariant. This follows because the Fermi surface of $h_1(\textbf{k})$ is hole-like and that of $h_2(\textbf{k})$ is electron-like, such that the Fermi velocity vectors $\textbf{v}_s=\nabla_{\textbf{k}}\epsilon_s(\textbf{k})$ will be oriented oppositely for both Fermi surfaces, leading to opposite Chern numbers (see Appendix \ref{continuum} for an alternative proof of this fact that does not rely on the weak pairing expression for $N$). 

Taking the parameters $a,b$ and $c$ to be non-zero will introduce nearest-neighbour terms that couple the superconductors.  One can check that for perturbatively small $a,b$ and $c$, the resulting BdG Hamiltonian by construction has following band indices:
\begin{eqnarray}
N_0=1\, \hspace{0.3 cm} \nu^{(1)}_0=1\, ,\nu^{(3)}_0=0\nonumber \\
N_\pi=0\, \hspace{0.3 cm} \nu^{(1)}_\pi=\nu^{(3)}_\pi=1\, ,
\end{eqnarray}
such that with open boundary conditions along the $x$ and $y$-directions we expect a hinge mode at $k_z=\pi$ and a Majorana cone at $k_z=0$.

We have studied this model numerically with open boundary conditions along the $x$ and $y$-directions, and infinite boundary conditions along the $z$ direction. The results are shown in Fig. \ref{fig:tightb}. The dispersion relation in Fig. \ref{fig:tightb}(a) clearly shows the Majorana surface cone at $k_z=0$ and the helical hinge modes at $k_z=\pi$, which disperse linearly upon moving away from $k_z=\pi$. From the local zero-energy density of states in Fig. \ref{fig:tightb}(b), one sees that the Majorana cone at $k_z=0$ is delocalized over the entire boundary, while the hinge modes at $k_z=\pi$ modes are localized at the four corners.

\begin{center}
\textbf{Discussion}
\end{center}

We have analyzed three dimensional BdG superconductors with $C_4$ rotation symmetry and time reversal with $\mathcal{T}^2 = -\mathds{1}$. We found that higher order topology characterized by helical hinge modes as recently discovered both theoretically and experimentally in three-dimensional insulators \cite{bismuth} can be generalized to such superconductors. It was shown that non-trivial higher order superconductors with even-parity pairing are closely related to three-dimensional topological insulators via a weak pairing condition. It remains to find a 3D superconducting material that realizes this type of higher order topology. 

The helical hinge modes can coexist with Majorana cones, leading to a new hybrid situation of strong topology combined with higher order topology. This was illustrated by a tight binding model, which was found numerically to have both types of zero modes with open boundary conditions. We also proposed a band invariant to identify such materials.  The coexistence of the two states could be  confirmed experimentally by the presence $k = \pi$ STM quasiparticle interference (Friedel oscillations) near the hinges. An interesting question is how robust this phenomenon is when disorder is introduced. Higher order topology relies on spatial symmetries such as mirror, inversion or rotation symmetry. The hybrid case discussed above also relies on translation symmetry for a clear distinction between hinge and boundary modes. It is not immediately clear whether this makes the zero-energy modes more susceptible to disorder.

Our results on hybrid higher order topology in superconductors are to be contrasted with the insulating case. One could imagine a 3D insulator where the momentum space Hamiltonian at, say, $k_z=0$ corresponds to a non-trivial quadrupole phase \cite{quadrupole}, but is trivial at $k_z=\pi$. This will not lead to hybrid higher order topology since the quadrupole phase is an obstructed atomic limit \cite{TQC}, and the quadrupole modes can be moved into the bulk by appropriate $C_4$ symmetric perturbations. Both of these statements do not apply for the superconducting models studied in this work.

To establish the non-trivial nature of the higher order phases we focussed on a bulk-boundary correspondence by connecting a band invariant to hinge modes. However, for crystalline superconductors one can also probe the non-trivial topological phases by studying lattice defects. For example, weak topological superconductors can be probed by introducing dislocations \cite{ran,miert}. In Refs. \cite{teo,bth}, two-dimensional crystalline superconductors with $C_4$ symmetry but without time reversal were studied and were found to bind Majorana modes to disclinations. Similarly, in the non-trivial 3D higher order time-reversal symmetric superconductors discussed in this work disclination lines will bind gapless Majorana modes. We leave the details of this open for future work. Finally, it would of course also be interesting to generalize the discussion presented here to other space groups. A natural first starting point would be to consider different $n$-fold rotation axis, or 3D systems with multiple rotation axis.

\emph{Acknowledgments --} We thank Jennifer Cano for helpful discussions, and two anonymous referees for helpful comments on the manuscript. B. A. B. was supported by the Department of Energy Grant No. DE-SC0016239, the National Science Foundation EAGER Grant No. NOA-AWD1004957, Simons Investigator Grants No. ONR-N00014-14-1-0330 and No. NSF-MRSEC DMR- 1420541, the Packard Foundation, the Schmidt Fund for Innovative Research.  M.P.Z. was funded by DOE BES Contract No. DE-AC02-05-CH11231, through the Scientific Discovery through Advanced Computing (SciDAC) program (KC23DAC
Topological and Correlated Matter via Tensor Networks and Quantum Monte Carlo). N.B. is supported by a BAEF Francqui fellowship.

\bibliography{3dHOTSC}

\begin{thebibliography}{99}%
\makeatletter
\providecommand \@ifxundefined [1]{%
 \@ifx{#1\undefined}
}%
\providecommand \@ifnum [1]{%
 \ifnum #1\expandafter \@firstoftwo
 \else \expandafter \@secondoftwo
 \fi
}%
\providecommand \@ifx [1]{%
 \ifx #1\expandafter \@firstoftwo
 \else \expandafter \@secondoftwo
 \fi
}%
\providecommand \natexlab [1]{#1}%
\providecommand \enquote  [1]{``#1''}%
\providecommand \bibnamefont  [1]{#1}%
\providecommand \bibfnamefont [1]{#1}%
\providecommand \citenamefont [1]{#1}%
\providecommand \href@noop [0]{\@secondoftwo}%
\providecommand \href [0]{\begingroup \@sanitize@url \@href}%
\providecommand \@href[1]{\@@startlink{#1}\@@href}%
\providecommand \@@href[1]{\endgroup#1\@@endlink}%
\providecommand \@sanitize@url [0]{\catcode `\\12\catcode `\$12\catcode
  `\&12\catcode `\#12\catcode `\^12\catcode `\_12\catcode `\%12\relax}%
\providecommand \@@startlink[1]{}%
\providecommand \@@endlink[0]{}%
\providecommand \url  [0]{\begingroup\@sanitize@url \@url }%
\providecommand \@url [1]{\endgroup\@href {#1}{\urlprefix }}%
\providecommand \urlprefix  [0]{URL }%
\providecommand \Eprint [0]{\href }%
\providecommand \doibase [0]{http://dx.doi.org/}%
\providecommand \selectlanguage [0]{\@gobble}%
\providecommand \bibinfo  [0]{\@secondoftwo}%
\providecommand \bibfield  [0]{\@secondoftwo}%
\providecommand \translation [1]{[#1]}%
\providecommand \BibitemOpen [0]{}%
\providecommand \bibitemStop [0]{}%
\providecommand \bibitemNoStop [0]{.\EOS\space}%
\providecommand \EOS [0]{\spacefactor3000\relax}%
\providecommand \BibitemShut  [1]{\csname bibitem#1\endcsname}%
\let\auto@bib@innerbib\@empty
\bibitem [{\citenamefont {Kane}\ and\ \citenamefont {Mele}(2005)}]{qsh}%
  \BibitemOpen
  \bibfield  {author} {\bibinfo {author} {\bibfnamefont {C.~L.}\ \bibnamefont
  {Kane}}\ and\ \bibinfo {author} {\bibfnamefont {E.~J.}\ \bibnamefont
  {Mele}},\ }\href {\doibase 10.1103/PhysRevLett.95.226801} {\bibfield
  {journal} {\bibinfo  {journal} {Phys. Rev. Lett.}\ }\textbf {\bibinfo
  {volume} {95}},\ \bibinfo {pages} {226801} (\bibinfo {year}
  {2005})}\BibitemShut {NoStop}%
\bibitem [{\citenamefont {Bernevig}\ \emph {et~al.}(2006)\citenamefont
  {Bernevig}, \citenamefont {Hughes},\ and\ \citenamefont {Zhang}}]{HgTe}%
  \BibitemOpen
  \bibfield  {author} {\bibinfo {author} {\bibfnamefont {B.~A.}\ \bibnamefont
  {Bernevig}}, \bibinfo {author} {\bibfnamefont {T.~L.}\ \bibnamefont
  {Hughes}}, \ and\ \bibinfo {author} {\bibfnamefont {S.-C.}\ \bibnamefont
  {Zhang}},\ }\href {\doibase 10.1126/science.1133734} {\bibfield  {journal}
  {\bibinfo  {journal} {Science}\ }\textbf {\bibinfo {volume} {314}},\ \bibinfo
  {pages} {1757} (\bibinfo {year} {2006})}\BibitemShut {NoStop}%
\bibitem [{\citenamefont {Kitaev}(2009)}]{kitaev2}%
  \BibitemOpen
  \bibfield  {author} {\bibinfo {author} {\bibfnamefont {A.}~\bibnamefont
  {Kitaev}},\ }\href {\doibase 10.1063/1.3149495} {\bibfield  {journal}
  {\bibinfo  {journal} {AIP Conference Proceedings}\ }\textbf {\bibinfo
  {volume} {1134}},\ \bibinfo {pages} {22} (\bibinfo {year}
  {2009})}\BibitemShut {NoStop}%
\bibitem [{\citenamefont {Schnyder}\ \emph {et~al.}(2008)\citenamefont
  {Schnyder}, \citenamefont {Ryu}, \citenamefont {Furusaki},\ and\
  \citenamefont {Ludwig}}]{ryu}%
  \BibitemOpen
  \bibfield  {author} {\bibinfo {author} {\bibfnamefont {A.~P.}\ \bibnamefont
  {Schnyder}}, \bibinfo {author} {\bibfnamefont {S.}~\bibnamefont {Ryu}},
  \bibinfo {author} {\bibfnamefont {A.}~\bibnamefont {Furusaki}}, \ and\
  \bibinfo {author} {\bibfnamefont {A.~W.~W.}\ \bibnamefont {Ludwig}},\ }\href
  {\doibase 10.1103/PhysRevB.78.195125} {\bibfield  {journal} {\bibinfo
  {journal} {Phys. Rev. B}\ }\textbf {\bibinfo {volume} {78}},\ \bibinfo
  {pages} {195125} (\bibinfo {year} {2008})}\BibitemShut {NoStop}%
\bibitem [{\citenamefont {Ryu}\ \emph {et~al.}(2010)\citenamefont {Ryu},
  \citenamefont {Schnyder}, \citenamefont {Furusaki},\ and\ \citenamefont
  {Ludwig}}]{ryu2}%
  \BibitemOpen
  \bibfield  {author} {\bibinfo {author} {\bibfnamefont {S.}~\bibnamefont
  {Ryu}}, \bibinfo {author} {\bibfnamefont {A.~P.}\ \bibnamefont {Schnyder}},
  \bibinfo {author} {\bibfnamefont {A.}~\bibnamefont {Furusaki}}, \ and\
  \bibinfo {author} {\bibfnamefont {A.~W.~W.}\ \bibnamefont {Ludwig}},\ }\href
  {http://stacks.iop.org/1367-2630/12/i=6/a=065010} {\bibfield  {journal}
  {\bibinfo  {journal} {New Journal of Physics}\ }\textbf {\bibinfo {volume}
  {12}},\ \bibinfo {pages} {065010} (\bibinfo {year} {2010})}\BibitemShut
  {NoStop}%
\bibitem [{\citenamefont {K{\"o}nig}\ \emph {et~al.}(2007)\citenamefont
  {K{\"o}nig}, \citenamefont {Wiedmann}, \citenamefont {Br{\"u}ne},
  \citenamefont {Roth}, \citenamefont {Buhmann}, \citenamefont {Molenkamp},
  \citenamefont {Qi},\ and\ \citenamefont {Zhang}}]{Molenkamp}%
  \BibitemOpen
  \bibfield  {author} {\bibinfo {author} {\bibfnamefont {M.}~\bibnamefont
  {K{\"o}nig}}, \bibinfo {author} {\bibfnamefont {S.}~\bibnamefont {Wiedmann}},
  \bibinfo {author} {\bibfnamefont {C.}~\bibnamefont {Br{\"u}ne}}, \bibinfo
  {author} {\bibfnamefont {A.}~\bibnamefont {Roth}}, \bibinfo {author}
  {\bibfnamefont {H.}~\bibnamefont {Buhmann}}, \bibinfo {author} {\bibfnamefont
  {L.~W.}\ \bibnamefont {Molenkamp}}, \bibinfo {author} {\bibfnamefont {X.-L.}\
  \bibnamefont {Qi}}, \ and\ \bibinfo {author} {\bibfnamefont {S.-C.}\
  \bibnamefont {Zhang}},\ }\href {\doibase 10.1126/science.1148047} {\bibfield
  {journal} {\bibinfo  {journal} {Science}\ }\textbf {\bibinfo {volume}
  {318}},\ \bibinfo {pages} {766} (\bibinfo {year} {2007})}\BibitemShut
  {NoStop}%
\bibitem [{\citenamefont {Hsieh}\ \emph
  {et~al.}(2009{\natexlab{a}})\citenamefont {Hsieh}, \citenamefont {Xia},
  \citenamefont {Wray}, \citenamefont {Qian}, \citenamefont {Pal},
  \citenamefont {Dil}, \citenamefont {Osterwalder}, \citenamefont {Meier},
  \citenamefont {Bihlmayer}, \citenamefont {Kane}, \citenamefont {Hor},
  \citenamefont {Cava},\ and\ \citenamefont {Hasan}}]{Hsieh}%
  \BibitemOpen
  \bibfield  {author} {\bibinfo {author} {\bibfnamefont {D.}~\bibnamefont
  {Hsieh}}, \bibinfo {author} {\bibfnamefont {Y.}~\bibnamefont {Xia}}, \bibinfo
  {author} {\bibfnamefont {L.}~\bibnamefont {Wray}}, \bibinfo {author}
  {\bibfnamefont {D.}~\bibnamefont {Qian}}, \bibinfo {author} {\bibfnamefont
  {A.}~\bibnamefont {Pal}}, \bibinfo {author} {\bibfnamefont {J.~H.}\
  \bibnamefont {Dil}}, \bibinfo {author} {\bibfnamefont {J.}~\bibnamefont
  {Osterwalder}}, \bibinfo {author} {\bibfnamefont {F.}~\bibnamefont {Meier}},
  \bibinfo {author} {\bibfnamefont {G.}~\bibnamefont {Bihlmayer}}, \bibinfo
  {author} {\bibfnamefont {C.~L.}\ \bibnamefont {Kane}}, \bibinfo {author}
  {\bibfnamefont {Y.~S.}\ \bibnamefont {Hor}}, \bibinfo {author} {\bibfnamefont
  {R.~J.}\ \bibnamefont {Cava}}, \ and\ \bibinfo {author} {\bibfnamefont
  {M.~Z.}\ \bibnamefont {Hasan}},\ }\href {\doibase 10.1126/science.1167733}
  {\bibfield  {journal} {\bibinfo  {journal} {Science}\ }\textbf {\bibinfo
  {volume} {323}},\ \bibinfo {pages} {919} (\bibinfo {year}
  {2009}{\natexlab{a}})}\BibitemShut {NoStop}%
\bibitem [{\citenamefont {Xia}\ \emph {et~al.}(2009)\citenamefont {Xia},
  \citenamefont {Qian}, \citenamefont {Hsieh}, \citenamefont {Wray},
  \citenamefont {Pal}, \citenamefont {Lin}, \citenamefont {Bansil},
  \citenamefont {Grauer}, \citenamefont {Hor}, \citenamefont {Cava},\ and\
  \citenamefont {Hasan}}]{Xia}%
  \BibitemOpen
  \bibfield  {author} {\bibinfo {author} {\bibfnamefont {Y.}~\bibnamefont
  {Xia}}, \bibinfo {author} {\bibfnamefont {D.}~\bibnamefont {Qian}}, \bibinfo
  {author} {\bibfnamefont {D.}~\bibnamefont {Hsieh}}, \bibinfo {author}
  {\bibfnamefont {L.}~\bibnamefont {Wray}}, \bibinfo {author} {\bibfnamefont
  {A.}~\bibnamefont {Pal}}, \bibinfo {author} {\bibfnamefont {H.}~\bibnamefont
  {Lin}}, \bibinfo {author} {\bibfnamefont {A.}~\bibnamefont {Bansil}},
  \bibinfo {author} {\bibfnamefont {D.}~\bibnamefont {Grauer}}, \bibinfo
  {author} {\bibfnamefont {Y.~S.}\ \bibnamefont {Hor}}, \bibinfo {author}
  {\bibfnamefont {R.~J.}\ \bibnamefont {Cava}}, \ and\ \bibinfo {author}
  {\bibfnamefont {M.~Z.}\ \bibnamefont {Hasan}},\ }\href
  {http://dx.doi.org/10.1038/nphys1274} {\bibfield  {journal} {\bibinfo
  {journal} {Nature Physics}\ }\textbf {\bibinfo {volume} {5}},\ \bibinfo
  {pages} {398 EP } (\bibinfo {year} {2009})}\BibitemShut {NoStop}%
\bibitem [{\citenamefont {Chen}\ \emph {et~al.}(2009)\citenamefont {Chen},
  \citenamefont {Analytis}, \citenamefont {Chu}, \citenamefont {Liu},
  \citenamefont {Mo}, \citenamefont {Qi}, \citenamefont {Zhang}, \citenamefont
  {Lu}, \citenamefont {Dai}, \citenamefont {Fang}, \citenamefont {Zhang},
  \citenamefont {Fisher}, \citenamefont {Hussain},\ and\ \citenamefont
  {Shen}}]{chu}%
  \BibitemOpen
  \bibfield  {author} {\bibinfo {author} {\bibfnamefont {Y.~L.}\ \bibnamefont
  {Chen}}, \bibinfo {author} {\bibfnamefont {J.~G.}\ \bibnamefont {Analytis}},
  \bibinfo {author} {\bibfnamefont {J.-H.}\ \bibnamefont {Chu}}, \bibinfo
  {author} {\bibfnamefont {Z.~K.}\ \bibnamefont {Liu}}, \bibinfo {author}
  {\bibfnamefont {S.-K.}\ \bibnamefont {Mo}}, \bibinfo {author} {\bibfnamefont
  {X.~L.}\ \bibnamefont {Qi}}, \bibinfo {author} {\bibfnamefont {H.~J.}\
  \bibnamefont {Zhang}}, \bibinfo {author} {\bibfnamefont {D.~H.}\ \bibnamefont
  {Lu}}, \bibinfo {author} {\bibfnamefont {X.}~\bibnamefont {Dai}}, \bibinfo
  {author} {\bibfnamefont {Z.}~\bibnamefont {Fang}}, \bibinfo {author}
  {\bibfnamefont {S.~C.}\ \bibnamefont {Zhang}}, \bibinfo {author}
  {\bibfnamefont {I.~R.}\ \bibnamefont {Fisher}}, \bibinfo {author}
  {\bibfnamefont {Z.}~\bibnamefont {Hussain}}, \ and\ \bibinfo {author}
  {\bibfnamefont {Z.-X.}\ \bibnamefont {Shen}},\ }\href {\doibase
  10.1126/science.1173034} {\bibfield  {journal} {\bibinfo  {journal}
  {Science}\ }\textbf {\bibinfo {volume} {325}},\ \bibinfo {pages} {178}
  (\bibinfo {year} {2009})}\BibitemShut {NoStop}%
\bibitem [{\citenamefont {Hor}\ \emph {et~al.}(2009)\citenamefont {Hor},
  \citenamefont {Richardella}, \citenamefont {Roushan}, \citenamefont {Xia},
  \citenamefont {Checkelsky}, \citenamefont {Yazdani}, \citenamefont {Hasan},
  \citenamefont {Ong},\ and\ \citenamefont {Cava}}]{hor}%
  \BibitemOpen
  \bibfield  {author} {\bibinfo {author} {\bibfnamefont {Y.~S.}\ \bibnamefont
  {Hor}}, \bibinfo {author} {\bibfnamefont {A.}~\bibnamefont {Richardella}},
  \bibinfo {author} {\bibfnamefont {P.}~\bibnamefont {Roushan}}, \bibinfo
  {author} {\bibfnamefont {Y.}~\bibnamefont {Xia}}, \bibinfo {author}
  {\bibfnamefont {J.~G.}\ \bibnamefont {Checkelsky}}, \bibinfo {author}
  {\bibfnamefont {A.}~\bibnamefont {Yazdani}}, \bibinfo {author} {\bibfnamefont
  {M.~Z.}\ \bibnamefont {Hasan}}, \bibinfo {author} {\bibfnamefont {N.~P.}\
  \bibnamefont {Ong}}, \ and\ \bibinfo {author} {\bibfnamefont {R.~J.}\
  \bibnamefont {Cava}},\ }\href {\doibase 10.1103/PhysRevB.79.195208}
  {\bibfield  {journal} {\bibinfo  {journal} {Phys. Rev. B}\ }\textbf {\bibinfo
  {volume} {79}},\ \bibinfo {pages} {195208} (\bibinfo {year}
  {2009})}\BibitemShut {NoStop}%
\bibitem [{\citenamefont {Park}\ \emph {et~al.}(2010)\citenamefont {Park},
  \citenamefont {Jung}, \citenamefont {Kim}, \citenamefont {Song},
  \citenamefont {Kim}, \citenamefont {Kimura}, \citenamefont {Lee},\ and\
  \citenamefont {Hur}}]{park}%
  \BibitemOpen
  \bibfield  {author} {\bibinfo {author} {\bibfnamefont {S.~R.}\ \bibnamefont
  {Park}}, \bibinfo {author} {\bibfnamefont {W.~S.}\ \bibnamefont {Jung}},
  \bibinfo {author} {\bibfnamefont {C.}~\bibnamefont {Kim}}, \bibinfo {author}
  {\bibfnamefont {D.~J.}\ \bibnamefont {Song}}, \bibinfo {author}
  {\bibfnamefont {C.}~\bibnamefont {Kim}}, \bibinfo {author} {\bibfnamefont
  {S.}~\bibnamefont {Kimura}}, \bibinfo {author} {\bibfnamefont {K.~D.}\
  \bibnamefont {Lee}}, \ and\ \bibinfo {author} {\bibfnamefont
  {N.}~\bibnamefont {Hur}},\ }\href {\doibase 10.1103/PhysRevB.81.041405}
  {\bibfield  {journal} {\bibinfo  {journal} {Phys. Rev. B}\ }\textbf {\bibinfo
  {volume} {81}},\ \bibinfo {pages} {041405} (\bibinfo {year}
  {2010})}\BibitemShut {NoStop}%
\bibitem [{\citenamefont {Roth}\ \emph {et~al.}(2009)\citenamefont {Roth},
  \citenamefont {Br{\"u}ne}, \citenamefont {Buhmann}, \citenamefont
  {Molenkamp}, \citenamefont {Maciejko}, \citenamefont {Qi},\ and\
  \citenamefont {Zhang}}]{Roth}%
  \BibitemOpen
  \bibfield  {author} {\bibinfo {author} {\bibfnamefont {A.}~\bibnamefont
  {Roth}}, \bibinfo {author} {\bibfnamefont {C.}~\bibnamefont {Br{\"u}ne}},
  \bibinfo {author} {\bibfnamefont {H.}~\bibnamefont {Buhmann}}, \bibinfo
  {author} {\bibfnamefont {L.~W.}\ \bibnamefont {Molenkamp}}, \bibinfo {author}
  {\bibfnamefont {J.}~\bibnamefont {Maciejko}}, \bibinfo {author}
  {\bibfnamefont {X.-L.}\ \bibnamefont {Qi}}, \ and\ \bibinfo {author}
  {\bibfnamefont {S.-C.}\ \bibnamefont {Zhang}},\ }\href {\doibase
  10.1126/science.1174736} {\bibfield  {journal} {\bibinfo  {journal}
  {Science}\ }\textbf {\bibinfo {volume} {325}},\ \bibinfo {pages} {294}
  (\bibinfo {year} {2009})}\BibitemShut {NoStop}%
\bibitem [{\citenamefont {Liu}\ \emph {et~al.}(2008)\citenamefont {Liu},
  \citenamefont {Hughes}, \citenamefont {Qi}, \citenamefont {Wang},\ and\
  \citenamefont {Zhang}}]{LiuHughes}%
  \BibitemOpen
  \bibfield  {author} {\bibinfo {author} {\bibfnamefont {C.}~\bibnamefont
  {Liu}}, \bibinfo {author} {\bibfnamefont {T.~L.}\ \bibnamefont {Hughes}},
  \bibinfo {author} {\bibfnamefont {X.-L.}\ \bibnamefont {Qi}}, \bibinfo
  {author} {\bibfnamefont {K.}~\bibnamefont {Wang}}, \ and\ \bibinfo {author}
  {\bibfnamefont {S.-C.}\ \bibnamefont {Zhang}},\ }\href {\doibase
  10.1103/PhysRevLett.100.236601} {\bibfield  {journal} {\bibinfo  {journal}
  {Phys. Rev. Lett.}\ }\textbf {\bibinfo {volume} {100}},\ \bibinfo {pages}
  {236601} (\bibinfo {year} {2008})}\BibitemShut {NoStop}%
\bibitem [{\citenamefont {Knez}\ \emph {et~al.}(2011)\citenamefont {Knez},
  \citenamefont {Du},\ and\ \citenamefont {Sullivan}}]{Knez}%
  \BibitemOpen
  \bibfield  {author} {\bibinfo {author} {\bibfnamefont {I.}~\bibnamefont
  {Knez}}, \bibinfo {author} {\bibfnamefont {R.-R.}\ \bibnamefont {Du}}, \ and\
  \bibinfo {author} {\bibfnamefont {G.}~\bibnamefont {Sullivan}},\ }\href
  {\doibase 10.1103/PhysRevLett.107.136603} {\bibfield  {journal} {\bibinfo
  {journal} {Phys. Rev. Lett.}\ }\textbf {\bibinfo {volume} {107}},\ \bibinfo
  {pages} {136603} (\bibinfo {year} {2011})}\BibitemShut {NoStop}%
\bibitem [{\citenamefont {Knez}\ \emph {et~al.}(2012)\citenamefont {Knez},
  \citenamefont {Du},\ and\ \citenamefont {Sullivan}}]{Knez2}%
  \BibitemOpen
  \bibfield  {author} {\bibinfo {author} {\bibfnamefont {I.}~\bibnamefont
  {Knez}}, \bibinfo {author} {\bibfnamefont {R.-R.}\ \bibnamefont {Du}}, \ and\
  \bibinfo {author} {\bibfnamefont {G.}~\bibnamefont {Sullivan}},\ }\href
  {\doibase 10.1103/PhysRevLett.109.186603} {\bibfield  {journal} {\bibinfo
  {journal} {Phys. Rev. Lett.}\ }\textbf {\bibinfo {volume} {109}},\ \bibinfo
  {pages} {186603} (\bibinfo {year} {2012})}\BibitemShut {NoStop}%
\bibitem [{\citenamefont {Qian}\ \emph {et~al.}(2014)\citenamefont {Qian},
  \citenamefont {Liu}, \citenamefont {Fu},\ and\ \citenamefont
  {Li}}]{QianXiaofeng}%
  \BibitemOpen
  \bibfield  {author} {\bibinfo {author} {\bibfnamefont {X.}~\bibnamefont
  {Qian}}, \bibinfo {author} {\bibfnamefont {J.}~\bibnamefont {Liu}}, \bibinfo
  {author} {\bibfnamefont {L.}~\bibnamefont {Fu}}, \ and\ \bibinfo {author}
  {\bibfnamefont {J.}~\bibnamefont {Li}},\ }\href {\doibase
  10.1126/science.1256815} {\bibfield  {journal} {\bibinfo  {journal}
  {Science}\ }\textbf {\bibinfo {volume} {346}},\ \bibinfo {pages} {1344}
  (\bibinfo {year} {2014})}\BibitemShut {NoStop}%
\bibitem [{\citenamefont {Hsieh}\ \emph
  {et~al.}(2009{\natexlab{b}})\citenamefont {Hsieh}, \citenamefont {Xia},
  \citenamefont {Qian}, \citenamefont {Wray}, \citenamefont {Dil},
  \citenamefont {Meier}, \citenamefont {Osterwalder}, \citenamefont {Patthey},
  \citenamefont {Checkelsky}, \citenamefont {Ong}, \citenamefont {Fedorov},
  \citenamefont {Lin}, \citenamefont {Bansil}, \citenamefont {Grauer},
  \citenamefont {Hor}, \citenamefont {Cava},\ and\ \citenamefont
  {Hasan}}]{CavaHasan}%
  \BibitemOpen
  \bibfield  {author} {\bibinfo {author} {\bibfnamefont {D.}~\bibnamefont
  {Hsieh}}, \bibinfo {author} {\bibfnamefont {Y.}~\bibnamefont {Xia}}, \bibinfo
  {author} {\bibfnamefont {D.}~\bibnamefont {Qian}}, \bibinfo {author}
  {\bibfnamefont {L.}~\bibnamefont {Wray}}, \bibinfo {author} {\bibfnamefont
  {J.~H.}\ \bibnamefont {Dil}}, \bibinfo {author} {\bibfnamefont
  {F.}~\bibnamefont {Meier}}, \bibinfo {author} {\bibfnamefont
  {J.}~\bibnamefont {Osterwalder}}, \bibinfo {author} {\bibfnamefont
  {L.}~\bibnamefont {Patthey}}, \bibinfo {author} {\bibfnamefont {J.~G.}\
  \bibnamefont {Checkelsky}}, \bibinfo {author} {\bibfnamefont {N.~P.}\
  \bibnamefont {Ong}}, \bibinfo {author} {\bibfnamefont {A.~V.}\ \bibnamefont
  {Fedorov}}, \bibinfo {author} {\bibfnamefont {H.}~\bibnamefont {Lin}},
  \bibinfo {author} {\bibfnamefont {A.}~\bibnamefont {Bansil}}, \bibinfo
  {author} {\bibfnamefont {D.}~\bibnamefont {Grauer}}, \bibinfo {author}
  {\bibfnamefont {Y.~S.}\ \bibnamefont {Hor}}, \bibinfo {author} {\bibfnamefont
  {R.~J.}\ \bibnamefont {Cava}}, \ and\ \bibinfo {author} {\bibfnamefont
  {M.~Z.}\ \bibnamefont {Hasan}},\ }\href
  {http://dx.doi.org/10.1038/nature08234} {\bibfield  {journal} {\bibinfo
  {journal} {Nature}\ }\textbf {\bibinfo {volume} {460}},\ \bibinfo {pages}
  {1101 EP } (\bibinfo {year} {2009}{\natexlab{b}})}\BibitemShut {NoStop}%
\bibitem [{\citenamefont {Hsieh}\ \emph
  {et~al.}(2009{\natexlab{c}})\citenamefont {Hsieh}, \citenamefont {Xia},
  \citenamefont {Qian}, \citenamefont {Wray}, \citenamefont {Meier},
  \citenamefont {Dil}, \citenamefont {Osterwalder}, \citenamefont {Patthey},
  \citenamefont {Fedorov}, \citenamefont {Lin}, \citenamefont {Bansil},
  \citenamefont {Grauer}, \citenamefont {Hor}, \citenamefont {Cava},\ and\
  \citenamefont {Hasan}}]{HorCava}%
  \BibitemOpen
  \bibfield  {author} {\bibinfo {author} {\bibfnamefont {D.}~\bibnamefont
  {Hsieh}}, \bibinfo {author} {\bibfnamefont {Y.}~\bibnamefont {Xia}}, \bibinfo
  {author} {\bibfnamefont {D.}~\bibnamefont {Qian}}, \bibinfo {author}
  {\bibfnamefont {L.}~\bibnamefont {Wray}}, \bibinfo {author} {\bibfnamefont
  {F.}~\bibnamefont {Meier}}, \bibinfo {author} {\bibfnamefont {J.~H.}\
  \bibnamefont {Dil}}, \bibinfo {author} {\bibfnamefont {J.}~\bibnamefont
  {Osterwalder}}, \bibinfo {author} {\bibfnamefont {L.}~\bibnamefont
  {Patthey}}, \bibinfo {author} {\bibfnamefont {A.~V.}\ \bibnamefont
  {Fedorov}}, \bibinfo {author} {\bibfnamefont {H.}~\bibnamefont {Lin}},
  \bibinfo {author} {\bibfnamefont {A.}~\bibnamefont {Bansil}}, \bibinfo
  {author} {\bibfnamefont {D.}~\bibnamefont {Grauer}}, \bibinfo {author}
  {\bibfnamefont {Y.~S.}\ \bibnamefont {Hor}}, \bibinfo {author} {\bibfnamefont
  {R.~J.}\ \bibnamefont {Cava}}, \ and\ \bibinfo {author} {\bibfnamefont
  {M.~Z.}\ \bibnamefont {Hasan}},\ }\href {\doibase
  10.1103/PhysRevLett.103.146401} {\bibfield  {journal} {\bibinfo  {journal}
  {Phys. Rev. Lett.}\ }\textbf {\bibinfo {volume} {103}},\ \bibinfo {pages}
  {146401} (\bibinfo {year} {2009}{\natexlab{c}})}\BibitemShut {NoStop}%
\bibitem [{\citenamefont {Roushan}\ \emph {et~al.}(2009)\citenamefont
  {Roushan}, \citenamefont {Seo}, \citenamefont {Parker}, \citenamefont {Hor},
  \citenamefont {Hsieh}, \citenamefont {Qian}, \citenamefont {Richardella},
  \citenamefont {Hasan}, \citenamefont {Cava},\ and\ \citenamefont
  {Yazdani}}]{Roushan}%
  \BibitemOpen
  \bibfield  {author} {\bibinfo {author} {\bibfnamefont {P.}~\bibnamefont
  {Roushan}}, \bibinfo {author} {\bibfnamefont {J.}~\bibnamefont {Seo}},
  \bibinfo {author} {\bibfnamefont {C.~V.}\ \bibnamefont {Parker}}, \bibinfo
  {author} {\bibfnamefont {Y.~S.}\ \bibnamefont {Hor}}, \bibinfo {author}
  {\bibfnamefont {D.}~\bibnamefont {Hsieh}}, \bibinfo {author} {\bibfnamefont
  {D.}~\bibnamefont {Qian}}, \bibinfo {author} {\bibfnamefont {A.}~\bibnamefont
  {Richardella}}, \bibinfo {author} {\bibfnamefont {M.~Z.}\ \bibnamefont
  {Hasan}}, \bibinfo {author} {\bibfnamefont {R.~J.}\ \bibnamefont {Cava}}, \
  and\ \bibinfo {author} {\bibfnamefont {A.}~\bibnamefont {Yazdani}},\ }\href
  {http://dx.doi.org/10.1038/nature08308} {\bibfield  {journal} {\bibinfo
  {journal} {Nature}\ }\textbf {\bibinfo {volume} {460}},\ \bibinfo {pages}
  {1106 EP } (\bibinfo {year} {2009})}\BibitemShut {NoStop}%
\bibitem [{\citenamefont {Alpichshev}\ \emph {et~al.}(2010)\citenamefont
  {Alpichshev}, \citenamefont {Analytis}, \citenamefont {Chu}, \citenamefont
  {Fisher}, \citenamefont {Chen}, \citenamefont {Shen}, \citenamefont {Fang},\
  and\ \citenamefont {Kapitulnik}}]{STM}%
  \BibitemOpen
  \bibfield  {author} {\bibinfo {author} {\bibfnamefont {Z.}~\bibnamefont
  {Alpichshev}}, \bibinfo {author} {\bibfnamefont {J.~G.}\ \bibnamefont
  {Analytis}}, \bibinfo {author} {\bibfnamefont {J.-H.}\ \bibnamefont {Chu}},
  \bibinfo {author} {\bibfnamefont {I.~R.}\ \bibnamefont {Fisher}}, \bibinfo
  {author} {\bibfnamefont {Y.~L.}\ \bibnamefont {Chen}}, \bibinfo {author}
  {\bibfnamefont {Z.~X.}\ \bibnamefont {Shen}}, \bibinfo {author}
  {\bibfnamefont {A.}~\bibnamefont {Fang}}, \ and\ \bibinfo {author}
  {\bibfnamefont {A.}~\bibnamefont {Kapitulnik}},\ }\href {\doibase
  10.1103/PhysRevLett.104.016401} {\bibfield  {journal} {\bibinfo  {journal}
  {Phys. Rev. Lett.}\ }\textbf {\bibinfo {volume} {104}},\ \bibinfo {pages}
  {016401} (\bibinfo {year} {2010})}\BibitemShut {NoStop}%
\bibitem [{\citenamefont {Chen}\ \emph {et~al.}(2010)\citenamefont {Chen},
  \citenamefont {Chu}, \citenamefont {Analytis}, \citenamefont {Liu},
  \citenamefont {Igarashi}, \citenamefont {Kuo}, \citenamefont {Qi},
  \citenamefont {Mo}, \citenamefont {Moore}, \citenamefont {Lu}, \citenamefont
  {Hashimoto}, \citenamefont {Sasagawa}, \citenamefont {Zhang}, \citenamefont
  {Fisher}, \citenamefont {Hussain},\ and\ \citenamefont {Shen}}]{Sasagawa}%
  \BibitemOpen
  \bibfield  {author} {\bibinfo {author} {\bibfnamefont {Y.~L.}\ \bibnamefont
  {Chen}}, \bibinfo {author} {\bibfnamefont {J.-H.}\ \bibnamefont {Chu}},
  \bibinfo {author} {\bibfnamefont {J.~G.}\ \bibnamefont {Analytis}}, \bibinfo
  {author} {\bibfnamefont {Z.~K.}\ \bibnamefont {Liu}}, \bibinfo {author}
  {\bibfnamefont {K.}~\bibnamefont {Igarashi}}, \bibinfo {author}
  {\bibfnamefont {H.-H.}\ \bibnamefont {Kuo}}, \bibinfo {author} {\bibfnamefont
  {X.~L.}\ \bibnamefont {Qi}}, \bibinfo {author} {\bibfnamefont {S.~K.}\
  \bibnamefont {Mo}}, \bibinfo {author} {\bibfnamefont {R.~G.}\ \bibnamefont
  {Moore}}, \bibinfo {author} {\bibfnamefont {D.~H.}\ \bibnamefont {Lu}},
  \bibinfo {author} {\bibfnamefont {M.}~\bibnamefont {Hashimoto}}, \bibinfo
  {author} {\bibfnamefont {T.}~\bibnamefont {Sasagawa}}, \bibinfo {author}
  {\bibfnamefont {S.~C.}\ \bibnamefont {Zhang}}, \bibinfo {author}
  {\bibfnamefont {I.~R.}\ \bibnamefont {Fisher}}, \bibinfo {author}
  {\bibfnamefont {Z.}~\bibnamefont {Hussain}}, \ and\ \bibinfo {author}
  {\bibfnamefont {Z.~X.}\ \bibnamefont {Shen}},\ }\href {\doibase
  10.1126/science.1189924} {\bibfield  {journal} {\bibinfo  {journal}
  {Science}\ }\textbf {\bibinfo {volume} {329}},\ \bibinfo {pages} {659}
  (\bibinfo {year} {2010})}\BibitemShut {NoStop}%
\bibitem [{\citenamefont {Seo}\ \emph {et~al.}(2010)\citenamefont {Seo},
  \citenamefont {Roushan}, \citenamefont {Beidenkopf}, \citenamefont {Hor},
  \citenamefont {Cava},\ and\ \citenamefont {Yazdani}}]{Seo}%
  \BibitemOpen
  \bibfield  {author} {\bibinfo {author} {\bibfnamefont {J.}~\bibnamefont
  {Seo}}, \bibinfo {author} {\bibfnamefont {P.}~\bibnamefont {Roushan}},
  \bibinfo {author} {\bibfnamefont {H.}~\bibnamefont {Beidenkopf}}, \bibinfo
  {author} {\bibfnamefont {Y.~S.}\ \bibnamefont {Hor}}, \bibinfo {author}
  {\bibfnamefont {R.~J.}\ \bibnamefont {Cava}}, \ and\ \bibinfo {author}
  {\bibfnamefont {A.}~\bibnamefont {Yazdani}},\ }\href
  {http://dx.doi.org/10.1038/nature09189} {\bibfield  {journal} {\bibinfo
  {journal} {Nature}\ }\textbf {\bibinfo {volume} {466}},\ \bibinfo {pages}
  {343 EP } (\bibinfo {year} {2010})}\BibitemShut {NoStop}%
\bibitem [{\citenamefont {Checkelsky}\ \emph {et~al.}(2011)\citenamefont
  {Checkelsky}, \citenamefont {Hor}, \citenamefont {Cava},\ and\ \citenamefont
  {Ong}}]{Ong}%
  \BibitemOpen
  \bibfield  {author} {\bibinfo {author} {\bibfnamefont {J.~G.}\ \bibnamefont
  {Checkelsky}}, \bibinfo {author} {\bibfnamefont {Y.~S.}\ \bibnamefont {Hor}},
  \bibinfo {author} {\bibfnamefont {R.~J.}\ \bibnamefont {Cava}}, \ and\
  \bibinfo {author} {\bibfnamefont {N.~P.}\ \bibnamefont {Ong}},\ }\href
  {\doibase 10.1103/PhysRevLett.106.196801} {\bibfield  {journal} {\bibinfo
  {journal} {Phys. Rev. Lett.}\ }\textbf {\bibinfo {volume} {106}},\ \bibinfo
  {pages} {196801} (\bibinfo {year} {2011})}\BibitemShut {NoStop}%
\bibitem [{\citenamefont {Okada}\ \emph {et~al.}(2011)\citenamefont {Okada},
  \citenamefont {Dhital}, \citenamefont {Zhou}, \citenamefont {Huemiller},
  \citenamefont {Lin}, \citenamefont {Basak}, \citenamefont {Bansil},
  \citenamefont {Huang}, \citenamefont {Ding}, \citenamefont {Wang},
  \citenamefont {Wilson},\ and\ \citenamefont {Madhavan}}]{Okada}%
  \BibitemOpen
  \bibfield  {author} {\bibinfo {author} {\bibfnamefont {Y.}~\bibnamefont
  {Okada}}, \bibinfo {author} {\bibfnamefont {C.}~\bibnamefont {Dhital}},
  \bibinfo {author} {\bibfnamefont {W.}~\bibnamefont {Zhou}}, \bibinfo {author}
  {\bibfnamefont {E.~D.}\ \bibnamefont {Huemiller}}, \bibinfo {author}
  {\bibfnamefont {H.}~\bibnamefont {Lin}}, \bibinfo {author} {\bibfnamefont
  {S.}~\bibnamefont {Basak}}, \bibinfo {author} {\bibfnamefont
  {A.}~\bibnamefont {Bansil}}, \bibinfo {author} {\bibfnamefont {Y.-B.}\
  \bibnamefont {Huang}}, \bibinfo {author} {\bibfnamefont {H.}~\bibnamefont
  {Ding}}, \bibinfo {author} {\bibfnamefont {Z.}~\bibnamefont {Wang}}, \bibinfo
  {author} {\bibfnamefont {S.~D.}\ \bibnamefont {Wilson}}, \ and\ \bibinfo
  {author} {\bibfnamefont {V.}~\bibnamefont {Madhavan}},\ }\href {\doibase
  10.1103/PhysRevLett.106.206805} {\bibfield  {journal} {\bibinfo  {journal}
  {Phys. Rev. Lett.}\ }\textbf {\bibinfo {volume} {106}},\ \bibinfo {pages}
  {206805} (\bibinfo {year} {2011})}\BibitemShut {NoStop}%
\bibitem [{\citenamefont {Chang}\ \emph {et~al.}(2013)\citenamefont {Chang},
  \citenamefont {Zhang}, \citenamefont {Feng}, \citenamefont {Shen},
  \citenamefont {Zhang}, \citenamefont {Guo}, \citenamefont {Li}, \citenamefont
  {Ou}, \citenamefont {Wei}, \citenamefont {Wang}, \citenamefont {Ji},
  \citenamefont {Feng}, \citenamefont {Ji}, \citenamefont {Chen}, \citenamefont
  {Jia}, \citenamefont {Dai}, \citenamefont {Fang}, \citenamefont {Zhang},
  \citenamefont {He}, \citenamefont {Wang}, \citenamefont {Lu}, \citenamefont
  {Ma},\ and\ \citenamefont {Xue}}]{QiKun}%
  \BibitemOpen
  \bibfield  {author} {\bibinfo {author} {\bibfnamefont {C.-Z.}\ \bibnamefont
  {Chang}}, \bibinfo {author} {\bibfnamefont {J.}~\bibnamefont {Zhang}},
  \bibinfo {author} {\bibfnamefont {X.}~\bibnamefont {Feng}}, \bibinfo {author}
  {\bibfnamefont {J.}~\bibnamefont {Shen}}, \bibinfo {author} {\bibfnamefont
  {Z.}~\bibnamefont {Zhang}}, \bibinfo {author} {\bibfnamefont
  {M.}~\bibnamefont {Guo}}, \bibinfo {author} {\bibfnamefont {K.}~\bibnamefont
  {Li}}, \bibinfo {author} {\bibfnamefont {Y.}~\bibnamefont {Ou}}, \bibinfo
  {author} {\bibfnamefont {P.}~\bibnamefont {Wei}}, \bibinfo {author}
  {\bibfnamefont {L.-L.}\ \bibnamefont {Wang}}, \bibinfo {author}
  {\bibfnamefont {Z.-Q.}\ \bibnamefont {Ji}}, \bibinfo {author} {\bibfnamefont
  {Y.}~\bibnamefont {Feng}}, \bibinfo {author} {\bibfnamefont {S.}~\bibnamefont
  {Ji}}, \bibinfo {author} {\bibfnamefont {X.}~\bibnamefont {Chen}}, \bibinfo
  {author} {\bibfnamefont {J.}~\bibnamefont {Jia}}, \bibinfo {author}
  {\bibfnamefont {X.}~\bibnamefont {Dai}}, \bibinfo {author} {\bibfnamefont
  {Z.}~\bibnamefont {Fang}}, \bibinfo {author} {\bibfnamefont {S.-C.}\
  \bibnamefont {Zhang}}, \bibinfo {author} {\bibfnamefont {K.}~\bibnamefont
  {He}}, \bibinfo {author} {\bibfnamefont {Y.}~\bibnamefont {Wang}}, \bibinfo
  {author} {\bibfnamefont {L.}~\bibnamefont {Lu}}, \bibinfo {author}
  {\bibfnamefont {X.-C.}\ \bibnamefont {Ma}}, \ and\ \bibinfo {author}
  {\bibfnamefont {Q.-K.}\ \bibnamefont {Xue}},\ }\href {\doibase
  10.1126/science.1234414} {\bibfield  {journal} {\bibinfo  {journal}
  {Science}\ }\textbf {\bibinfo {volume} {340}},\ \bibinfo {pages} {167}
  (\bibinfo {year} {2013})}\BibitemShut {NoStop}%
\bibitem [{\citenamefont {Altland}\ and\ \citenamefont {Zirnbauer}(1997)}]{az}%
  \BibitemOpen
  \bibfield  {author} {\bibinfo {author} {\bibfnamefont {A.}~\bibnamefont
  {Altland}}\ and\ \bibinfo {author} {\bibfnamefont {M.~R.}\ \bibnamefont
  {Zirnbauer}},\ }\href {\doibase 10.1103/PhysRevB.55.1142} {\bibfield
  {journal} {\bibinfo  {journal} {Phys. Rev. B}\ }\textbf {\bibinfo {volume}
  {55}},\ \bibinfo {pages} {1142} (\bibinfo {year} {1997})}\BibitemShut
  {NoStop}%
\bibitem [{\citenamefont {Fu}(2011)}]{fucrystalline}%
  \BibitemOpen
  \bibfield  {author} {\bibinfo {author} {\bibfnamefont {L.}~\bibnamefont
  {Fu}},\ }\href {\doibase 10.1103/PhysRevLett.106.106802} {\bibfield
  {journal} {\bibinfo  {journal} {Phys. Rev. Lett.}\ }\textbf {\bibinfo
  {volume} {106}},\ \bibinfo {pages} {106802} (\bibinfo {year}
  {2011})}\BibitemShut {NoStop}%
\bibitem [{\citenamefont {Ando}\ and\ \citenamefont {Fu}(2015)}]{fureview}%
  \BibitemOpen
  \bibfield  {author} {\bibinfo {author} {\bibfnamefont {Y.}~\bibnamefont
  {Ando}}\ and\ \bibinfo {author} {\bibfnamefont {L.}~\bibnamefont {Fu}},\
  }\href {\doibase 10.1146/annurev-conmatphys-031214-014501} {\bibfield
  {journal} {\bibinfo  {journal} {Annual Review of Condensed Matter Physics}\
  }\textbf {\bibinfo {volume} {6}},\ \bibinfo {pages} {361} (\bibinfo {year}
  {2015})}\BibitemShut {NoStop}%
\bibitem [{\citenamefont {Hsieh}\ \emph {et~al.}(2012)\citenamefont {Hsieh},
  \citenamefont {Lin}, \citenamefont {Liu}, \citenamefont {Duan}, \citenamefont
  {Bansil},\ and\ \citenamefont {Fu}}]{hsin}%
  \BibitemOpen
  \bibfield  {author} {\bibinfo {author} {\bibfnamefont {T.~H.}\ \bibnamefont
  {Hsieh}}, \bibinfo {author} {\bibfnamefont {H.}~\bibnamefont {Lin}}, \bibinfo
  {author} {\bibfnamefont {J.}~\bibnamefont {Liu}}, \bibinfo {author}
  {\bibfnamefont {W.}~\bibnamefont {Duan}}, \bibinfo {author} {\bibfnamefont
  {A.}~\bibnamefont {Bansil}}, \ and\ \bibinfo {author} {\bibfnamefont
  {L.}~\bibnamefont {Fu}},\ }\href {http://dx.doi.org/10.1038/ncomms1969}
  {\bibfield  {journal} {\bibinfo  {journal} {Nature Communications}\ }\textbf
  {\bibinfo {volume} {3}},\ \bibinfo {pages} {982 EP } (\bibinfo {year}
  {2012})}\BibitemShut {NoStop}%
\bibitem [{\citenamefont {Hughes}\ \emph {et~al.}(2011)\citenamefont {Hughes},
  \citenamefont {Prodan},\ and\ \citenamefont {Bernevig}}]{ProdanBernevig}%
  \BibitemOpen
  \bibfield  {author} {\bibinfo {author} {\bibfnamefont {T.~L.}\ \bibnamefont
  {Hughes}}, \bibinfo {author} {\bibfnamefont {E.}~\bibnamefont {Prodan}}, \
  and\ \bibinfo {author} {\bibfnamefont {B.~A.}\ \bibnamefont {Bernevig}},\
  }\href {\doibase 10.1103/PhysRevB.83.245132} {\bibfield  {journal} {\bibinfo
  {journal} {Phys. Rev. B}\ }\textbf {\bibinfo {volume} {83}},\ \bibinfo
  {pages} {245132} (\bibinfo {year} {2011})}\BibitemShut {NoStop}%
\bibitem [{\citenamefont {Turner}\ \emph {et~al.}(2010)\citenamefont {Turner},
  \citenamefont {Zhang},\ and\ \citenamefont {Vishwanath}}]{VishwanathYi}%
  \BibitemOpen
  \bibfield  {author} {\bibinfo {author} {\bibfnamefont {A.~M.}\ \bibnamefont
  {Turner}}, \bibinfo {author} {\bibfnamefont {Y.}~\bibnamefont {Zhang}}, \
  and\ \bibinfo {author} {\bibfnamefont {A.}~\bibnamefont {Vishwanath}},\
  }\href {\doibase 10.1103/PhysRevB.82.241102} {\bibfield  {journal} {\bibinfo
  {journal} {Phys. Rev. B}\ }\textbf {\bibinfo {volume} {82}},\ \bibinfo
  {pages} {241102} (\bibinfo {year} {2010})}\BibitemShut {NoStop}%
\bibitem [{\citenamefont {Turner}\ \emph {et~al.}(2012)\citenamefont {Turner},
  \citenamefont {Zhang}, \citenamefont {Mong},\ and\ \citenamefont
  {Vishwanath}}]{TurnerMong}%
  \BibitemOpen
  \bibfield  {author} {\bibinfo {author} {\bibfnamefont {A.~M.}\ \bibnamefont
  {Turner}}, \bibinfo {author} {\bibfnamefont {Y.}~\bibnamefont {Zhang}},
  \bibinfo {author} {\bibfnamefont {R.~S.~K.}\ \bibnamefont {Mong}}, \ and\
  \bibinfo {author} {\bibfnamefont {A.}~\bibnamefont {Vishwanath}},\ }\href
  {\doibase 10.1103/PhysRevB.85.165120} {\bibfield  {journal} {\bibinfo
  {journal} {Phys. Rev. B}\ }\textbf {\bibinfo {volume} {85}},\ \bibinfo
  {pages} {165120} (\bibinfo {year} {2012})}\BibitemShut {NoStop}%
\bibitem [{\citenamefont {Mong}\ \emph {et~al.}(2010)\citenamefont {Mong},
  \citenamefont {Essin},\ and\ \citenamefont {Moore}}]{mong}%
  \BibitemOpen
  \bibfield  {author} {\bibinfo {author} {\bibfnamefont {R.~S.~K.}\
  \bibnamefont {Mong}}, \bibinfo {author} {\bibfnamefont {A.~M.}\ \bibnamefont
  {Essin}}, \ and\ \bibinfo {author} {\bibfnamefont {J.~E.}\ \bibnamefont
  {Moore}},\ }\href {\doibase 10.1103/PhysRevB.81.245209} {\bibfield  {journal}
  {\bibinfo  {journal} {Phys. Rev. B}\ }\textbf {\bibinfo {volume} {81}},\
  \bibinfo {pages} {245209} (\bibinfo {year} {2010})}\BibitemShut {NoStop}%
\bibitem [{\citenamefont {Fang}\ \emph
  {et~al.}(2012{\natexlab{a}})\citenamefont {Fang}, \citenamefont {Gilbert},\
  and\ \citenamefont {Bernevig}}]{FangGilbertBernevig}%
  \BibitemOpen
  \bibfield  {author} {\bibinfo {author} {\bibfnamefont {C.}~\bibnamefont
  {Fang}}, \bibinfo {author} {\bibfnamefont {M.~J.}\ \bibnamefont {Gilbert}}, \
  and\ \bibinfo {author} {\bibfnamefont {B.~A.}\ \bibnamefont {Bernevig}},\
  }\href {\doibase 10.1103/PhysRevB.86.115112} {\bibfield  {journal} {\bibinfo
  {journal} {Phys. Rev. B}\ }\textbf {\bibinfo {volume} {86}},\ \bibinfo
  {pages} {115112} (\bibinfo {year} {2012}{\natexlab{a}})}\BibitemShut
  {NoStop}%
\bibitem [{\citenamefont {Liu}\ \emph {et~al.}(2013{\natexlab{a}})\citenamefont
  {Liu}, \citenamefont {Duan},\ and\ \citenamefont {Fu}}]{Duan}%
  \BibitemOpen
  \bibfield  {author} {\bibinfo {author} {\bibfnamefont {J.}~\bibnamefont
  {Liu}}, \bibinfo {author} {\bibfnamefont {W.}~\bibnamefont {Duan}}, \ and\
  \bibinfo {author} {\bibfnamefont {L.}~\bibnamefont {Fu}},\ }\href {\doibase
  10.1103/PhysRevB.88.241303} {\bibfield  {journal} {\bibinfo  {journal} {Phys.
  Rev. B}\ }\textbf {\bibinfo {volume} {88}},\ \bibinfo {pages} {241303}
  (\bibinfo {year} {2013}{\natexlab{a}})}\BibitemShut {NoStop}%
\bibitem [{\citenamefont {Fang}\ \emph
  {et~al.}(2014{\natexlab{a}})\citenamefont {Fang}, \citenamefont {Gilbert},\
  and\ \citenamefont {Bernevig}}]{FangChen}%
  \BibitemOpen
  \bibfield  {author} {\bibinfo {author} {\bibfnamefont {C.}~\bibnamefont
  {Fang}}, \bibinfo {author} {\bibfnamefont {M.~J.}\ \bibnamefont {Gilbert}}, \
  and\ \bibinfo {author} {\bibfnamefont {B.~A.}\ \bibnamefont {Bernevig}},\
  }\href {\doibase 10.1103/PhysRevLett.112.046801} {\bibfield  {journal}
  {\bibinfo  {journal} {Phys. Rev. Lett.}\ }\textbf {\bibinfo {volume} {112}},\
  \bibinfo {pages} {046801} (\bibinfo {year} {2014}{\natexlab{a}})}\BibitemShut
  {NoStop}%
\bibitem [{\citenamefont {Fang}\ \emph
  {et~al.}(2014{\natexlab{b}})\citenamefont {Fang}, \citenamefont {Gilbert},\
  and\ \citenamefont {Bernevig}}]{Matthew}%
  \BibitemOpen
  \bibfield  {author} {\bibinfo {author} {\bibfnamefont {C.}~\bibnamefont
  {Fang}}, \bibinfo {author} {\bibfnamefont {M.~J.}\ \bibnamefont {Gilbert}}, \
  and\ \bibinfo {author} {\bibfnamefont {B.~A.}\ \bibnamefont {Bernevig}},\
  }\href {\doibase 10.1103/PhysRevLett.112.106401} {\bibfield  {journal}
  {\bibinfo  {journal} {Phys. Rev. Lett.}\ }\textbf {\bibinfo {volume} {112}},\
  \bibinfo {pages} {106401} (\bibinfo {year} {2014}{\natexlab{b}})}\BibitemShut
  {NoStop}%
\bibitem [{\citenamefont {Liu}\ \emph {et~al.}(2014)\citenamefont {Liu},
  \citenamefont {Zhang},\ and\ \citenamefont {VanLeeuwen}}]{VanLeeuwen}%
  \BibitemOpen
  \bibfield  {author} {\bibinfo {author} {\bibfnamefont {C.-X.}\ \bibnamefont
  {Liu}}, \bibinfo {author} {\bibfnamefont {R.-X.}\ \bibnamefont {Zhang}}, \
  and\ \bibinfo {author} {\bibfnamefont {B.~K.}\ \bibnamefont {VanLeeuwen}},\
  }\href {\doibase 10.1103/PhysRevB.90.085304} {\bibfield  {journal} {\bibinfo
  {journal} {Phys. Rev. B}\ }\textbf {\bibinfo {volume} {90}},\ \bibinfo
  {pages} {085304} (\bibinfo {year} {2014})}\BibitemShut {NoStop}%
\bibitem [{\citenamefont {Fang}\ and\ \citenamefont {Fu}(2015)}]{FangFu}%
  \BibitemOpen
  \bibfield  {author} {\bibinfo {author} {\bibfnamefont {C.}~\bibnamefont
  {Fang}}\ and\ \bibinfo {author} {\bibfnamefont {L.}~\bibnamefont {Fu}},\
  }\href {\doibase 10.1103/PhysRevB.91.161105} {\bibfield  {journal} {\bibinfo
  {journal} {Phys. Rev. B}\ }\textbf {\bibinfo {volume} {91}},\ \bibinfo
  {pages} {161105} (\bibinfo {year} {2015})}\BibitemShut {NoStop}%
\bibitem [{\citenamefont {Zhang}\ \emph {et~al.}(2015)\citenamefont {Zhang},
  \citenamefont {Cheng},\ and\ \citenamefont
  {Schwingenschl{\"o}gl}}]{yingchun}%
  \BibitemOpen
  \bibfield  {author} {\bibinfo {author} {\bibfnamefont {Q.}~\bibnamefont
  {Zhang}}, \bibinfo {author} {\bibfnamefont {Y.}~\bibnamefont {Cheng}}, \ and\
  \bibinfo {author} {\bibfnamefont {U.}~\bibnamefont {Schwingenschl{\"o}gl}},\
  }\href {http://dx.doi.org/10.1038/srep08379} {\bibfield  {journal} {\bibinfo
  {journal} {Scientific Reports}\ }\textbf {\bibinfo {volume} {5}},\ \bibinfo
  {pages} {8379 EP } (\bibinfo {year} {2015})}\BibitemShut {NoStop}%
\bibitem [{\citenamefont {Slager}\ \emph {et~al.}(2012)\citenamefont {Slager},
  \citenamefont {Mesaros}, \citenamefont {Juri{\v c}i{\'c}},\ and\
  \citenamefont {Zaanen}}]{slager}%
  \BibitemOpen
  \bibfield  {author} {\bibinfo {author} {\bibfnamefont {R.-J.}\ \bibnamefont
  {Slager}}, \bibinfo {author} {\bibfnamefont {A.}~\bibnamefont {Mesaros}},
  \bibinfo {author} {\bibfnamefont {V.}~\bibnamefont {Juri{\v c}i{\'c}}}, \
  and\ \bibinfo {author} {\bibfnamefont {J.}~\bibnamefont {Zaanen}},\ }\href
  {http://dx.doi.org/10.1038/nphys2513} {\bibfield  {journal} {\bibinfo
  {journal} {Nature Physics}\ }\textbf {\bibinfo {volume} {9}},\ \bibinfo
  {pages} {98 EP } (\bibinfo {year} {2012})}\BibitemShut {NoStop}%
\bibitem [{\citenamefont {Wang}\ \emph {et~al.}(2016)\citenamefont {Wang},
  \citenamefont {Alexandradinata}, \citenamefont {Cava},\ and\ \citenamefont
  {Bernevig}}]{Hourglass}%
  \BibitemOpen
  \bibfield  {author} {\bibinfo {author} {\bibfnamefont {Z.}~\bibnamefont
  {Wang}}, \bibinfo {author} {\bibfnamefont {A.}~\bibnamefont
  {Alexandradinata}}, \bibinfo {author} {\bibfnamefont {R.~J.}\ \bibnamefont
  {Cava}}, \ and\ \bibinfo {author} {\bibfnamefont {B.~A.}\ \bibnamefont
  {Bernevig}},\ }\href {http://dx.doi.org/10.1038/nature17410} {\bibfield
  {journal} {\bibinfo  {journal} {Nature}\ }\textbf {\bibinfo {volume} {532}},\
  \bibinfo {pages} {189 EP } (\bibinfo {year} {2016})}\BibitemShut {NoStop}%
\bibitem [{\citenamefont {Dziawa}\ \emph {et~al.}(2012)\citenamefont {Dziawa},
  \citenamefont {Kowalski}, \citenamefont {Dybko}, \citenamefont {Buczko},
  \citenamefont {Szczerbakow}, \citenamefont {Szot}, \citenamefont
  {{\L}usakowska}, \citenamefont {Balasubramanian}, \citenamefont {Wojek},
  \citenamefont {Berntsen}, \citenamefont {Tjernberg},\ and\ \citenamefont
  {Story}}]{Dziawa}%
  \BibitemOpen
  \bibfield  {author} {\bibinfo {author} {\bibfnamefont {P.}~\bibnamefont
  {Dziawa}}, \bibinfo {author} {\bibfnamefont {B.~J.}\ \bibnamefont
  {Kowalski}}, \bibinfo {author} {\bibfnamefont {K.}~\bibnamefont {Dybko}},
  \bibinfo {author} {\bibfnamefont {R.}~\bibnamefont {Buczko}}, \bibinfo
  {author} {\bibfnamefont {A.}~\bibnamefont {Szczerbakow}}, \bibinfo {author}
  {\bibfnamefont {M.}~\bibnamefont {Szot}}, \bibinfo {author} {\bibfnamefont
  {E.}~\bibnamefont {{\L}usakowska}}, \bibinfo {author} {\bibfnamefont
  {T.}~\bibnamefont {Balasubramanian}}, \bibinfo {author} {\bibfnamefont
  {B.~M.}\ \bibnamefont {Wojek}}, \bibinfo {author} {\bibfnamefont {M.~H.}\
  \bibnamefont {Berntsen}}, \bibinfo {author} {\bibfnamefont {O.}~\bibnamefont
  {Tjernberg}}, \ and\ \bibinfo {author} {\bibfnamefont {T.}~\bibnamefont
  {Story}},\ }\href {http://dx.doi.org/10.1038/nmat3449} {\bibfield  {journal}
  {\bibinfo  {journal} {Nature Materials}\ }\textbf {\bibinfo {volume} {11}},\
  \bibinfo {pages} {1023 EP } (\bibinfo {year} {2012})}\BibitemShut {NoStop}%
\bibitem [{\citenamefont {Tanaka}\ \emph {et~al.}(2012)\citenamefont {Tanaka},
  \citenamefont {Ren}, \citenamefont {Sato}, \citenamefont {Nakayama},
  \citenamefont {Souma}, \citenamefont {Takahashi}, \citenamefont {Segawa},\
  and\ \citenamefont {Ando}}]{Tanaka}%
  \BibitemOpen
  \bibfield  {author} {\bibinfo {author} {\bibfnamefont {Y.}~\bibnamefont
  {Tanaka}}, \bibinfo {author} {\bibfnamefont {Z.}~\bibnamefont {Ren}},
  \bibinfo {author} {\bibfnamefont {T.}~\bibnamefont {Sato}}, \bibinfo {author}
  {\bibfnamefont {K.}~\bibnamefont {Nakayama}}, \bibinfo {author}
  {\bibfnamefont {S.}~\bibnamefont {Souma}}, \bibinfo {author} {\bibfnamefont
  {T.}~\bibnamefont {Takahashi}}, \bibinfo {author} {\bibfnamefont
  {K.}~\bibnamefont {Segawa}}, \ and\ \bibinfo {author} {\bibfnamefont
  {Y.}~\bibnamefont {Ando}},\ }\href {http://dx.doi.org/10.1038/nphys2442}
  {\bibfield  {journal} {\bibinfo  {journal} {Nature Physics}\ }\textbf
  {\bibinfo {volume} {8}},\ \bibinfo {pages} {800 EP } (\bibinfo {year}
  {2012})}\BibitemShut {NoStop}%
\bibitem [{\citenamefont {Xu}\ \emph {et~al.}(2012)\citenamefont {Xu},
  \citenamefont {Liu}, \citenamefont {Alidoust}, \citenamefont {Neupane},
  \citenamefont {Qian}, \citenamefont {Belopolski}, \citenamefont {Denlinger},
  \citenamefont {Wang}, \citenamefont {Lin}, \citenamefont {Wray},
  \citenamefont {Landolt}, \citenamefont {Slomski}, \citenamefont {Dil},
  \citenamefont {Marcinkova}, \citenamefont {Morosan}, \citenamefont {Gibson},
  \citenamefont {Sankar}, \citenamefont {Chou}, \citenamefont {Cava},
  \citenamefont {Bansil},\ and\ \citenamefont {Hasan}}]{Xu}%
  \BibitemOpen
  \bibfield  {author} {\bibinfo {author} {\bibfnamefont {S.-Y.}\ \bibnamefont
  {Xu}}, \bibinfo {author} {\bibfnamefont {C.}~\bibnamefont {Liu}}, \bibinfo
  {author} {\bibfnamefont {N.}~\bibnamefont {Alidoust}}, \bibinfo {author}
  {\bibfnamefont {M.}~\bibnamefont {Neupane}}, \bibinfo {author} {\bibfnamefont
  {D.}~\bibnamefont {Qian}}, \bibinfo {author} {\bibfnamefont {I.}~\bibnamefont
  {Belopolski}}, \bibinfo {author} {\bibfnamefont {J.~D.}\ \bibnamefont
  {Denlinger}}, \bibinfo {author} {\bibfnamefont {Y.~J.}\ \bibnamefont {Wang}},
  \bibinfo {author} {\bibfnamefont {H.}~\bibnamefont {Lin}}, \bibinfo {author}
  {\bibfnamefont {L.~A.}\ \bibnamefont {Wray}}, \bibinfo {author}
  {\bibfnamefont {G.}~\bibnamefont {Landolt}}, \bibinfo {author} {\bibfnamefont
  {B.}~\bibnamefont {Slomski}}, \bibinfo {author} {\bibfnamefont {J.~H.}\
  \bibnamefont {Dil}}, \bibinfo {author} {\bibfnamefont {A.}~\bibnamefont
  {Marcinkova}}, \bibinfo {author} {\bibfnamefont {E.}~\bibnamefont {Morosan}},
  \bibinfo {author} {\bibfnamefont {Q.}~\bibnamefont {Gibson}}, \bibinfo
  {author} {\bibfnamefont {R.}~\bibnamefont {Sankar}}, \bibinfo {author}
  {\bibfnamefont {F.~C.}\ \bibnamefont {Chou}}, \bibinfo {author}
  {\bibfnamefont {R.~J.}\ \bibnamefont {Cava}}, \bibinfo {author}
  {\bibfnamefont {A.}~\bibnamefont {Bansil}}, \ and\ \bibinfo {author}
  {\bibfnamefont {M.~Z.}\ \bibnamefont {Hasan}},\ }\href
  {http://dx.doi.org/10.1038/ncomms2191} {\bibfield  {journal} {\bibinfo
  {journal} {Nature Communications}\ }\textbf {\bibinfo {volume} {3}},\
  \bibinfo {pages} {1192 EP } (\bibinfo {year} {2012})}\BibitemShut {NoStop}%
\bibitem [{\citenamefont {Liu}\ \emph {et~al.}(2013{\natexlab{b}})\citenamefont
  {Liu}, \citenamefont {Hsieh}, \citenamefont {Wei}, \citenamefont {Duan},
  \citenamefont {Moodera},\ and\ \citenamefont {Fu}}]{Wei}%
  \BibitemOpen
  \bibfield  {author} {\bibinfo {author} {\bibfnamefont {J.}~\bibnamefont
  {Liu}}, \bibinfo {author} {\bibfnamefont {T.~H.}\ \bibnamefont {Hsieh}},
  \bibinfo {author} {\bibfnamefont {P.}~\bibnamefont {Wei}}, \bibinfo {author}
  {\bibfnamefont {W.}~\bibnamefont {Duan}}, \bibinfo {author} {\bibfnamefont
  {J.}~\bibnamefont {Moodera}}, \ and\ \bibinfo {author} {\bibfnamefont
  {L.}~\bibnamefont {Fu}},\ }\href {http://dx.doi.org/10.1038/nmat3828}
  {\bibfield  {journal} {\bibinfo  {journal} {Nature Materials}\ }\textbf
  {\bibinfo {volume} {13}},\ \bibinfo {pages} {178 EP } (\bibinfo {year}
  {2013}{\natexlab{b}})}\BibitemShut {NoStop}%
\bibitem [{\citenamefont {Tanaka}\ \emph {et~al.}(2013)\citenamefont {Tanaka},
  \citenamefont {Sato}, \citenamefont {Nakayama}, \citenamefont {Souma},
  \citenamefont {Takahashi}, \citenamefont {Ren}, \citenamefont {Novak},
  \citenamefont {Segawa},\ and\ \citenamefont {Ando}}]{sato}%
  \BibitemOpen
  \bibfield  {author} {\bibinfo {author} {\bibfnamefont {Y.}~\bibnamefont
  {Tanaka}}, \bibinfo {author} {\bibfnamefont {T.}~\bibnamefont {Sato}},
  \bibinfo {author} {\bibfnamefont {K.}~\bibnamefont {Nakayama}}, \bibinfo
  {author} {\bibfnamefont {S.}~\bibnamefont {Souma}}, \bibinfo {author}
  {\bibfnamefont {T.}~\bibnamefont {Takahashi}}, \bibinfo {author}
  {\bibfnamefont {Z.}~\bibnamefont {Ren}}, \bibinfo {author} {\bibfnamefont
  {M.}~\bibnamefont {Novak}}, \bibinfo {author} {\bibfnamefont
  {K.}~\bibnamefont {Segawa}}, \ and\ \bibinfo {author} {\bibfnamefont
  {Y.}~\bibnamefont {Ando}},\ }\href {\doibase 10.1103/PhysRevB.87.155105}
  {\bibfield  {journal} {\bibinfo  {journal} {Phys. Rev. B}\ }\textbf {\bibinfo
  {volume} {87}},\ \bibinfo {pages} {155105} (\bibinfo {year}
  {2013})}\BibitemShut {NoStop}%
\bibitem [{\citenamefont {Ma}\ \emph {et~al.}(2017)\citenamefont {Ma},
  \citenamefont {Yi}, \citenamefont {Lv}, \citenamefont {Wang}, \citenamefont
  {Nie}, \citenamefont {Wang}, \citenamefont {Kong}, \citenamefont {Huang},
  \citenamefont {Richard}, \citenamefont {Zhang}, \citenamefont {Yaji},
  \citenamefont {Kuroda}, \citenamefont {Shin}, \citenamefont {Weng},
  \citenamefont {Bernevig}, \citenamefont {Shi}, \citenamefont {Qian},\ and\
  \citenamefont {Ding}}]{MaYi}%
  \BibitemOpen
  \bibfield  {author} {\bibinfo {author} {\bibfnamefont {J.}~\bibnamefont
  {Ma}}, \bibinfo {author} {\bibfnamefont {C.}~\bibnamefont {Yi}}, \bibinfo
  {author} {\bibfnamefont {B.}~\bibnamefont {Lv}}, \bibinfo {author}
  {\bibfnamefont {Z.}~\bibnamefont {Wang}}, \bibinfo {author} {\bibfnamefont
  {S.}~\bibnamefont {Nie}}, \bibinfo {author} {\bibfnamefont {L.}~\bibnamefont
  {Wang}}, \bibinfo {author} {\bibfnamefont {L.}~\bibnamefont {Kong}}, \bibinfo
  {author} {\bibfnamefont {Y.}~\bibnamefont {Huang}}, \bibinfo {author}
  {\bibfnamefont {P.}~\bibnamefont {Richard}}, \bibinfo {author} {\bibfnamefont
  {P.}~\bibnamefont {Zhang}}, \bibinfo {author} {\bibfnamefont
  {K.}~\bibnamefont {Yaji}}, \bibinfo {author} {\bibfnamefont {K.}~\bibnamefont
  {Kuroda}}, \bibinfo {author} {\bibfnamefont {S.}~\bibnamefont {Shin}},
  \bibinfo {author} {\bibfnamefont {H.}~\bibnamefont {Weng}}, \bibinfo {author}
  {\bibfnamefont {B.~A.}\ \bibnamefont {Bernevig}}, \bibinfo {author}
  {\bibfnamefont {Y.}~\bibnamefont {Shi}}, \bibinfo {author} {\bibfnamefont
  {T.}~\bibnamefont {Qian}}, \ and\ \bibinfo {author} {\bibfnamefont
  {H.}~\bibnamefont {Ding}},\ }\href {\doibase 10.1126/sciadv.1602415}
  {\bibfield  {journal} {\bibinfo  {journal} {Science Advances}\ }\textbf
  {\bibinfo {volume} {3}} (\bibinfo {year} {2017}),\
  10.1126/sciadv.1602415}\BibitemShut {NoStop}%
\bibitem [{\citenamefont {Benalcazar}\ \emph
  {et~al.}(2017{\natexlab{a}})\citenamefont {Benalcazar}, \citenamefont
  {Bernevig},\ and\ \citenamefont {Hughes}}]{quadrupole}%
  \BibitemOpen
  \bibfield  {author} {\bibinfo {author} {\bibfnamefont {W.~A.}\ \bibnamefont
  {Benalcazar}}, \bibinfo {author} {\bibfnamefont {B.~A.}\ \bibnamefont
  {Bernevig}}, \ and\ \bibinfo {author} {\bibfnamefont {T.~L.}\ \bibnamefont
  {Hughes}},\ }\href {\doibase 10.1126/science.aah6442} {\bibfield  {journal}
  {\bibinfo  {journal} {Science}\ }\textbf {\bibinfo {volume} {357}},\ \bibinfo
  {pages} {61} (\bibinfo {year} {2017}{\natexlab{a}})}\BibitemShut {NoStop}%
\bibitem [{\citenamefont {Benalcazar}\ \emph
  {et~al.}(2017{\natexlab{b}})\citenamefont {Benalcazar}, \citenamefont
  {Bernevig},\ and\ \citenamefont {Hughes}}]{hughes}%
  \BibitemOpen
  \bibfield  {author} {\bibinfo {author} {\bibfnamefont {W.~A.}\ \bibnamefont
  {Benalcazar}}, \bibinfo {author} {\bibfnamefont {B.~A.}\ \bibnamefont
  {Bernevig}}, \ and\ \bibinfo {author} {\bibfnamefont {T.~L.}\ \bibnamefont
  {Hughes}},\ }\href {\doibase 10.1103/PhysRevB.96.245115} {\bibfield
  {journal} {\bibinfo  {journal} {Phys. Rev. B}\ }\textbf {\bibinfo {volume}
  {96}},\ \bibinfo {pages} {245115} (\bibinfo {year}
  {2017}{\natexlab{b}})}\BibitemShut {NoStop}%
\bibitem [{\citenamefont {{Schindler}}\ \emph {et~al.}(2017)\citenamefont
  {{Schindler}}, \citenamefont {{Cook}}, \citenamefont {{Vergniory}},
  \citenamefont {{Wang}}, \citenamefont {{Parkin}}, \citenamefont
  {{Bernevig}},\ and\ \citenamefont {{Neupert}}}]{hoti}%
  \BibitemOpen
  \bibfield  {author} {\bibinfo {author} {\bibfnamefont {F.}~\bibnamefont
  {{Schindler}}}, \bibinfo {author} {\bibfnamefont {A.~M.}\ \bibnamefont
  {{Cook}}}, \bibinfo {author} {\bibfnamefont {M.~G.}\ \bibnamefont
  {{Vergniory}}}, \bibinfo {author} {\bibfnamefont {Z.}~\bibnamefont {{Wang}}},
  \bibinfo {author} {\bibfnamefont {S.~S.~P.}\ \bibnamefont {{Parkin}}},
  \bibinfo {author} {\bibfnamefont {B.~A.}\ \bibnamefont {{Bernevig}}}, \ and\
  \bibinfo {author} {\bibfnamefont {T.}~\bibnamefont {{Neupert}}},\ }\href@noop
  {} {\bibfield  {journal} {\bibinfo  {journal} {ArXiv e-prints}\ } (\bibinfo
  {year} {2017})},\ \Eprint {http://arxiv.org/abs/1708.03636} {arXiv:1708.03636
  [cond-mat.mes-hall]} \BibitemShut {NoStop}%
\bibitem [{\citenamefont {Teo}\ and\ \citenamefont {Hughes}(2013)}]{teo}%
  \BibitemOpen
  \bibfield  {author} {\bibinfo {author} {\bibfnamefont {J.~C.~Y.}\
  \bibnamefont {Teo}}\ and\ \bibinfo {author} {\bibfnamefont {T.~L.}\
  \bibnamefont {Hughes}},\ }\href {\doibase 10.1103/PhysRevLett.111.047006}
  {\bibfield  {journal} {\bibinfo  {journal} {Phys. Rev. Lett.}\ }\textbf
  {\bibinfo {volume} {111}},\ \bibinfo {pages} {047006} (\bibinfo {year}
  {2013})}\BibitemShut {NoStop}%
\bibitem [{\citenamefont {Benalcazar}\ \emph {et~al.}(2014)\citenamefont
  {Benalcazar}, \citenamefont {Teo},\ and\ \citenamefont {Hughes}}]{bth}%
  \BibitemOpen
  \bibfield  {author} {\bibinfo {author} {\bibfnamefont {W.~A.}\ \bibnamefont
  {Benalcazar}}, \bibinfo {author} {\bibfnamefont {J.~C.~Y.}\ \bibnamefont
  {Teo}}, \ and\ \bibinfo {author} {\bibfnamefont {T.~L.}\ \bibnamefont
  {Hughes}},\ }\href {\doibase 10.1103/PhysRevB.89.224503} {\bibfield
  {journal} {\bibinfo  {journal} {Phys. Rev. B}\ }\textbf {\bibinfo {volume}
  {89}},\ \bibinfo {pages} {224503} (\bibinfo {year} {2014})}\BibitemShut
  {NoStop}%
\bibitem [{\citenamefont {{Zhu}}(2018)}]{zhu}%
  \BibitemOpen
  \bibfield  {author} {\bibinfo {author} {\bibfnamefont {X.}~\bibnamefont
  {{Zhu}}},\ }\href@noop {} {\bibfield  {journal} {\bibinfo  {journal} {ArXiv
  e-prints}\ } (\bibinfo {year} {2018})},\ \Eprint
  {http://arxiv.org/abs/1802.00270} {arXiv:1802.00270 [cond-mat.mes-hall]}
  \BibitemShut {NoStop}%
\bibitem [{\citenamefont {Schindler}\ \emph {et~al.}(2018)\citenamefont
  {Schindler}, \citenamefont {Wang}, \citenamefont {Vergniory}, \citenamefont
  {Cook}, \citenamefont {Murani}, \citenamefont {Sengupta}, \citenamefont
  {Kasumov}, \citenamefont {Deblock}, \citenamefont {Jeon}, \citenamefont
  {Drozdov}, \citenamefont {Bouchiat}, \citenamefont {Gu{\'e}ron},
  \citenamefont {Yazdani}, \citenamefont {Bernevig},\ and\ \citenamefont
  {Neupert}}]{bismuth}%
  \BibitemOpen
  \bibfield  {author} {\bibinfo {author} {\bibfnamefont {F.}~\bibnamefont
  {Schindler}}, \bibinfo {author} {\bibfnamefont {Z.}~\bibnamefont {Wang}},
  \bibinfo {author} {\bibfnamefont {M.~G.}\ \bibnamefont {Vergniory}}, \bibinfo
  {author} {\bibfnamefont {A.~M.}\ \bibnamefont {Cook}}, \bibinfo {author}
  {\bibfnamefont {A.}~\bibnamefont {Murani}}, \bibinfo {author} {\bibfnamefont
  {S.}~\bibnamefont {Sengupta}}, \bibinfo {author} {\bibfnamefont {A.~Y.}\
  \bibnamefont {Kasumov}}, \bibinfo {author} {\bibfnamefont {R.}~\bibnamefont
  {Deblock}}, \bibinfo {author} {\bibfnamefont {S.}~\bibnamefont {Jeon}},
  \bibinfo {author} {\bibfnamefont {I.}~\bibnamefont {Drozdov}}, \bibinfo
  {author} {\bibfnamefont {H.}~\bibnamefont {Bouchiat}}, \bibinfo {author}
  {\bibfnamefont {S.}~\bibnamefont {Gu{\'e}ron}}, \bibinfo {author}
  {\bibfnamefont {A.}~\bibnamefont {Yazdani}}, \bibinfo {author} {\bibfnamefont
  {B.~A.}\ \bibnamefont {Bernevig}}, \ and\ \bibinfo {author} {\bibfnamefont
  {T.}~\bibnamefont {Neupert}},\ }\href {\doibase 10.1038/s41567-018-0224-7}
  {\bibfield  {journal} {\bibinfo  {journal} {Nature Physics}\ } (\bibinfo
  {year} {2018}),\ 10.1038/s41567-018-0224-7}\BibitemShut {NoStop}%
\bibitem [{\citenamefont {Serra-Garcia}\ \emph {et~al.}(2018)\citenamefont
  {Serra-Garcia}, \citenamefont {Peri}, \citenamefont {S{\"u}sstrunk},
  \citenamefont {Bilal}, \citenamefont {Larsen}, \citenamefont {Villanueva},\
  and\ \citenamefont {Huber}}]{SerraGarcia}%
  \BibitemOpen
  \bibfield  {author} {\bibinfo {author} {\bibfnamefont {M.}~\bibnamefont
  {Serra-Garcia}}, \bibinfo {author} {\bibfnamefont {V.}~\bibnamefont {Peri}},
  \bibinfo {author} {\bibfnamefont {R.}~\bibnamefont {S{\"u}sstrunk}}, \bibinfo
  {author} {\bibfnamefont {O.~R.}\ \bibnamefont {Bilal}}, \bibinfo {author}
  {\bibfnamefont {T.}~\bibnamefont {Larsen}}, \bibinfo {author} {\bibfnamefont
  {L.~G.}\ \bibnamefont {Villanueva}}, \ and\ \bibinfo {author} {\bibfnamefont
  {S.~D.}\ \bibnamefont {Huber}},\ }\href
  {http://dx.doi.org/10.1038/nature25156} {\bibfield  {journal} {\bibinfo
  {journal} {Nature}\ }\textbf {\bibinfo {volume} {555}},\ \bibinfo {pages}
  {342 EP } (\bibinfo {year} {2018})}\BibitemShut {NoStop}%
\bibitem [{\citenamefont {Peterson}\ \emph {et~al.}(2018)\citenamefont
  {Peterson}, \citenamefont {Benalcazar}, \citenamefont {Hughes},\ and\
  \citenamefont {Bahl}}]{Peterson}%
  \BibitemOpen
  \bibfield  {author} {\bibinfo {author} {\bibfnamefont {C.~W.}\ \bibnamefont
  {Peterson}}, \bibinfo {author} {\bibfnamefont {W.~A.}\ \bibnamefont
  {Benalcazar}}, \bibinfo {author} {\bibfnamefont {T.~L.}\ \bibnamefont
  {Hughes}}, \ and\ \bibinfo {author} {\bibfnamefont {G.}~\bibnamefont
  {Bahl}},\ }\href {http://dx.doi.org/10.1038/nature25777} {\bibfield
  {journal} {\bibinfo  {journal} {Nature}\ }\textbf {\bibinfo {volume} {555}},\
  \bibinfo {pages} {346 EP } (\bibinfo {year} {2018})}\BibitemShut {NoStop}%
\bibitem [{\citenamefont {{Serra-Garcia}}\ \emph {et~al.}(2018)\citenamefont
  {{Serra-Garcia}}, \citenamefont {{S{\"u}sstrunk}},\ and\ \citenamefont
  {{Huber}}}]{huber}%
  \BibitemOpen
  \bibfield  {author} {\bibinfo {author} {\bibfnamefont {M.}~\bibnamefont
  {{Serra-Garcia}}}, \bibinfo {author} {\bibfnamefont {R.}~\bibnamefont
  {{S{\"u}sstrunk}}}, \ and\ \bibinfo {author} {\bibfnamefont {S.~D.}\
  \bibnamefont {{Huber}}},\ }\href@noop {} {\bibfield  {journal} {\bibinfo
  {journal} {ArXiv e-prints}\ } (\bibinfo {year} {2018})},\ \Eprint
  {http://arxiv.org/abs/1806.07367} {arXiv:1806.07367 [cond-mat.mes-hall]}
  \BibitemShut {NoStop}%
\bibitem [{\citenamefont {{Imhof}}\ \emph {et~al.}(2017)\citenamefont
  {{Imhof}}, \citenamefont {{Berger}}, \citenamefont {{Bayer}}, \citenamefont
  {{Brehm}}, \citenamefont {{Molenkamp}}, \citenamefont {{Kiessling}},
  \citenamefont {{Schindler}}, \citenamefont {{Lee}}, \citenamefont
  {{Greiter}}, \citenamefont {{Neupert}},\ and\ \citenamefont
  {{Thomale}}}]{thomale}%
  \BibitemOpen
  \bibfield  {author} {\bibinfo {author} {\bibfnamefont {S.}~\bibnamefont
  {{Imhof}}}, \bibinfo {author} {\bibfnamefont {C.}~\bibnamefont {{Berger}}},
  \bibinfo {author} {\bibfnamefont {F.}~\bibnamefont {{Bayer}}}, \bibinfo
  {author} {\bibfnamefont {J.}~\bibnamefont {{Brehm}}}, \bibinfo {author}
  {\bibfnamefont {L.}~\bibnamefont {{Molenkamp}}}, \bibinfo {author}
  {\bibfnamefont {T.}~\bibnamefont {{Kiessling}}}, \bibinfo {author}
  {\bibfnamefont {F.}~\bibnamefont {{Schindler}}}, \bibinfo {author}
  {\bibfnamefont {C.~H.}\ \bibnamefont {{Lee}}}, \bibinfo {author}
  {\bibfnamefont {M.}~\bibnamefont {{Greiter}}}, \bibinfo {author}
  {\bibfnamefont {T.}~\bibnamefont {{Neupert}}}, \ and\ \bibinfo {author}
  {\bibfnamefont {R.}~\bibnamefont {{Thomale}}},\ }\href@noop {} {\bibfield
  {journal} {\bibinfo  {journal} {ArXiv e-prints}\ } (\bibinfo {year}
  {2017})},\ \Eprint {http://arxiv.org/abs/1708.03647} {arXiv:1708.03647
  [cond-mat.mes-hall]} \BibitemShut {NoStop}%
\bibitem [{\citenamefont {{Wang}}\ \emph
  {et~al.}(2018{\natexlab{a}})\citenamefont {{Wang}}, \citenamefont {{Wieder}},
  \citenamefont {{Li}}, \citenamefont {{Yan}},\ and\ \citenamefont
  {{Bernevig}}}]{wieder}%
  \BibitemOpen
  \bibfield  {author} {\bibinfo {author} {\bibfnamefont {Z.}~\bibnamefont
  {{Wang}}}, \bibinfo {author} {\bibfnamefont {B.~J.}\ \bibnamefont
  {{Wieder}}}, \bibinfo {author} {\bibfnamefont {J.}~\bibnamefont {{Li}}},
  \bibinfo {author} {\bibfnamefont {B.}~\bibnamefont {{Yan}}}, \ and\ \bibinfo
  {author} {\bibfnamefont {B.~A.}\ \bibnamefont {{Bernevig}}},\ }\href@noop {}
  {\bibfield  {journal} {\bibinfo  {journal} {ArXiv e-prints}\ } (\bibinfo
  {year} {2018}{\natexlab{a}})},\ \Eprint {http://arxiv.org/abs/1806.11116}
  {arXiv:1806.11116 [cond-mat.mtrl-sci]} \BibitemShut {NoStop}%
\bibitem [{\citenamefont {Bradlyn}\ \emph {et~al.}(2017)\citenamefont
  {Bradlyn}, \citenamefont {Elcoro}, \citenamefont {Cano}, \citenamefont
  {Vergniory}, \citenamefont {Wang}, \citenamefont {Felser}, \citenamefont
  {Aroyo},\ and\ \citenamefont {Bernevig}}]{TQC}%
  \BibitemOpen
  \bibfield  {author} {\bibinfo {author} {\bibfnamefont {B.}~\bibnamefont
  {Bradlyn}}, \bibinfo {author} {\bibfnamefont {L.}~\bibnamefont {Elcoro}},
  \bibinfo {author} {\bibfnamefont {J.}~\bibnamefont {Cano}}, \bibinfo {author}
  {\bibfnamefont {M.~G.}\ \bibnamefont {Vergniory}}, \bibinfo {author}
  {\bibfnamefont {Z.}~\bibnamefont {Wang}}, \bibinfo {author} {\bibfnamefont
  {C.}~\bibnamefont {Felser}}, \bibinfo {author} {\bibfnamefont {M.~I.}\
  \bibnamefont {Aroyo}}, \ and\ \bibinfo {author} {\bibfnamefont {B.~A.}\
  \bibnamefont {Bernevig}},\ }\href {http://dx.doi.org/10.1038/nature23268}
  {\bibfield  {journal} {\bibinfo  {journal} {Nature}\ }\textbf {\bibinfo
  {volume} {547}},\ \bibinfo {pages} {298 EP } (\bibinfo {year}
  {2017})}\BibitemShut {NoStop}%
\bibitem [{\citenamefont {Cano}\ \emph {et~al.}(2018)\citenamefont {Cano},
  \citenamefont {Bradlyn}, \citenamefont {Wang}, \citenamefont {Elcoro},
  \citenamefont {Vergniory}, \citenamefont {Felser}, \citenamefont {Aroyo},\
  and\ \citenamefont {Bernevig}}]{cano}%
  \BibitemOpen
  \bibfield  {author} {\bibinfo {author} {\bibfnamefont {J.}~\bibnamefont
  {Cano}}, \bibinfo {author} {\bibfnamefont {B.}~\bibnamefont {Bradlyn}},
  \bibinfo {author} {\bibfnamefont {Z.}~\bibnamefont {Wang}}, \bibinfo {author}
  {\bibfnamefont {L.}~\bibnamefont {Elcoro}}, \bibinfo {author} {\bibfnamefont
  {M.~G.}\ \bibnamefont {Vergniory}}, \bibinfo {author} {\bibfnamefont
  {C.}~\bibnamefont {Felser}}, \bibinfo {author} {\bibfnamefont {M.~I.}\
  \bibnamefont {Aroyo}}, \ and\ \bibinfo {author} {\bibfnamefont {B.~A.}\
  \bibnamefont {Bernevig}},\ }\href {\doibase 10.1103/PhysRevB.97.035139}
  {\bibfield  {journal} {\bibinfo  {journal} {Phys. Rev. B}\ }\textbf {\bibinfo
  {volume} {97}},\ \bibinfo {pages} {035139} (\bibinfo {year}
  {2018})}\BibitemShut {NoStop}%
\bibitem [{\citenamefont {Vergniory}\ \emph {et~al.}(2017)\citenamefont
  {Vergniory}, \citenamefont {Elcoro}, \citenamefont {Wang}, \citenamefont
  {Cano}, \citenamefont {Felser}, \citenamefont {Aroyo}, \citenamefont
  {Bernevig},\ and\ \citenamefont {Bradlyn}}]{bradlyn}%
  \BibitemOpen
  \bibfield  {author} {\bibinfo {author} {\bibfnamefont {M.~G.}\ \bibnamefont
  {Vergniory}}, \bibinfo {author} {\bibfnamefont {L.}~\bibnamefont {Elcoro}},
  \bibinfo {author} {\bibfnamefont {Z.}~\bibnamefont {Wang}}, \bibinfo {author}
  {\bibfnamefont {J.}~\bibnamefont {Cano}}, \bibinfo {author} {\bibfnamefont
  {C.}~\bibnamefont {Felser}}, \bibinfo {author} {\bibfnamefont {M.~I.}\
  \bibnamefont {Aroyo}}, \bibinfo {author} {\bibfnamefont {B.~A.}\ \bibnamefont
  {Bernevig}}, \ and\ \bibinfo {author} {\bibfnamefont {B.}~\bibnamefont
  {Bradlyn}},\ }\href {\doibase 10.1103/PhysRevE.96.023310} {\bibfield
  {journal} {\bibinfo  {journal} {Phys. Rev. E}\ }\textbf {\bibinfo {volume}
  {96}},\ \bibinfo {pages} {023310} (\bibinfo {year} {2017})}\BibitemShut
  {NoStop}%
\bibitem [{\citenamefont {Po}\ \emph {et~al.}(2017)\citenamefont {Po},
  \citenamefont {Vishwanath},\ and\ \citenamefont {Watanabe}}]{po2}%
  \BibitemOpen
  \bibfield  {author} {\bibinfo {author} {\bibfnamefont {H.~C.}\ \bibnamefont
  {Po}}, \bibinfo {author} {\bibfnamefont {A.}~\bibnamefont {Vishwanath}}, \
  and\ \bibinfo {author} {\bibfnamefont {H.}~\bibnamefont {Watanabe}},\ }\href
  {\doibase 10.1038/s41467-017-00133-2} {\bibfield  {journal} {\bibinfo
  {journal} {Nature Communications}\ }\textbf {\bibinfo {volume} {8}},\
  \bibinfo {pages} {50} (\bibinfo {year} {2017})}\BibitemShut {NoStop}%
\bibitem [{\citenamefont {Watanabe}\ \emph {et~al.}(2018)\citenamefont
  {Watanabe}, \citenamefont {Po},\ and\ \citenamefont {Vishwanath}}]{watanabe}%
  \BibitemOpen
  \bibfield  {author} {\bibinfo {author} {\bibfnamefont {H.}~\bibnamefont
  {Watanabe}}, \bibinfo {author} {\bibfnamefont {H.~C.}\ \bibnamefont {Po}}, \
  and\ \bibinfo {author} {\bibfnamefont {A.}~\bibnamefont {Vishwanath}},\
  }\href@noop {} {\bibfield  {journal} {\bibinfo  {journal} {Science Advances}\
  }\textbf {\bibinfo {volume} {4}} (\bibinfo {year} {2018})}\BibitemShut
  {NoStop}%
\bibitem [{\citenamefont {Song}\ \emph {et~al.}(2018)\citenamefont {Song},
  \citenamefont {Zhang}, \citenamefont {Fang},\ and\ \citenamefont
  {Fang}}]{Tiantian}%
  \BibitemOpen
  \bibfield  {author} {\bibinfo {author} {\bibfnamefont {Z.}~\bibnamefont
  {Song}}, \bibinfo {author} {\bibfnamefont {T.}~\bibnamefont {Zhang}},
  \bibinfo {author} {\bibfnamefont {Z.}~\bibnamefont {Fang}}, \ and\ \bibinfo
  {author} {\bibfnamefont {C.}~\bibnamefont {Fang}},\ }\href {\doibase
  10.1038/s41467-018-06010-w} {\bibfield  {journal} {\bibinfo  {journal}
  {Nature Communications}\ }\textbf {\bibinfo {volume} {9}},\ \bibinfo {pages}
  {3530} (\bibinfo {year} {2018})}\BibitemShut {NoStop}%
\bibitem [{\citenamefont {Alexandradinata}\ \emph {et~al.}(2014)\citenamefont
  {Alexandradinata}, \citenamefont {Fang}, \citenamefont {Gilbert},\ and\
  \citenamefont {Bernevig}}]{Alexandradinata}%
  \BibitemOpen
  \bibfield  {author} {\bibinfo {author} {\bibfnamefont {A.}~\bibnamefont
  {Alexandradinata}}, \bibinfo {author} {\bibfnamefont {C.}~\bibnamefont
  {Fang}}, \bibinfo {author} {\bibfnamefont {M.~J.}\ \bibnamefont {Gilbert}}, \
  and\ \bibinfo {author} {\bibfnamefont {B.~A.}\ \bibnamefont {Bernevig}},\
  }\href {\doibase 10.1103/PhysRevLett.113.116403} {\bibfield  {journal}
  {\bibinfo  {journal} {Phys. Rev. Lett.}\ }\textbf {\bibinfo {volume} {113}},\
  \bibinfo {pages} {116403} (\bibinfo {year} {2014})}\BibitemShut {NoStop}%
\bibitem [{\citenamefont {Fang}\ \emph
  {et~al.}(2012{\natexlab{b}})\citenamefont {Fang}, \citenamefont {Gilbert},\
  and\ \citenamefont {Bernevig}}]{Gilbert}%
  \BibitemOpen
  \bibfield  {author} {\bibinfo {author} {\bibfnamefont {C.}~\bibnamefont
  {Fang}}, \bibinfo {author} {\bibfnamefont {M.~J.}\ \bibnamefont {Gilbert}}, \
  and\ \bibinfo {author} {\bibfnamefont {B.~A.}\ \bibnamefont {Bernevig}},\
  }\href {\doibase 10.1103/PhysRevB.86.115112} {\bibfield  {journal} {\bibinfo
  {journal} {Phys. Rev. B}\ }\textbf {\bibinfo {volume} {86}},\ \bibinfo
  {pages} {115112} (\bibinfo {year} {2012}{\natexlab{b}})}\BibitemShut
  {NoStop}%
\bibitem [{\citenamefont {{Khalaf}}\ \emph {et~al.}(2017)\citenamefont
  {{Khalaf}}, \citenamefont {{Po}}, \citenamefont {{Vishwanath}},\ and\
  \citenamefont {{Watanabe}}}]{Po}%
  \BibitemOpen
  \bibfield  {author} {\bibinfo {author} {\bibfnamefont {E.}~\bibnamefont
  {{Khalaf}}}, \bibinfo {author} {\bibfnamefont {H.~C.}\ \bibnamefont {{Po}}},
  \bibinfo {author} {\bibfnamefont {A.}~\bibnamefont {{Vishwanath}}}, \ and\
  \bibinfo {author} {\bibfnamefont {H.}~\bibnamefont {{Watanabe}}},\
  }\href@noop {} {\bibfield  {journal} {\bibinfo  {journal} {ArXiv e-prints}\ }
  (\bibinfo {year} {2017})},\ \Eprint {http://arxiv.org/abs/1711.11589}
  {arXiv:1711.11589 [cond-mat.str-el]} \BibitemShut {NoStop}%
\bibitem [{\citenamefont {{Fang}}\ and\ \citenamefont {{Fu}}(2017)}]{Fang2}%
  \BibitemOpen
  \bibfield  {author} {\bibinfo {author} {\bibfnamefont {C.}~\bibnamefont
  {{Fang}}}\ and\ \bibinfo {author} {\bibfnamefont {L.}~\bibnamefont {{Fu}}},\
  }\href@noop {} {\bibfield  {journal} {\bibinfo  {journal} {ArXiv e-prints}\ }
  (\bibinfo {year} {2017})},\ \Eprint {http://arxiv.org/abs/1709.01929}
  {arXiv:1709.01929} \BibitemShut {NoStop}%
\bibitem [{\citenamefont {Song}\ \emph
  {et~al.}(2017{\natexlab{a}})\citenamefont {Song}, \citenamefont {Fang},\ and\
  \citenamefont {Fang}}]{song}%
  \BibitemOpen
  \bibfield  {author} {\bibinfo {author} {\bibfnamefont {Z.}~\bibnamefont
  {Song}}, \bibinfo {author} {\bibfnamefont {Z.}~\bibnamefont {Fang}}, \ and\
  \bibinfo {author} {\bibfnamefont {C.}~\bibnamefont {Fang}},\ }\href {\doibase
  10.1103/PhysRevLett.119.246402} {\bibfield  {journal} {\bibinfo  {journal}
  {Phys. Rev. Lett.}\ }\textbf {\bibinfo {volume} {119}},\ \bibinfo {pages}
  {246402} (\bibinfo {year} {2017}{\natexlab{a}})}\BibitemShut {NoStop}%
\bibitem [{\citenamefont {{Xu}}\ \emph {et~al.}(2017)\citenamefont {{Xu}},
  \citenamefont {{Xue}},\ and\ \citenamefont {{Wan}}}]{xue}%
  \BibitemOpen
  \bibfield  {author} {\bibinfo {author} {\bibfnamefont {Y.}~\bibnamefont
  {{Xu}}}, \bibinfo {author} {\bibfnamefont {R.}~\bibnamefont {{Xue}}}, \ and\
  \bibinfo {author} {\bibfnamefont {S.}~\bibnamefont {{Wan}}},\ }\href@noop {}
  {\bibfield  {journal} {\bibinfo  {journal} {ArXiv e-prints}\ } (\bibinfo
  {year} {2017})},\ \Eprint {http://arxiv.org/abs/1711.09202} {arXiv:1711.09202
  [cond-mat.str-el]} \BibitemShut {NoStop}%
\bibitem [{\citenamefont {{Benalcazar}}\ \emph {et~al.}(2018)\citenamefont
  {{Benalcazar}}, \citenamefont {{Li}},\ and\ \citenamefont
  {{Hughes}}}]{benalcazarli}%
  \BibitemOpen
  \bibfield  {author} {\bibinfo {author} {\bibfnamefont {W.~A.}\ \bibnamefont
  {{Benalcazar}}}, \bibinfo {author} {\bibfnamefont {T.}~\bibnamefont {{Li}}},
  \ and\ \bibinfo {author} {\bibfnamefont {T.~L.}\ \bibnamefont {{Hughes}}},\
  }\href@noop {} {\bibfield  {journal} {\bibinfo  {journal} {ArXiv e-prints}\ }
  (\bibinfo {year} {2018})},\ \Eprint {http://arxiv.org/abs/1809.02142}
  {arXiv:1809.02142 [cond-mat.str-el]} \BibitemShut {NoStop}%
\bibitem [{\citenamefont {Shiozaki}\ and\ \citenamefont
  {Sato}(2014)}]{Shiozaki}%
  \BibitemOpen
  \bibfield  {author} {\bibinfo {author} {\bibfnamefont {K.}~\bibnamefont
  {Shiozaki}}\ and\ \bibinfo {author} {\bibfnamefont {M.}~\bibnamefont
  {Sato}},\ }\href {\doibase 10.1103/PhysRevB.90.165114} {\bibfield  {journal}
  {\bibinfo  {journal} {Phys. Rev. B}\ }\textbf {\bibinfo {volume} {90}},\
  \bibinfo {pages} {165114} (\bibinfo {year} {2014})}\BibitemShut {NoStop}%
\bibitem [{\citenamefont {Langbehn}\ \emph {et~al.}(2017)\citenamefont
  {Langbehn}, \citenamefont {Peng}, \citenamefont {Trifunovic}, \citenamefont
  {von Oppen},\ and\ \citenamefont {Brouwer}}]{langbehn}%
  \BibitemOpen
  \bibfield  {author} {\bibinfo {author} {\bibfnamefont {J.}~\bibnamefont
  {Langbehn}}, \bibinfo {author} {\bibfnamefont {Y.}~\bibnamefont {Peng}},
  \bibinfo {author} {\bibfnamefont {L.}~\bibnamefont {Trifunovic}}, \bibinfo
  {author} {\bibfnamefont {F.}~\bibnamefont {von Oppen}}, \ and\ \bibinfo
  {author} {\bibfnamefont {P.~W.}\ \bibnamefont {Brouwer}},\ }\href {\doibase
  10.1103/PhysRevLett.119.246401} {\bibfield  {journal} {\bibinfo  {journal}
  {Phys. Rev. Lett.}\ }\textbf {\bibinfo {volume} {119}},\ \bibinfo {pages}
  {246401} (\bibinfo {year} {2017})}\BibitemShut {NoStop}%
\bibitem [{\citenamefont {Shapourian}\ \emph {et~al.}(2018)\citenamefont
  {Shapourian}, \citenamefont {Wang},\ and\ \citenamefont {Ryu}}]{Shapourian}%
  \BibitemOpen
  \bibfield  {author} {\bibinfo {author} {\bibfnamefont {H.}~\bibnamefont
  {Shapourian}}, \bibinfo {author} {\bibfnamefont {Y.}~\bibnamefont {Wang}}, \
  and\ \bibinfo {author} {\bibfnamefont {S.}~\bibnamefont {Ryu}},\ }\href
  {\doibase 10.1103/PhysRevB.97.094508} {\bibfield  {journal} {\bibinfo
  {journal} {Phys. Rev. B}\ }\textbf {\bibinfo {volume} {97}},\ \bibinfo
  {pages} {094508} (\bibinfo {year} {2018})}\BibitemShut {NoStop}%
\bibitem [{\citenamefont {{Geier}}\ \emph {et~al.}(2018)\citenamefont
  {{Geier}}, \citenamefont {{Trifunovic}}, \citenamefont {{Hoskam}},\ and\
  \citenamefont {{Brouwer}}}]{Geier}%
  \BibitemOpen
  \bibfield  {author} {\bibinfo {author} {\bibfnamefont {M.}~\bibnamefont
  {{Geier}}}, \bibinfo {author} {\bibfnamefont {L.}~\bibnamefont
  {{Trifunovic}}}, \bibinfo {author} {\bibfnamefont {M.}~\bibnamefont
  {{Hoskam}}}, \ and\ \bibinfo {author} {\bibfnamefont {P.~W.}\ \bibnamefont
  {{Brouwer}}},\ }\href@noop {} {\bibfield  {journal} {\bibinfo  {journal}
  {ArXiv e-prints}\ } (\bibinfo {year} {2018})},\ \Eprint
  {http://arxiv.org/abs/1801.10053} {arXiv:1801.10053 [cond-mat.mes-hall]}
  \BibitemShut {NoStop}%
\bibitem [{\citenamefont {{Khalaf}}(2018)}]{Khalaf}%
  \BibitemOpen
  \bibfield  {author} {\bibinfo {author} {\bibfnamefont {E.}~\bibnamefont
  {{Khalaf}}},\ }\href@noop {} {\bibfield  {journal} {\bibinfo  {journal}
  {ArXiv e-prints}\ } (\bibinfo {year} {2018})},\ \Eprint
  {http://arxiv.org/abs/1801.10050} {arXiv:1801.10050 [cond-mat.mes-hall]}
  \BibitemShut {NoStop}%
\bibitem [{\citenamefont {{Fang}}\ \emph {et~al.}(2017)\citenamefont {{Fang}},
  \citenamefont {{Bernevig}},\ and\ \citenamefont {{Gilbert}}}]{fang}%
  \BibitemOpen
  \bibfield  {author} {\bibinfo {author} {\bibfnamefont {C.}~\bibnamefont
  {{Fang}}}, \bibinfo {author} {\bibfnamefont {B.~A.}\ \bibnamefont
  {{Bernevig}}}, \ and\ \bibinfo {author} {\bibfnamefont {M.~J.}\ \bibnamefont
  {{Gilbert}}},\ }\href@noop {} {\bibfield  {journal} {\bibinfo  {journal}
  {ArXiv e-prints}\ } (\bibinfo {year} {2017})},\ \Eprint
  {http://arxiv.org/abs/1701.01944} {arXiv:1701.01944 [cond-mat.supr-con]}
  \BibitemShut {NoStop}%
\bibitem [{\citenamefont {Song}\ \emph
  {et~al.}(2017{\natexlab{b}})\citenamefont {Song}, \citenamefont {Huang},
  \citenamefont {Fu},\ and\ \citenamefont {Hermele}}]{hermele1}%
  \BibitemOpen
  \bibfield  {author} {\bibinfo {author} {\bibfnamefont {H.}~\bibnamefont
  {Song}}, \bibinfo {author} {\bibfnamefont {S.-J.}\ \bibnamefont {Huang}},
  \bibinfo {author} {\bibfnamefont {L.}~\bibnamefont {Fu}}, \ and\ \bibinfo
  {author} {\bibfnamefont {M.}~\bibnamefont {Hermele}},\ }\href {\doibase
  10.1103/PhysRevX.7.011020} {\bibfield  {journal} {\bibinfo  {journal} {Phys.
  Rev. X}\ }\textbf {\bibinfo {volume} {7}},\ \bibinfo {pages} {011020}
  (\bibinfo {year} {2017}{\natexlab{b}})}\BibitemShut {NoStop}%
\bibitem [{\citenamefont {Huang}\ \emph {et~al.}(2017)\citenamefont {Huang},
  \citenamefont {Song}, \citenamefont {Huang},\ and\ \citenamefont
  {Hermele}}]{hermele2}%
  \BibitemOpen
  \bibfield  {author} {\bibinfo {author} {\bibfnamefont {S.-J.}\ \bibnamefont
  {Huang}}, \bibinfo {author} {\bibfnamefont {H.}~\bibnamefont {Song}},
  \bibinfo {author} {\bibfnamefont {Y.-P.}\ \bibnamefont {Huang}}, \ and\
  \bibinfo {author} {\bibfnamefont {M.}~\bibnamefont {Hermele}},\ }\href
  {\doibase 10.1103/PhysRevB.96.205106} {\bibfield  {journal} {\bibinfo
  {journal} {Phys. Rev. B}\ }\textbf {\bibinfo {volume} {96}},\ \bibinfo
  {pages} {205106} (\bibinfo {year} {2017})}\BibitemShut {NoStop}%
\bibitem [{\citenamefont {Cheng}\ \emph {et~al.}(2016)\citenamefont {Cheng},
  \citenamefont {Zaletel}, \citenamefont {Barkeshli}, \citenamefont
  {Vishwanath},\ and\ \citenamefont {Bonderson}}]{Cheng}%
  \BibitemOpen
  \bibfield  {author} {\bibinfo {author} {\bibfnamefont {M.}~\bibnamefont
  {Cheng}}, \bibinfo {author} {\bibfnamefont {M.}~\bibnamefont {Zaletel}},
  \bibinfo {author} {\bibfnamefont {M.}~\bibnamefont {Barkeshli}}, \bibinfo
  {author} {\bibfnamefont {A.}~\bibnamefont {Vishwanath}}, \ and\ \bibinfo
  {author} {\bibfnamefont {P.}~\bibnamefont {Bonderson}},\ }\href {\doibase
  10.1103/PhysRevX.6.041068} {\bibfield  {journal} {\bibinfo  {journal} {Phys.
  Rev. X}\ }\textbf {\bibinfo {volume} {6}},\ \bibinfo {pages} {041068}
  (\bibinfo {year} {2016})}\BibitemShut {NoStop}%
\bibitem [{\citenamefont {Thorngren}\ and\ \citenamefont {Else}(2018)}]{else}%
  \BibitemOpen
  \bibfield  {author} {\bibinfo {author} {\bibfnamefont {R.}~\bibnamefont
  {Thorngren}}\ and\ \bibinfo {author} {\bibfnamefont {D.~V.}\ \bibnamefont
  {Else}},\ }\href {\doibase 10.1103/PhysRevX.8.011040} {\bibfield  {journal}
  {\bibinfo  {journal} {Phys. Rev. X}\ }\textbf {\bibinfo {volume} {8}},\
  \bibinfo {pages} {011040} (\bibinfo {year} {2018})}\BibitemShut {NoStop}%
\bibitem [{\citenamefont {{You}}\ \emph
  {et~al.}(2018{\natexlab{a}})\citenamefont {{You}}, \citenamefont {{Devakul}},
  \citenamefont {{Burnell}},\ and\ \citenamefont {{Neupert}}}]{devakul}%
  \BibitemOpen
  \bibfield  {author} {\bibinfo {author} {\bibfnamefont {Y.}~\bibnamefont
  {{You}}}, \bibinfo {author} {\bibfnamefont {T.}~\bibnamefont {{Devakul}}},
  \bibinfo {author} {\bibfnamefont {F.~J.}\ \bibnamefont {{Burnell}}}, \ and\
  \bibinfo {author} {\bibfnamefont {T.}~\bibnamefont {{Neupert}}},\ }\href@noop
  {} {\bibfield  {journal} {\bibinfo  {journal} {ArXiv e-prints}\ } (\bibinfo
  {year} {2018}{\natexlab{a}})},\ \Eprint {http://arxiv.org/abs/1807.09788}
  {arXiv:1807.09788 [cond-mat.str-el]} \BibitemShut {NoStop}%
\bibitem [{\citenamefont {{Dubinkin}}\ and\ \citenamefont
  {{Hughes}}(2018)}]{dubinkin}%
  \BibitemOpen
  \bibfield  {author} {\bibinfo {author} {\bibfnamefont {O.}~\bibnamefont
  {{Dubinkin}}}\ and\ \bibinfo {author} {\bibfnamefont {T.~L.}\ \bibnamefont
  {{Hughes}}},\ }\href@noop {} {\bibfield  {journal} {\bibinfo  {journal}
  {ArXiv e-prints}\ } (\bibinfo {year} {2018})},\ \Eprint
  {http://arxiv.org/abs/1807.09781} {arXiv:1807.09781 [cond-mat.str-el]}
  \BibitemShut {NoStop}%
\bibitem [{\citenamefont {{You}}\ \emph
  {et~al.}(2018{\natexlab{b}})\citenamefont {{You}}, \citenamefont
  {{Litinski}},\ and\ \citenamefont {{von Oppen}}}]{Litinski}%
  \BibitemOpen
  \bibfield  {author} {\bibinfo {author} {\bibfnamefont {Y.}~\bibnamefont
  {{You}}}, \bibinfo {author} {\bibfnamefont {D.}~\bibnamefont {{Litinski}}}, \
  and\ \bibinfo {author} {\bibfnamefont {F.}~\bibnamefont {{von Oppen}}},\
  }\href@noop {} {\bibfield  {journal} {\bibinfo  {journal} {ArXiv e-prints}\ }
  (\bibinfo {year} {2018}{\natexlab{b}})},\ \Eprint
  {http://arxiv.org/abs/1810.10556} {arXiv:1810.10556 [cond-mat.str-el]}
  \BibitemShut {NoStop}%
\bibitem [{\citenamefont {{Wang}}\ \emph
  {et~al.}(2018{\natexlab{b}})\citenamefont {{Wang}}, \citenamefont {{Lin}},\
  and\ \citenamefont {{Hughes}}}]{hughes2}%
  \BibitemOpen
  \bibfield  {author} {\bibinfo {author} {\bibfnamefont {Y.}~\bibnamefont
  {{Wang}}}, \bibinfo {author} {\bibfnamefont {M.}~\bibnamefont {{Lin}}}, \
  and\ \bibinfo {author} {\bibfnamefont {T.~L.}\ \bibnamefont {{Hughes}}},\
  }\href@noop {} {\bibfield  {journal} {\bibinfo  {journal} {ArXiv e-prints}\ }
  (\bibinfo {year} {2018}{\natexlab{b}})},\ \Eprint
  {http://arxiv.org/abs/1804.01531} {arXiv:1804.01531 [cond-mat.supr-con]}
  \BibitemShut {NoStop}%
\bibitem [{bil()}]{bilbao}%
  \BibitemOpen
  \href {http://www.cryst.ehu.es/cryst} {\bibfield  {journal} {\bibinfo
  {journal} {Bilbao Crystallographic Server}\ }}\Eprint
  {http://arxiv.org/abs/http://www.cryst.ehu.es/cryst/get\_wp.html}
  {http://www.cryst.ehu.es/cryst/get\_wp.html} \BibitemShut {NoStop}%
\bibitem [{\citenamefont {Fu}\ and\ \citenamefont {Kane}(2007)}]{fukane}%
  \BibitemOpen
  \bibfield  {author} {\bibinfo {author} {\bibfnamefont {L.}~\bibnamefont
  {Fu}}\ and\ \bibinfo {author} {\bibfnamefont {C.~L.}\ \bibnamefont {Kane}},\
  }\href {\doibase 10.1103/PhysRevB.76.045302} {\bibfield  {journal} {\bibinfo
  {journal} {Phys. Rev. B}\ }\textbf {\bibinfo {volume} {76}},\ \bibinfo
  {pages} {045302} (\bibinfo {year} {2007})}\BibitemShut {NoStop}%
\bibitem [{\citenamefont {Fu}\ and\ \citenamefont {Berg}(2010)}]{fuberg}%
  \BibitemOpen
  \bibfield  {author} {\bibinfo {author} {\bibfnamefont {L.}~\bibnamefont
  {Fu}}\ and\ \bibinfo {author} {\bibfnamefont {E.}~\bibnamefont {Berg}},\
  }\href {\doibase 10.1103/PhysRevLett.105.097001} {\bibfield  {journal}
  {\bibinfo  {journal} {Phys. Rev. Lett.}\ }\textbf {\bibinfo {volume} {105}},\
  \bibinfo {pages} {097001} (\bibinfo {year} {2010})}\BibitemShut {NoStop}%
\bibitem [{\citenamefont {{Wang}}\ \emph
  {et~al.}(2018{\natexlab{c}})\citenamefont {{Wang}}, \citenamefont {{Liu}},
  \citenamefont {{Lu}},\ and\ \citenamefont {{Zhang}}}]{lu}%
  \BibitemOpen
  \bibfield  {author} {\bibinfo {author} {\bibfnamefont {Q.}~\bibnamefont
  {{Wang}}}, \bibinfo {author} {\bibfnamefont {C.-C.}\ \bibnamefont {{Liu}}},
  \bibinfo {author} {\bibfnamefont {Y.-M.}\ \bibnamefont {{Lu}}}, \ and\
  \bibinfo {author} {\bibfnamefont {F.}~\bibnamefont {{Zhang}}},\ }\href@noop
  {} {\bibfield  {journal} {\bibinfo  {journal} {ArXiv e-prints}\ } (\bibinfo
  {year} {2018}{\natexlab{c}})},\ \Eprint {http://arxiv.org/abs/1804.04711}
  {arXiv:1804.04711 [cond-mat.mes-hall]} \BibitemShut {NoStop}%
\bibitem [{\citenamefont {{Yan}}\ \emph {et~al.}(2018)\citenamefont {{Yan}},
  \citenamefont {{Song}},\ and\ \citenamefont {{Wang}}}]{yan}%
  \BibitemOpen
  \bibfield  {author} {\bibinfo {author} {\bibfnamefont {Z.}~\bibnamefont
  {{Yan}}}, \bibinfo {author} {\bibfnamefont {F.}~\bibnamefont {{Song}}}, \
  and\ \bibinfo {author} {\bibfnamefont {Z.}~\bibnamefont {{Wang}}},\
  }\href@noop {} {\bibfield  {journal} {\bibinfo  {journal} {ArXiv e-prints}\ }
  (\bibinfo {year} {2018})},\ \Eprint {http://arxiv.org/abs/1803.08545}
  {arXiv:1803.08545 [cond-mat.mes-hall]} \BibitemShut {NoStop}%
\bibitem [{\citenamefont {{Liu}}\ \emph {et~al.}(2018)\citenamefont {{Liu}},
  \citenamefont {{Jun He}},\ and\ \citenamefont {{Nori}}}]{liu}%
  \BibitemOpen
  \bibfield  {author} {\bibinfo {author} {\bibfnamefont {T.}~\bibnamefont
  {{Liu}}}, \bibinfo {author} {\bibfnamefont {J.}~\bibnamefont {{Jun He}}}, \
  and\ \bibinfo {author} {\bibfnamefont {F.}~\bibnamefont {{Nori}}},\
  }\href@noop {} {\bibfield  {journal} {\bibinfo  {journal} {ArXiv e-prints}\ }
  (\bibinfo {year} {2018})},\ \Eprint {http://arxiv.org/abs/1806.07002}
  {arXiv:1806.07002 [cond-mat.mes-hall]} \BibitemShut {NoStop}%
\bibitem [{\citenamefont {Qi}\ \emph {et~al.}(2010)\citenamefont {Qi},
  \citenamefont {Hughes},\ and\ \citenamefont {Zhang}}]{qi}%
  \BibitemOpen
  \bibfield  {author} {\bibinfo {author} {\bibfnamefont {X.-L.}\ \bibnamefont
  {Qi}}, \bibinfo {author} {\bibfnamefont {T.~L.}\ \bibnamefont {Hughes}}, \
  and\ \bibinfo {author} {\bibfnamefont {S.-C.}\ \bibnamefont {Zhang}},\ }\href
  {\doibase 10.1103/PhysRevB.81.134508} {\bibfield  {journal} {\bibinfo
  {journal} {Phys. Rev. B}\ }\textbf {\bibinfo {volume} {81}},\ \bibinfo
  {pages} {134508} (\bibinfo {year} {2010})}\BibitemShut {NoStop}%
\bibitem [{\citenamefont {Weng}\ \emph {et~al.}(2015)\citenamefont {Weng},
  \citenamefont {Fang}, \citenamefont {Fang}, \citenamefont {Bernevig},\ and\
  \citenamefont {Dai}}]{weyl1}%
  \BibitemOpen
  \bibfield  {author} {\bibinfo {author} {\bibfnamefont {H.}~\bibnamefont
  {Weng}}, \bibinfo {author} {\bibfnamefont {C.}~\bibnamefont {Fang}}, \bibinfo
  {author} {\bibfnamefont {Z.}~\bibnamefont {Fang}}, \bibinfo {author}
  {\bibfnamefont {B.~A.}\ \bibnamefont {Bernevig}}, \ and\ \bibinfo {author}
  {\bibfnamefont {X.}~\bibnamefont {Dai}},\ }\href {\doibase
  10.1103/PhysRevX.5.011029} {\bibfield  {journal} {\bibinfo  {journal} {Phys.
  Rev. X}\ }\textbf {\bibinfo {volume} {5}},\ \bibinfo {pages} {011029}
  (\bibinfo {year} {2015})}\BibitemShut {NoStop}%
\bibitem [{\citenamefont {Huang}\ \emph {et~al.}(2015)\citenamefont {Huang},
  \citenamefont {Xu}, \citenamefont {Belopolski}, \citenamefont {Lee},
  \citenamefont {Chang}, \citenamefont {Wang}, \citenamefont {Alidoust},
  \citenamefont {Bian}, \citenamefont {Neupane}, \citenamefont {Zhang},
  \citenamefont {Jia}, \citenamefont {Bansil}, \citenamefont {Lin},\ and\
  \citenamefont {Hasan}}]{weyl2}%
  \BibitemOpen
  \bibfield  {author} {\bibinfo {author} {\bibfnamefont {S.-M.}\ \bibnamefont
  {Huang}}, \bibinfo {author} {\bibfnamefont {S.-Y.}\ \bibnamefont {Xu}},
  \bibinfo {author} {\bibfnamefont {I.}~\bibnamefont {Belopolski}}, \bibinfo
  {author} {\bibfnamefont {C.-C.}\ \bibnamefont {Lee}}, \bibinfo {author}
  {\bibfnamefont {G.}~\bibnamefont {Chang}}, \bibinfo {author} {\bibfnamefont
  {B.}~\bibnamefont {Wang}}, \bibinfo {author} {\bibfnamefont {N.}~\bibnamefont
  {Alidoust}}, \bibinfo {author} {\bibfnamefont {G.}~\bibnamefont {Bian}},
  \bibinfo {author} {\bibfnamefont {M.}~\bibnamefont {Neupane}}, \bibinfo
  {author} {\bibfnamefont {C.}~\bibnamefont {Zhang}}, \bibinfo {author}
  {\bibfnamefont {S.}~\bibnamefont {Jia}}, \bibinfo {author} {\bibfnamefont
  {A.}~\bibnamefont {Bansil}}, \bibinfo {author} {\bibfnamefont
  {H.}~\bibnamefont {Lin}}, \ and\ \bibinfo {author} {\bibfnamefont {M.~Z.}\
  \bibnamefont {Hasan}},\ }\href {http://dx.doi.org/10.1038/ncomms8373}
  {\bibfield  {journal} {\bibinfo  {journal} {Nature Communications}\ }\textbf
  {\bibinfo {volume} {6}},\ \bibinfo {pages} {7373 EP } (\bibinfo {year}
  {2015})}\BibitemShut {NoStop}%
\bibitem [{\citenamefont {Ran}\ \emph {et~al.}(2009)\citenamefont {Ran},
  \citenamefont {Zhang},\ and\ \citenamefont {Vishwanath}}]{ran}%
  \BibitemOpen
  \bibfield  {author} {\bibinfo {author} {\bibfnamefont {Y.}~\bibnamefont
  {Ran}}, \bibinfo {author} {\bibfnamefont {Y.}~\bibnamefont {Zhang}}, \ and\
  \bibinfo {author} {\bibfnamefont {A.}~\bibnamefont {Vishwanath}},\ }\href
  {http://dx.doi.org/10.1038/nphys1220} {\bibfield  {journal} {\bibinfo
  {journal} {Nature Physics}\ }\textbf {\bibinfo {volume} {5}},\ \bibinfo
  {pages} {298 EP } (\bibinfo {year} {2009})}\BibitemShut {NoStop}%
\bibitem [{\citenamefont {{van Miert}}\ and\ \citenamefont
  {{Ortix}}(2018)}]{miert}%
  \BibitemOpen
  \bibfield  {author} {\bibinfo {author} {\bibfnamefont {G.}~\bibnamefont {{van
  Miert}}}\ and\ \bibinfo {author} {\bibfnamefont {C.}~\bibnamefont
  {{Ortix}}},\ }\href@noop {} {\bibfield  {journal} {\bibinfo  {journal} {ArXiv
  e-prints}\ } (\bibinfo {year} {2018})},\ \Eprint
  {http://arxiv.org/abs/1802.00715} {arXiv:1802.00715 [cond-mat.mes-hall]}
  \BibitemShut {NoStop}%
\bibitem [{\citenamefont {Bernevig}\ and\ \citenamefont
  {Hughes}(2013)}]{bernevigbook}%
  \BibitemOpen
  \bibfield  {author} {\bibinfo {author} {\bibfnamefont {B.}~\bibnamefont
  {Bernevig}}\ and\ \bibinfo {author} {\bibfnamefont {T.}~\bibnamefont
  {Hughes}},\ }\href {https://books.google.com/books?id=wOn7JHSSxrsC} {\emph
  {\bibinfo {title} {Topological Insulators and Topological Superconductors}}}\
  (\bibinfo  {publisher} {Princeton University Press},\ \bibinfo {year}
  {2013})\BibitemShut {NoStop}%
\end{thebibliography}%

\appendix
\onecolumngrid
\vspace{1 cm}
\begin{center}
\large{\textbf{Supplementary material}}
\end{center}

\section{Helical hinge modes in a continuum model}\label{continuum}

In this appendix we show that the 3D higher order band invariant in Eq. \eqref{invariant1} detects the presence of helical hinge modes. We will do this by explicitely constructing a model that has a non-trivial higher order invariant and solving for the hinge modes. We start with following continuum model in the basis $\Psi_{\textbf{k}}=\left(c_{\textbf{k}\uparrow} , c_{\textbf{k}\downarrow} ,c^\dagger_{-\textbf{k}\uparrow} , c^\dagger_{-\textbf{k}\downarrow} \right)^T$, realizing a strong TSC :
\begin{equation}\label{strongTSC}
H_1(\textbf{k})=\left( \begin{matrix}\frac{\textbf{k}^2}{2m}-\mu & 0 & \Delta k_- & i\Delta k _z \\ 0 & \frac{\textbf{k}^2}{2m}-\mu & i\Delta k_z & \Delta k_+ \\ \Delta k_+ & -i\Delta k_z & -\frac{\textbf{k}^2}{2m}+\mu & 0 \\ -i\Delta k_z & \Delta k_- & 0 &- \frac{\textbf{k}^2}{2m}+\mu \end{matrix}\right)\, ,
\end{equation}
with $\Delta > 0$ and $k_\pm = k_x \pm i k_y$. For $\mu>0$ it has a non-trivial strong invariant $N=-1$ \cite{bernevigbook}. The continuum Hamiltonian is time reversal symmetric with $\mathcal{T}=(\tau^z\otimes\sigma^y)K$ (The standard action of time reversal $\mathcal{T}=i\sigma^yK$ can be recovered by a U$(1)$ gauge transformation $c_{\textbf{k},\uparrow(\downarrow)}\rightarrow e^{i\pi/4}c_{\textbf{k},\uparrow(\downarrow)}$, which would also multiply the pairing with $i$).

To construct a model for the higher order TSC, we proceed in analogy to the continuum model construction in appendix A of Ref. \cite{bismuth}. We start with a block-diagonal BdG Hamiltonian:
\begin{equation}
H(\textbf{k}) = H_1(\textbf{k})\oplus H_3(\textbf{k})\, ,
\end{equation}
where $H_1(\textbf{k})$ is given by the strong TSC \eqref{strongTSC}, and is defined using the orbitals $\{|p_+\,\downarrow\rangle\,, |p_-\,\uparrow\rangle\}$. $C_4$ acts on these orbitals as $U_{R_1}=e^{i\pi\sigma^z/4}$. The BdG Hamiltonian $H_1(\textbf{k})$ satisfies $\mathcal{R}_1^\dagger H_1(\textbf{k})\mathcal{R}_1=H_1(\textbf{Rk})$, with $\mathcal{R}_1=U_{R_1}\oplus U_{R_1}^*$.

To construct a model with non-trivial higher order invariants and trivial strong invariant, the Hamiltonian $H_3(\textbf{k})$ has to differ from $H_1(\textbf{k})$ in two ways: it needs have an opposite strong invariant, i.e. $N=1$, and it should live in the $(e^{3i\pi/4},e^{-3i\pi/4})$ rotation subspace. To achieve the latter, we define $H_3(\textbf{k})$ using the orbitals $\{|p_+\,\uparrow\rangle\,,|p_-\,\downarrow\rangle\}$. $H_3(\textbf{k})$ is given explicitly by
\begin{equation}
H_3(\textbf{k}) = \left( \begin{matrix}-\frac{\textbf{k}^2}{2m}+\mu & 0 & \Delta k_{-}^3 & i\Delta k _z \\ 0 & -\frac{\textbf{k}^2}{2m}+\mu & i\Delta k_z & \Delta k_{+}^3 \\ \Delta k^3_+ & -i\Delta k_z & \frac{\textbf{k}^2}{2m}-\mu & 0 \\ -i\Delta k_z & \Delta k_{-}^3 & 0 & \frac{\textbf{k}^2}{2m}-\mu \end{matrix}\right)\, .
\end{equation}
The $C_4$ rotation symmetry is realized as $\mathcal{R}_3^\dagger H_3(\textbf{k})\mathcal{R}_3=H_3(\textbf{Rk})$, with $\mathcal{R}_3=U_{R_3}\oplus U_{R_3}^*$ and $U_{R_3}=e^{i3\pi\sigma^z/4}$. It is clear that the $C_4$ symmetry matrix of the complete Hamiltonian $H(\textbf{k})=H_1(\textbf{k})\oplus H_3(\textbf{k})$, given by $\mathcal{R} = \mathcal{R}_1\oplus \mathcal{R}_3$, satisfies $\mathcal{R}^4 = -\mathds{1}$ and commutes with the particle-hole symmetry operator $\mathcal{C} = (\mathds{1}\otimes\tau^x\otimes\mathds{1})K$. Furthermore, it also holds that $[\mathcal{T},\mathcal{R}]=[\mathcal{C},\mathcal{T}]=0$.

To see that $H_3(\textbf{k})$ has a strong invariant $N=1$, consider the Hamiltonian
\begin{equation}
\tilde{H}_\alpha(\textbf{k}) = \left( \begin{matrix}-\frac{\textbf{k}^2}{2m}+\mu & 0 & \Delta(\alpha k_{-}+(1-\alpha)k_{-}^3) & i\Delta k _z \\ 0 & -\frac{\textbf{k}^2}{2m}+\mu & i\Delta k_z & \Delta(\alpha k_{+}+(1-\alpha)k_+^3) \\ \Delta(\alpha k_+ +(1-\alpha)k_+^3)& -i\Delta k_z & \frac{\textbf{k}^2}{2m}-\mu & 0 \\ -i\Delta k_z & \Delta (\alpha k_{-}+(1-\alpha)k_-^3) & 0 & \frac{\textbf{k}^2}{2m}-\mu \end{matrix}\right)\, .
\end{equation}
First, we note that $\alpha$ can be continuously interpolated from $0$ to $1$, without closing the energy gap or breaking time reversal symmetry. We have that $\tilde{H}_0(\textbf{k})=H_3(\textbf{k})$, and $\tilde{H}_1(\textbf{k})$ differs from $H_1(\textbf{k})$ only by the sign of the kinetic terms and chemical potential. Now define $\tilde{h}_1(\textbf{k})=-h_1(\textbf{k})=\left(-\frac{\textbf{k}^2}{2m}+\mu\right)\mathds{1}$ and $\tilde{\Delta}_1(\textbf{k})=\Delta_1(\textbf{k})=\Delta(k_x\mathds{1}-ik_y\sigma^z+ik_z\sigma^x)$. To calculate the strong invariant of $\tilde{H}_1(\textbf{k})$, one considers the matrix $\tilde{h}_1(\textbf{k})+i\sigma^y\tilde{\Delta}_1(\textbf{k})^\dagger$. Taking the singular value decomposition of this matrix gives $\tilde{h}_1(\textbf{k})+i\sigma^y\tilde{\Delta}_1(\textbf{k})^\dagger=\tilde{U}_1^\dagger(\textbf{k})\tilde{D}_1(\textbf{k})\tilde{V}_1\textbf{k})$, where $\tilde{U}_1(\textbf{k})$ and $\tilde{V}_1(\textbf{k})$ are unitary matrices. $\tilde{D}_1(\textbf{k})$ is a diagonal matrix which is strictly positive iff the BdG Hamiltonian is gapped. Defining $\tilde{Q}_1(\textbf{k})\equiv \tilde{U}_1^\dagger(\textbf{k})\tilde{V}_1(\textbf{k})$, the strong invariant of $\tilde{H}_1(\textbf{k})$ is given by \cite{ryu}
\begin{equation}\label{integralexpr}
\tilde{N}_1=\frac{1}{24\pi^2}\int \mathrm{d}^3\textbf{k}\,\epsilon^{mnl}\text{tr}\left(\tilde{Q}_1(\textbf{k})^\dagger\partial_m \tilde{Q}_1(\textbf{k})\tilde{Q}_1(\textbf{k})^\dagger\partial_n\tilde{Q}_1(\textbf{k})\tilde{Q}_1(\textbf{k})^\dagger\partial_l\tilde{Q}_1(\textbf{k}) \right)\, .
\end{equation} 
To calculate the strong invariant of $H_1(\textbf{k})$, we now define $Q_1(\textbf{k})=U_1^\dagger(\textbf{k})V_1(\textbf{k})$ via the singular value decomposition of $h_1(\textbf{k})+i\sigma^y\Delta_1(\textbf{k})=U_1^\dagger(\textbf{k})D_1(\textbf{k})V_1(\textbf{k})$. Because $\sigma^y\Delta_1(\textbf{k})$ is Hermitian it follows that $Q_1(\textbf{k})=-\tilde{Q}_1(\textbf{k})^\dagger$. Plugging this into the integral expression \eqref{integralexpr}, we immediately see that $N_1=-\tilde{N}_1$. Since $\tilde{H}_1$ and $H_3$ have the same invariant, his completes the proof that $H_3(\textbf{k})$ and $H_1(\textbf{k})$ have opposite strong invariants.

The Hamiltonian $H_1(\textbf{k})\oplus H_3(\textbf{k})$ has two boundary Majorana cones with opposite chirality. We now want to add terms to the Hamiltonian that gap out these Majorana cones. Such terms should should couple the different $C_4$ subspaces, i.e. be of the form: $\mu^{x/y}\otimes\tau^\mu\otimes\sigma^\nu$; be time reversal symmetric; preserve particle-hole symmetry; anti-commute with the kinetic terms that involve momenta tangent to the boundary and finally anti-commute with the bulk chemical potential in order not to result in a bulk gap closing. Concretely, these conditions imply that the mass terms have to commute with $(\mathds{1}\otimes\tau^z\otimes\sigma^y)K$, and anti-commute with $(\mathds{1}\otimes \tau^x\otimes\mathds{1})K$, $\mathds{1}\otimes\tau^y\otimes\sigma^x$ and $\mu^z\otimes\tau^z\otimes\mathds{1}$. Out of the 32 terms of the form $\mu^{x/y}\otimes\tau^\mu\otimes\sigma^\nu$, there are two terms satisfying all these criterea:
\begin{equation}
\mu^y\otimes\mathds{1}\otimes \sigma^z\, , \hspace{0,5 cm} \mu^x\otimes \tau^z\otimes\mathds{1}\, . 
\end{equation}
Under $C_4$ they transform into each other as
\begin{eqnarray}\label{massrot}
\mathcal{R}(\mu^y\otimes\mathds{1}\otimes \sigma^z)\mathcal{R}^\dagger =  \mu^x \otimes\tau^z\otimes\mathds{1}\, , \hspace{1 cm}
\mathcal{R}(\mu^x \otimes\tau^z\otimes\mathds{1})\mathcal{R}^\dagger = -\mu^y\otimes\mathds{1}\otimes\sigma^z\, ,
\end{eqnarray}
where as before $\mathcal{R}=\mathcal{R}_1\oplus\mathcal{R}_3=U_{R_1}\oplus U_{R_1}^*\oplus U_{R_3}\oplus U_{R_3}^*$, with $U_{R_1}=e^{i\pi\sigma^z/4}$ and $U_{R_3}=e^{i3\pi\sigma^z/4}$.

Now we explicitly solve for the surface zero energy modes of the Hamiltonian $H_1(\textbf{k})\oplus H_3(\textbf{k})$ on a solid cylinder defined by $r\leq r_0$ in cylindrical coordinates $(r,\phi,z)$. To find the surface modes, we define $\rho=r-r_0$ and set all momenta tangent to the surface $\rho=0$ to zero to look for the zero energy boundary states of following Hamiltonians:
\begin{equation}
\left(\begin{matrix}-\mu\text{sgn}(\rho) & 0 & \Delta (-i)^je^{-i\phi j}\partial_\rho^j & 0 \\ 0 & -\mu\text{sgn}(\rho) & 0 & \Delta (-i)^je^{i\phi j}\partial_\rho^j  \\ \Delta (-i)^je^{i\phi j}\partial_\rho^j  & 0 & \mu\text{sgn}(\rho) & 0 \\ 0 & \Delta(-i)^je^{-i\phi j}\partial^j_\rho & 0 & \mu \text{sgn}(\rho)\end{matrix}\right)\, ,
\end{equation}
where either $j=1$ (corresponding to $H_1(\textbf{k})$) or $j=3$ (corresponding to $H_3(\textbf{k})$). We have also taken the limit $m\rightarrow \infty$, which can be done for the purpose of finding the surface zero modes. Each of these Hamiltonians has two normalizable zero modes, given by
\begin{equation}
|v_j\rangle = a_j\left(\begin{matrix}1 \\ 0 \\ (-i)^je^{i\phi j}\text{sgn}\left(\frac{\mu}{\Delta} \right) \\ 0 \end{matrix}\right)\;\;\;\;\;\;|w_j\rangle =  a_j\left(\begin{matrix} 0 \\ 1 \\ 0 \\ (-i)^je^{-i\phi j}\text{sgn}\left(\frac{\mu}{\Delta} \right) \end{matrix}\right)
\end{equation}
with 
\begin{equation}
a_j = e^{-\left|\frac{\mu}{\Delta}\right|^{1/j} |\rho|}\, .
\end{equation}
Because of the transformation properties \eqref{massrot} we can without loss of generality write a generic $C_4$ symmetic surface mass term as:
\begin{equation}\label{massterm}
M_s(\phi) = M(\cos(\phi) \mu^x\otimes\tau^z\otimes\mathds{1} - \sin(\phi)\mu^y\otimes\mathds{1}\otimes\sigma^z)\, ,
\end{equation}
We now calculate the following matrix elements (matrix elements within the same $C_4$ sector always vanish by construction since the mass term in Eq. \eqref{massterm} only contains off-diagonal matrices $\mu^x$ and $\mu^y$)
\begin{eqnarray}
\langle v_1|\mu^y\otimes\mathds{1}\otimes\sigma^z|v_3\rangle & \propto &  i(1-e^{2i\phi}) \\ 
\langle v_1|\mu^x\otimes\tau^z\otimes\mathds{1}|v_3\rangle & \propto &  1+e^{2i\phi} \\
\langle v_1|\mu^y\otimes\mathds{1}\otimes\sigma^z|w_3\rangle & = &  0 \\
\langle v_1|\mu^x\otimes\tau^z\otimes\mathds{1}|w_3\rangle & = &  0 \\
\langle w_1|\mu^y\otimes\mathds{1}\otimes\sigma^z|w_3\rangle & \propto & -i(1-e^{-i2\phi})  \\
\langle w_1|\mu^x\otimes\tau^z\otimes\mathds{1}|w_3\rangle & \propto & 1+e^{-2i\phi}
\end{eqnarray}
From these matrix elements we see that the surface mass term projected onto the zero energy sector satisfies
\begin{eqnarray}
\langle v_1|M_s(\phi)|v_3\rangle \propto e^{-i\phi}(1+e^{i4\phi}) \\
\langle w_1|M_s(\phi)|w_3\rangle \propto e^{-3i\phi}(1+e^{i4\phi})
\end{eqnarray}
This shows what we expected: at four values for $\phi$ the surface mass gap vanishes, leading to zero energy states that disperse into helical hinge states when adding $k_z$-dependent terms.

\section{Even-parity $C_4$ symmetric higher order superconductors from 3D topological insulators}

Here we present the details about the construction of $C_4$ symmetric higher order topological superconductors in the symmetry class $\{\mathcal{T},\mathcal{R}\}=0$ with non-trivial band indices as in Eq. \eqref{TIcase}, by combining a topological insulator with the appropriate pairing. We start with a three-dimensional particle number conserving Hamiltonian $h(\textbf{k})$ satisfying
\begin{eqnarray}
U_T^\dagger h(-\textbf{k})^*U_T & = & h(\textbf{k})\\
U_I^\dagger h(-\textbf{k})U_I & = & h(\textbf{k})\\
U_R^\dagger h(\textbf{k})U_R & = & h(\textbf{Rk})\, ,
\end{eqnarray}
such that $U_R^4=-\mathds{1}$, $U_I^2=\mathds{1}$, $U_TU_T^*=-\mathds{1}$, $[U_I,U_R]=0$, $U_TU_I=U_I^*U_T$ and $U_TU_R=U_R^*U_T$. We take $h(\textbf{k})$ to be a topological insulator, implying that there is a band gap containing the Fermi level and that the number of occupied Kramers pairs at the TRIM with negative inversion eigenvalues is odd \cite{fukane}. Under $C_4$, Kramers pairs at $\textbf{k}=(0,\pi,0)$ and $\textbf{k}=(\pi,0,0)$ (and at $\textbf{k}=(0,\pi,\pi)$ and $\textbf{k}=(\pi,0,\pi)$) are interchanged. Because $U_R$ commutes with $U_I$, this implies that the inversion eigenvalues of the occupied Kramers pairs at $(0,\pi,0)$ are the same as those at $(\pi,0,0)$ (and at $(0,\pi,\pi)$ and $(\pi,0,\pi)$), so they do not contibute to the Fu-Kane invariant.

\begin{center}
\begin{figure}[h]
a)
\includegraphics[scale=0.465]{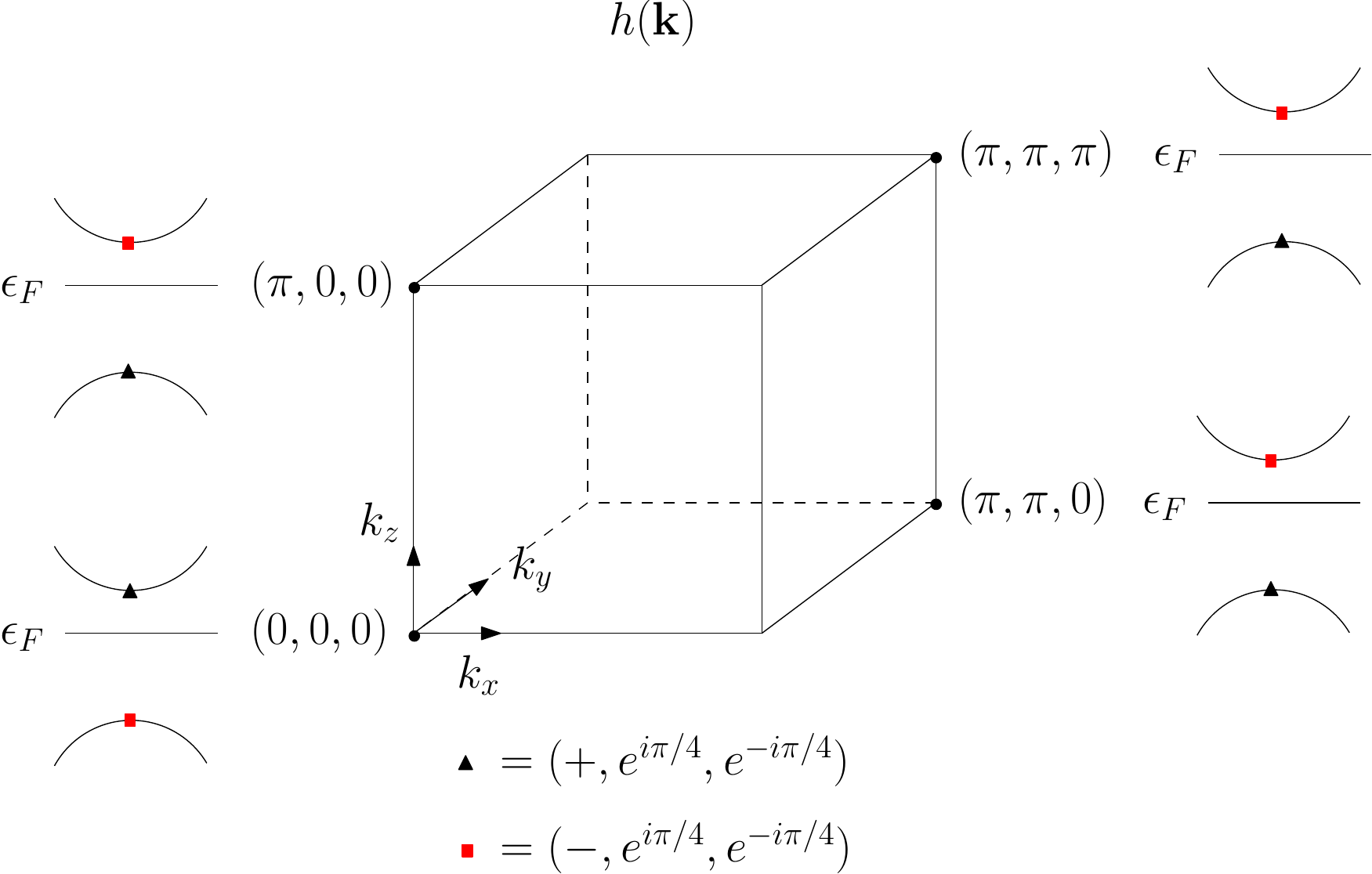}
\hspace{0.5 cm}b)
\includegraphics[scale=0.465]{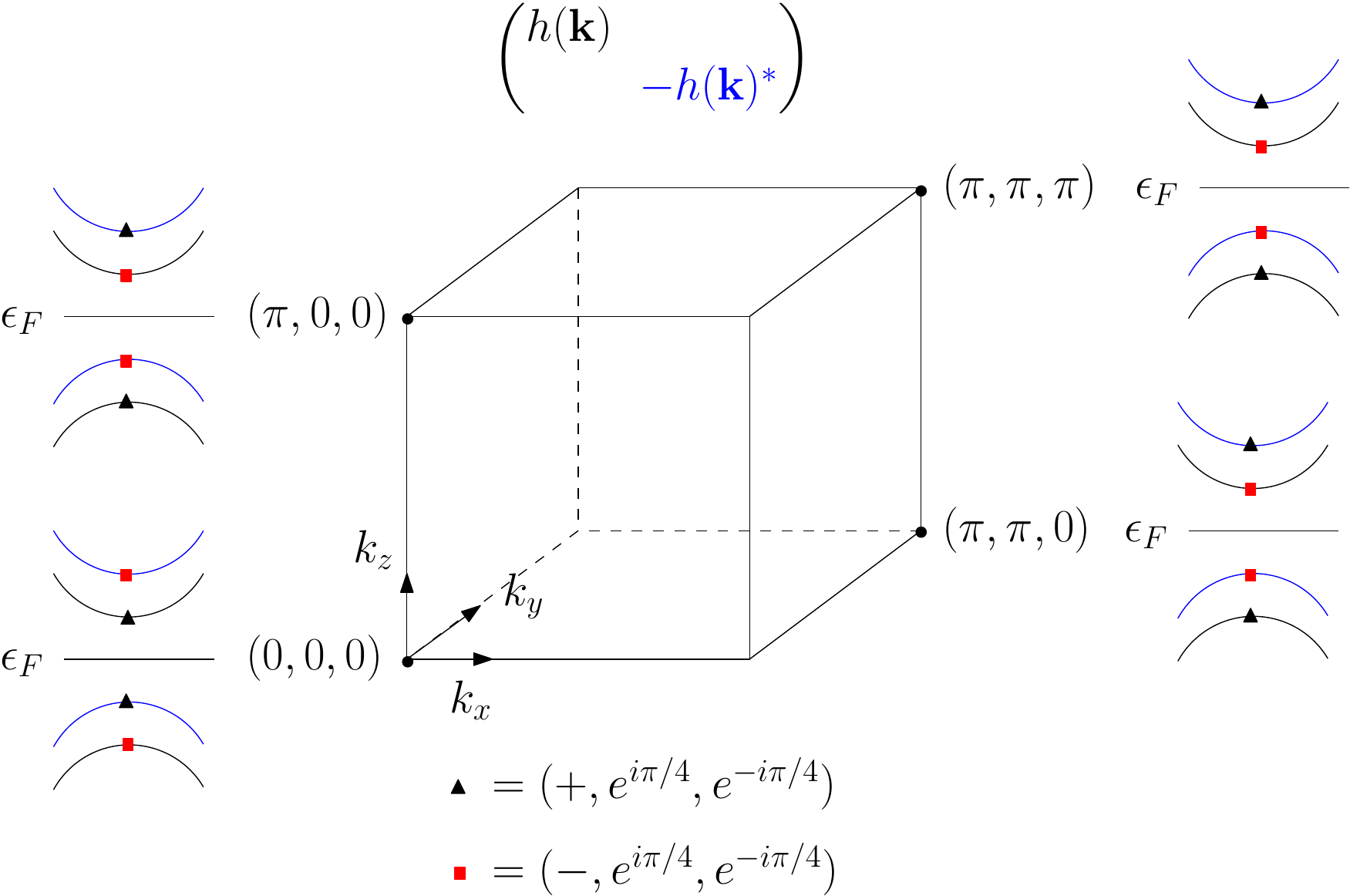}\\
\vspace{1 cm}c)
\includegraphics[scale=0.465]{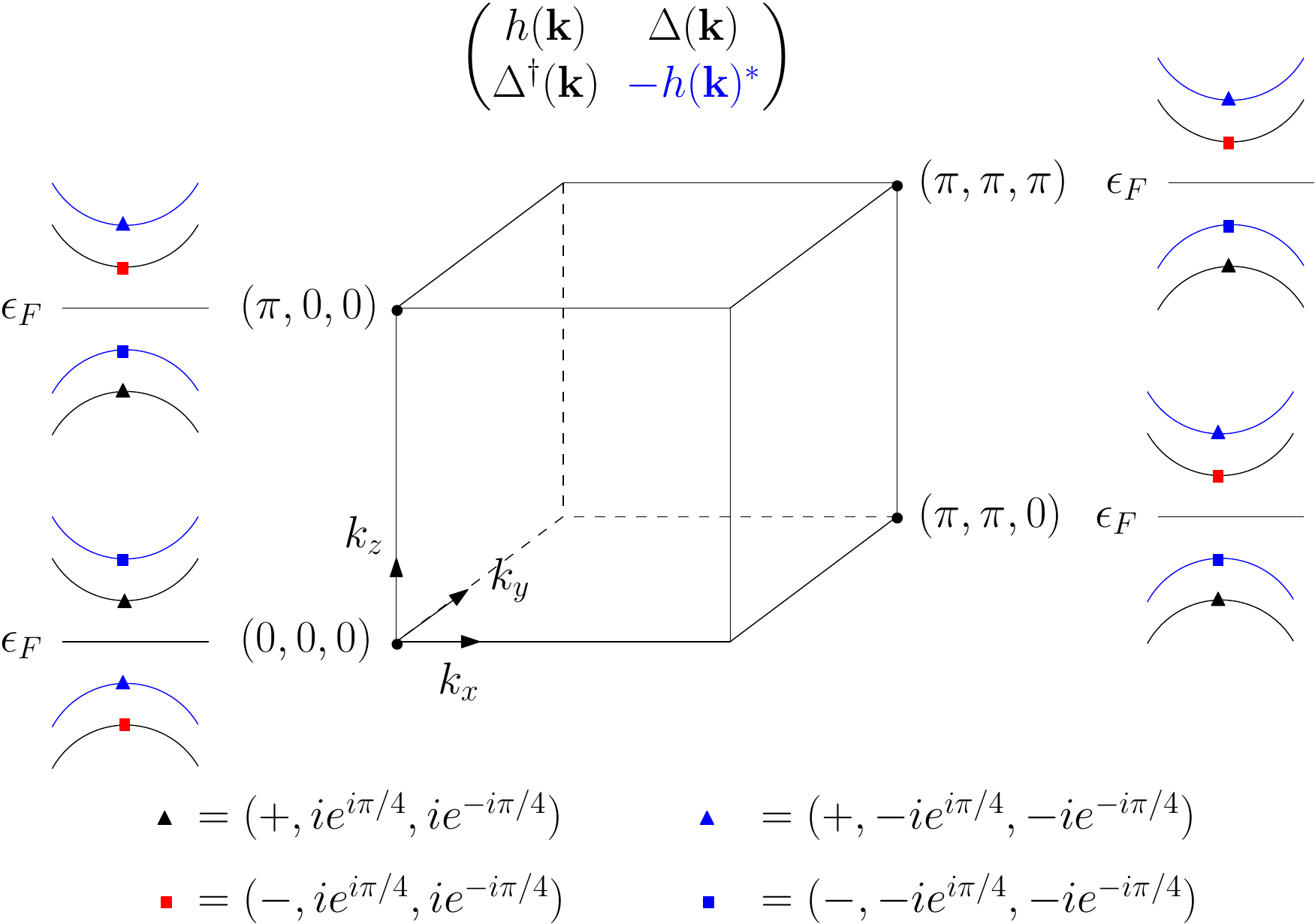}
\caption{a) Schematic representation of the occupied Kramers pairs and their symmetry eigenvalues at the four TRIM fixed under $C_4$ of a simple four-band --the bands drawn are doubly degenerate-- topological insulator model $h(\textbf{k})$. $\epsilon_F$ is the Fermi energy. $(+,e^{i\pi/4},e^{-i\pi/4})$ and $(-,e^{i\pi/4},e^{-i\pi/4})$ denote the inversion and $C_4$ eigenvalues. We consider the case where in total there is a single occupied Kramers pair with negative inversion eigenvalues at the four TRIM fixed under $C_4$, which is located at $\Gamma=(0,0,0)$. As explained in the text, Kramers pairs at TRIM exchanged by $C_4$ can be ignored. Inversion and $C_4$ are implemented as $U_I^\dagger h(\textbf{k})U_I=h(-\textbf{k})$ and $U_R^\dagger h(\textbf{k})U_R=h(\textbf{Rk})$, where there exists a basis such that $U_I=\text{diag}(1,1,-1,-1)$ and $U_R=\text{diag}(e^{i\pi/4},e^{-i\pi/4},e^{i\pi/4},e^{-i\pi/4})$. Since $U_I$ and $U_R$ are independent of $\textbf{k}$ the Kramers pairs (both occupied and unoccupied) have the same symmetry eigenvalues at all TRIM fixed under $C_4$. This implies that there is an unoccupied Kramers pair with negative inversion eigenvalues and $C_4$ eigenvalues $(e^{i\pi/4},e^{-i\pi/4})$ at the three TRIM $(\pi,0,0), (\pi,\pi,0)$ and $(\pi,\pi,\pi)$. b) The particle-hole symmetric band spectrum before the addition of $\Delta(\textbf{k})$. c) We add the small pairing $\Delta(\textbf{k})$. Because $U_R^\dagger\Delta(\textbf{k})U_R^*=-\Delta(\textbf{Rk})$, the $C_4$ eigenvalues of the bands corresponding to $h(\textbf{k})$ pick up an additional factor of $i$, and those of $-h(\textbf{k})^*$ a factor of $-i$. Because the pairing is parity even, the inversion eigenvalues remain unchanged. Using $(ie^{i\pi/4},ie^{-i\pi/4})=(e^{i3\pi/4},e^{i\pi/4})$ and $(-ie^{i\pi/4},-ie^{-i\pi/4})=(e^{-i\pi/4},e^{-i3\pi/4})$, we see that there is a single occupied Kramers pair with negative inversion eigenvalues and $C_4$ eigenvalues $(e^{i3\pi/4},e^{i\pi/4})$, and three occupied Kramers pairs with negative inversion eigenvalues and $C_4$ eigenvalues $(e^{-i\pi/4},e^{-i3\pi/4})$. This gives a non-trivial higher order band invariant for the superconductor $\nu^{(-)}=\nu^{(+)}=1$.}\label{figapp0}
\end{figure}
\end{center}

At the four TRIM fixed under $C_4$, which are $(0,0,0)$, $(\pi,\pi,0)$, $(0,0,\pi)$ and $(\pi,\pi,\pi)$, we can label the occupied Kramers pairs of $h(\textbf{k})$ with their $C_4$ eigenvalues $(e^{i\pi/4},e^{-i\pi/4})$ or $(e^{3i\pi/4},e^{-3i\pi/4})$. Assume that the number of occupied Kramers pairs at TRIM fixed under $C_4$ with negative inversion eigenvalues and $C_4$ eigenvalues $(e^{i\pi/4},e^{-i\pi/4})$ is odd. Because $h(\textbf{k})$ is a topological insulator, this automatically implies that the number of occupied Kramers pairs with negative inversion eigenvalues and $C_4$ eigenvalues $(e^{3i\pi/4},e^{-3i\pi/4})$ is even (the reverse case where the number of occupied Kramers with negative inversion eigenvalues and $C_4$ eigenvalues $(e^{3i\pi/4},e^{-3i\pi/4})$ is odd, and the number of occupied negative inversion eigenvalue Kramers pairs with $C_4$ eigenvalues $(e^{i\pi/4},e^{-i\pi/4})$ is even can be done by analogy and will not be considered explicitly here). Note that if there is an odd number of \emph{occupied} Kramers pairs of $h(\textbf{k})$ with negative inversion eigenvalues and $C_4$ eigenvalues $(e^{i\pi/4},e^{-i\pi/4})$ at the TRIM fixed under $C_4$, then there is necessarily also and odd number of \emph{unoccupied} Kramers pairs of $h(\textbf{k})$ with negative inversion eigenvalues and $C_4$ eigenvalues $(e^{i\pi/4},e^{-i\pi/4})$ at the TRIM fixed under $C_4$. This is because $U_I$ and $U_R$ do not depend on momentum (recall that all orbitals are on the vertices of the cubic lattice and there are no orbitals on other Wyckoff positions), and the total number of TRIM fixed under $C_4$ is even. We illustrate this with a simple example in figure \ref{figapp0}.

Now we add a small pairing with the properties
\begin{eqnarray}
U_I^\dagger\Delta(-\textbf{k})U_I^* & = & \Delta(\textbf{k})\\
U_R^\dagger\Delta(\textbf{k})U_R^* & = & -\Delta(\textbf{Rk})\,,
\end{eqnarray}
whose lowest harmonic corresponds to $d$-wave pairing. Because of the transformation property of the pairing under $C_4$, the rotation symmetry of the resulting BdG Hamiltonian is implemented by $\mathcal{R}=iU_R\oplus(iU_R)^*$. At the TRIM fixed by $C_4$, the BdG Hamiltonian is of the form $H(\textbf{k}) = h(\textbf{k})\oplus(-h(\textbf{k})^*)$, where we ignore the pairing since it is assumed to be small and does not cause a band inversion. There are two types of occupied BdG Kramers pairs at the TRIM fixed under $C_4$. The first type are simply the occupied Kramers pairs of $h(\textbf{k})$. Under the BdG rotation operator $\mathcal{R}$ these acquire rotation eigenvalues $(ie^{i\pi/4},ie^{-i\pi/4})=(e^{i3\pi/4},e^{i\pi/4})$ and $(ie^{i3\pi/4},ie^{-i3\pi/4}) = (e^{-i3\pi/4},e^{-i\pi/4})$. The second type of occupied BdG Kramers pairs come from $-h(\textbf{k})^*$ and correspond to the originally unoccupied Kramers pairs. They acquire $C_4$ eigenvalues $(-ie^{i\pi/4},-ie^{-i\pi/4})=(e^{-i\pi/4},e^{-i3\pi/4})$ and $(-ie^{i3\pi/4},-ie^{-i3\pi/4})=(e^{i\pi/4},e^{i3\pi/4})$. Because the pairing is parity even, the Kramers pairs keep their inversion eigenvalues after the pairing is introduced. In Fig. \ref{figapp0}, we illustrate how the Kramers pairs transform before and after adding the small pairing $\Delta(\textbf{k})$ via an explicit example.

From the properties of the particle number conserving Hamiltonian, we see that after introducing the small pairing there will be an odd number of occupied Kramers pairs with negative inversion eigenvalues and $C_4$ eigenvalues $(e^{i3\pi/4},e^{i\pi/4})$ at TRIM fixed under $C_4$, and likewise also an odd number of occupied Kramers pairs with negative inversion eigenvalues and with $C_4$ eigenvalues $(e^{-i3\pi/4},e^{-i\pi/4})$. So we conclude that the BdG superconductor indeed has a non-trivial higher order band invariant $\nu^{(+)}=\nu^{(-)}=1$.

\section{Hermiticity of $U_T^\dagger\Delta(\textbf{k})^\dagger$}
To show the Hermiticity of $U_T^\dagger\Delta(\textbf{k})^\dagger$ we need following properties of the pairing:
\begin{eqnarray}
\Delta(-\textbf{k})^T&=&-\Delta(\textbf{k})\, ,\\
U_T^\dagger\Delta(-\textbf{k})^*U_T^*&=&\Delta(\textbf{k})\,,
\end{eqnarray}
where the first property holds for any nonvanishing pairing and the second follows from time reversal symmetry. We now do following manipulations
\begin{eqnarray}
\left(U_T^\dagger\Delta(\textbf{k})^\dagger \right)^\dagger & = & \Delta(\textbf{k})U_T\\
& = & -\Delta(\textbf{k})U_T^T\\
 & = & -U_T^\dagger \Delta(-\textbf{k})^*\\
  & = &U_T^\dagger\Delta(\textbf{k})^\dagger\, ,
\end{eqnarray}
where we also used that $U_T$ is anti-symmetric, which follows from $U_TU_T^*=-\mathds{1}$. Another important feature of the matrix $U_T^\dagger\Delta(\textbf{k})^\dagger$ that makes the weak pairing invariant \eqref{weakpairing} well defined is that it is invariant under global U$(1)$ transformations $c_{\alpha,\textbf{r}}\rightarrow e^{i\theta}c_{\alpha,\textbf{r}}$. After the U$(1)$ transformation, time reversal acts as $e^{i\theta}c_{\alpha,\textbf{r}}\rightarrow \sum_\beta (U_T)_{\alpha\beta}e^{-i\theta}c_{\beta,\textbf{r}}=\sum_\beta \left(e^{-i2\theta}(U_T)_{\alpha\beta}\right)\left(e^{i\theta}c_{\beta,\textbf{r}}\right)$. The U$(1)$ action also transforms the pairing as $\Delta(\textbf{k})\rightarrow e^{i2\theta}\Delta(\textbf{k})$, so the combination $U_T^\dagger\Delta(\textbf{k})^\dagger$ is invariant.

\section{Real space tight binding model realizing hybrid higher order topology}\label{realspace}

The tight binding model described by Eqs. \eqref{eq:normal} and \eqref{eq:pairing2} and exhibiting hybrid higher order topology is defined on a cubic lattice, with a $p_x$, $p_y$ and $s$ orbital on every site. The model is defined in the orbital basis $\{|p_+\,\downarrow\rangle\,,|p_-\,\uparrow\rangle\,,\,|p_-\,\downarrow\rangle\,,\,|p_+\,\uparrow\rangle\,,\,|s \,\downarrow\rangle\,,\,|s\uparrow\rangle \}$, where $p_\pm = p_x\pm ip_y$. $C_4$ acts on these orbitals as
\begin{equation}
U_R = e^{i\pi\sigma^z/4}\oplus e^{-i3\pi\sigma^z/4}\oplus e^{i\pi\sigma^z/4}\, .
\end{equation}
Note that the momentum space BdG Hamiltonian described in Eqs. \eqref{eq:normal} and \eqref{eq:pairing2} is inversion symmetric with $\mathcal{I}=\tau^z$, while the inversion operator on the orbitals above takes the form $U'_I=\text{diag}(-1,-1,-1,-1,1,1)$. However, the fact that the BdG Hamiltonian breaks $U'_I$ does not matter for the calculation of the band invariants indicating non-trivial hybrid higher order topology, as long as $\mathcal{I}$ is preserved. We fix the phase of the orbitals such that time reversal acts as $\mathcal{T}=\sigma^yK$. The standard expression for time reversal $\mathcal{T}=i\sigma^yK$ can be recovered by a gauge transformation where the orbitals are multiplied by $e^{-i\pi/4}$. Indeed, as was shown in the previous appendix, under a gauge tranformation $c_{\alpha,\textbf{r}}\rightarrow e^{-i\pi/4}c_{\alpha,\textbf{r}}$ the unitary part of the time reversal operator transforms as $U_T\rightarrow e^{i2\pi/4}U_T=iU_T$.

\begin{center}
\begin{figure}[h]\label{figrealspace}
\includegraphics[scale=0.55]{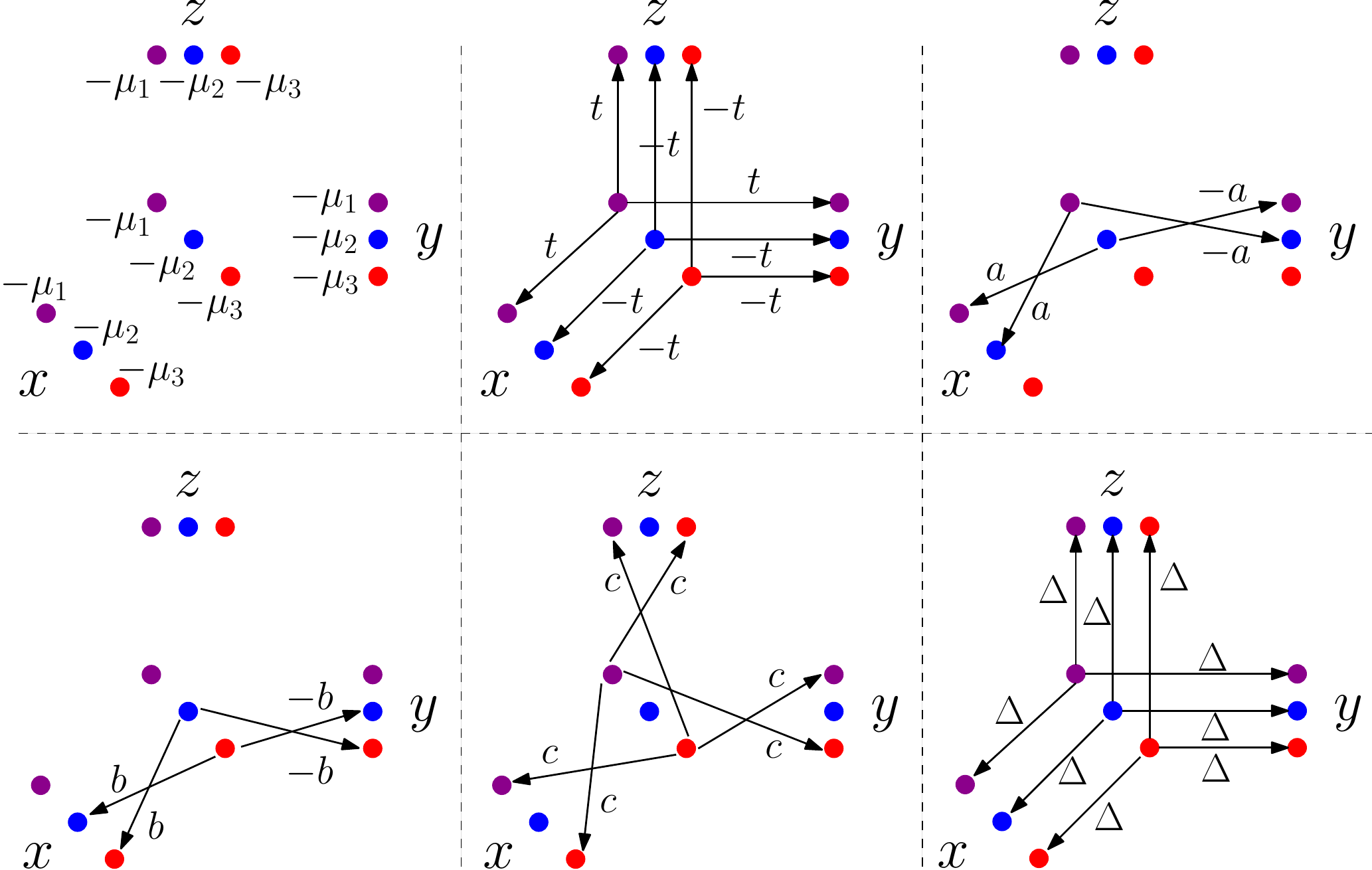}
\caption{Schematic representation of the terms in the real space tight binding model. We group the six orbitals per site of the cubic lattice in three groups of two, labeled by $1,2$ and $3$ as in Eqs. \eqref{g1}, \eqref{g2} and \eqref{g3}. The orbitals in group one are denoted with a purple dot, those in group two with a blue dot and the orbitals in group three with a red dot. The on-site terms for the three groups of orbitals consist of the chemical potentials $-\mu_1$, $-\mu_2$ and $-\mu_3$. We schematically show the nearest-neighbour terms along the $x$, $y$ and $z$-directions (the tight binding model does not contain longer-range terms). Along the $x$-direction, there is a diagonal hopping parameter $t$ for orbitals in group $1$ and a diagonal hopping parameter $-t$ for orbitals in groups $2$ and $3$. There is hopping between orbitals $2$ and $3$ along the $x$-direction with hopping parameter $a$, and along the $y$-direction with hopping parameter $-a$. The hopping between $2$ and $3$ happens with strength $b$ along the $x$-direction and $-b$ along the $y$-direction. Finally, the hopping between orbitals in group $1$ and $3$ has the same hopping parameter $c$ in both the $x$, $y$ and $z$-direction. The pairing is isotropic with strength $\Delta$ and only couples orbitals within the same group labeled by $1$, $2$ and $3$.}
\end{figure}
\end{center}

To write down the real space Hamiltonian, we first define
\begin{eqnarray}
\psi_{1,\textbf{r}} & = & \left(c_{p_+,\downarrow,\textbf{r}}\, , \, c_{p_-,\uparrow,\textbf{r}}\right)^T \label{g1}\\
\psi_{2,\textbf{r}} & = & \left(c_{p_-,\downarrow,\textbf{r}}\, , \, c_{p_+,\uparrow,\textbf{r}}\right) ^T\label{g2}\\
\psi_{3,\textbf{r}} & = & \left(c_{s,\downarrow,\textbf{r}}\, , \, c_{s,\uparrow,\textbf{r}}\right)^T\, , \label{g3}
\end{eqnarray}
where $\textbf{r}$ denotes the sites on the cubic lattice. The Hamiltonian is then
\begin{eqnarray}
\hat{H}& = & \sum_{\textbf{r}} t\,\psi_{1,\textbf{r}}^\dagger \psi_{1,\textbf{r}+\hat{x}} - t\,\psi_{2,\textbf{r}}^\dagger \psi_{2,\textbf{r}+\hat{x}} - t\,\psi_{3,\textbf{r}}^\dagger \psi_{3,\textbf{r}+\hat{x}} + t\,\psi_{1,\textbf{r}}^\dagger \psi_{1,\textbf{r}+\hat{y}} - t\,\psi_{2,\textbf{r}}^\dagger \psi_{2,\textbf{r}+\hat{y}} - t\,\psi_{3,\textbf{r}}^\dagger \psi_{3,\textbf{r}+\hat{y}}  \nonumber\\
& & \hspace{0.5 cm} + t\,\psi_{1,\textbf{r}}^\dagger \psi_{1,\textbf{r}+\hat{z}} - t\,\psi_{2,\textbf{r}}^\dagger \psi_{2,\textbf{r}+\hat{z}} - t\,\psi_{3,\textbf{r}}^\dagger \psi_{3,\textbf{r}+\hat{z}} - \mu_1 \psi_{1,\textbf{r}}^\dagger \psi_{1,\textbf{r}}  - \mu_2 \psi_{2,\textbf{r}}^\dagger \psi_{2,\textbf{r}}  - \mu_3 \psi_{3,\textbf{r}}^\dagger \psi_{3,\textbf{r}}   \nonumber\\
& & \hspace{0.5 cm}+ a\, \psi^\dagger_{1,\textbf{r}}\psi_{2,\textbf{r}+\hat{x}} + a\, \psi^\dagger_{2,\textbf{r}}\psi_{1,\textbf{r}+\hat{x}} - a\, \psi^\dagger_{1,\textbf{r}}\psi_{2,\textbf{r}+\hat{y}} - a\, \psi^\dagger_{2,\textbf{r}}\psi_{1,\textbf{r}+\hat{y}}\nonumber\\
& & \hspace{0.5 cm}+ b\, \psi^\dagger_{2,\textbf{r}}\psi_{3,\textbf{r}+\hat{x}} + b\, \psi^\dagger_{3,\textbf{r}}\psi_{2,\textbf{r}+\hat{x}} - b\, \psi^\dagger_{2,\textbf{r}}\psi_{3,\textbf{r}+\hat{y}} - b\, \psi^\dagger_{3,\textbf{r}}\psi_{2,\textbf{r}+\hat{y}}\nonumber\\
& & \hspace{0.5 cm} + c\, \psi^\dagger_{1,\textbf{r}}\psi_{3,\textbf{r}+\hat{x}} + c\, \psi^\dagger_{3,\textbf{r}}\psi_{1,\textbf{r}+\hat{x}}+c\, \psi^\dagger_{1,\textbf{r}}\psi_{3,\textbf{r}+\hat{y}}+c\, \psi^\dagger_{3,\textbf{r}}\psi_{1,\textbf{r}+\hat{y}}+ c\, \psi^\dagger_{1,\textbf{r}}\psi_{3,\textbf{r}+\hat{z}}+ c\, \psi^\dagger_{3,\textbf{r}}\psi_{1,\textbf{r}+\hat{z}} + h.c.\nonumber \\
& & +\Delta\sum_{\textbf{r}}\sum_{\alpha=1}^3 i\psi_{\alpha,\textbf{r}}\psi_{\alpha,\textbf{r}+\hat{x}} + \psi_{\alpha,\textbf{r}}\sigma^z\psi_{\alpha,\textbf{r}+\hat{y}}+ \psi_{\alpha,\textbf{r}}\sigma^x\psi_{\alpha,\textbf{r}+\hat{z}}+ h.c,
\end{eqnarray}
where $\hat{x}$, $\hat{y}$ and $\hat{z}$ denote unit vectors in respectively the $x$, $y$ and $z$ directions. The parameters $t,\mu_1,\mu_2,\mu_3,a,b,c$ and $\Delta$ are all real, which is required for the Hamiltonian to be invariant under time reversal. When $a=b=c=0$, $\hat{H}$ can be written as $\hat{H}=\hat{H}_1+\hat{H}_2+\hat{H}_3$, where $\hat{H}_\alpha$ is defined using the annihilation operators $\psi_{\alpha,\textbf{r}}$. We choose the parameters $t=\Delta=1$ and $\mu_1=-\mu_2=-\mu_3=2$ such that every $\hat{H}_\alpha$ realizes a strong TSC, with $\hat{H}_1$ and $\hat{H}_2$ having opposite strong invariants. When $a,b$ and $c$ are non-zero, the three TSCs $\hat{H}_\alpha$ are coupled in a way that respects $C_4$ and time reversal. The result is a superconductor with hybrid higher order topology, as detailed in the main text.

\end{document}